\documentclass[aps,10pt,prc,tightenlines,twocolumn,twoside,showpacs,nofootinbib,superscriptaddress]{revtex4-1}

\usepackage{amsmath,amssymb,amsthm,slashed}   
\usepackage{hyperref}  
\usepackage{graphicx,color}
\usepackage{verbatim}  
\usepackage{tabularx,multirow,makecell} 
\usepackage{soul}   
\usepackage{siunitx}  

\newcommand{\eg}{\textit{e.g.}}  

\newcommand{\ie}{\textit{i.e.}} 
\newcommand{\etc}{\textit{etc.}}
\newcommand{\bra}[1]{\langle\,#1\,|}          
\newcommand{\ket}[1]{|\,#1\,\rangle}          
\newcommand{\braket}[2]{\bra{#1}\,#2\rangle} 
\newcommand{\rmd}{\mathrm{d}}     
\newcommand{\fig}[1]{Fig.~#1}
\newcommand{\figs}[1]{Figs.~#1}

\newcommand{\eq}[1]{Eq.~(#1)}
\newcommand{\eqs}[1]{Eqs.~(#1)}
\newcommand{\paper}[1]{Ref.~#1}
\newcommand{\papers}[1]{Refs.~#1}

\newcommand{\LCm}{{\scriptscriptstyle -}} 
\newcommand{\LCp}{{\scriptscriptstyle +}}

\newcommand{\LCperp}{{\scriptscriptstyle \perp}}

\def\be{\begin{eqnarray}}
\def\ee{\end{eqnarray}}

\begin{document}
\title{	
Transverse structure of electron in momentum space in basis light-front quantization}

\author{Zhi Hu}
\email{huzhi@impcas.ac.cn}
\affiliation{Institute of Modern Physics, Chinese Academy of Sciences, Lanzhou, Gansu, 730000, China}
\affiliation{School of Nuclear Physics, University of Chinese Academy of Sciences, Beijing, 100049, China}
\affiliation{CAS Key Laboratory of High Precision Nuclear Spectroscopy, Institute of Modern Physics, Chinese Academy of Sciences, Lanzhou 730000, China}

\author{Siqi Xu} 
\email{xsq234@impcas.ac.cn}
\affiliation{Institute of Modern Physics, Chinese Academy of Sciences, Lanzhou, Gansu, 730000, China}
\affiliation{School of Nuclear Physics, University of Chinese Academy of Sciences, Beijing, 100049, China}
\affiliation{CAS Key Laboratory of High Precision Nuclear Spectroscopy, Institute of Modern Physics, Chinese Academy of Sciences, Lanzhou 730000, China}

\author{Chandan Mondal} 
\email{mondal@impcas.ac.cn}
\affiliation{Institute of Modern Physics, Chinese Academy of Sciences, Lanzhou, Gansu, 730000, China}
\affiliation{School of Nuclear Physics, University of Chinese Academy of Sciences, Beijing, 100049, China}
\affiliation{CAS Key Laboratory of High Precision Nuclear Spectroscopy, Institute of Modern Physics, Chinese Academy of Sciences, Lanzhou 730000, China}

\author{Xingbo Zhao}
\email{xbzhao@impcas.ac.cn}
\affiliation{Institute of Modern Physics, Chinese Academy of Sciences, Lanzhou, Gansu, 730000, China}
\affiliation{School of Nuclear Physics, University of Chinese Academy of Sciences, Beijing, 100049, China}
\affiliation{CAS Key Laboratory of High Precision Nuclear Spectroscopy, Institute of Modern Physics, Chinese Academy of Sciences, Lanzhou 730000, China}

\author{James P. Vary}
\email{jvary@iastate.edu}
\affiliation{Department of Physics and Astronomy, Iowa State University, Ames, IA 50011, USA}

\date{\today}

\begin{abstract}
    We investigate the leading-twist transverse momentum-dependent distribution functions (TMDs) for a physical electron, a spin-$1/2$ composite system consisting of a bare electron and a photon, using the Basis Light-front Quantization (BLFQ) framework. The light-front wave functions of the physical electron are obtained from the eigenvectors of the light-front QED Hamiltonian. We evaluate the TMDs using the overlaps of the light-front wave functions. The BLFQ results are found to be in excellent agreement with those TMDs calculated using lowest-order perturbation theory.
\end{abstract}
\maketitle

\section{Introduction}
A key tool for studying hadron structure is the deep inelastic scattering (DIS) process, where
individual partons are resolved. One can extract the Parton Distribution Functions (PDFs)~\cite{Collins:1981uw, Martin:1998sq, Gluck:1994uf, Gluck:1998xa} from such experiments. PDFs encode the distribution of longitudinal momenta and polarizations carried by the partons. Being functions of longitudinal momentum fraction $(x)$ only, they do not provide knowledge about the transverse spatial location and motion of the constituents inside the hadrons. On the other hand, the Transverse Momentum-dependent Parton Distribution Functions (TMDs)~\cite{Angeles-Martinez:2015sea,Collins:2011ca,Pasquini:2012jm} provide essential information about the distributions of both the longitudinal momentum fraction $ (x) $ and the relative transverse momentum $(k^\LCperp)$ as well as the orbital motion of partons inside hadrons, allowing us to draw three-dimensional pictures of the hadrons. They appear in the description of semi-inclusive reactions like the semi-inclusive deep inelastic scattering (SIDIS)~\cite{Brodsky:2002cx,Bacchetta:2017gcc} and Drell-Yan process~\cite{Ralston:1979ys,Donohue:1980tn,Tangerman:1994eh,Zhou:2009jm,Collins:2002kn}. Quantum Chromodynamics (QCD) factorization theorems allow one to relate physical observables, such as cross sections, to TMDs via perturbatively calculable kernels. Many theoretical studies have concentrated on the formalism and the accompanying factorization theorem of the TMDs~\cite{Collins1984TMD, Collins1982factorization, Catani1990gluon, Catani1990smallx, Mulders1995treelevel,Ji:2004wu}. 

At leading-twist, there are eight TMDs for a composite spin-$ 1/2 $ system. They are characterized by different spin-spin and spin-orbit correlations of the partons and the system. Three of them are generalizations of the three familiar collinear PDFs, namely the unpolarized $f_1(x)$, helicity $g_1(x)$, and the transversity $h_1(x)$ distributions. TMDs reduce to collinear PDFs by integrating over the transverse momenta. Basically, TMDs are the extended version of collinear PDFs and one can extract the parton densities for various configurations of the target and the parton polarizations from TMDs. TMDs are very useful to explain a range of phenomena even when we only consider the leading-twist factorization. For example, one can explain the Collins asymmetry using transversity TMD $h_1(x,k^{\perp})$~\cite{Anselmino:2007fs,Anselmino:2011ch,Anselmino:2013vqa}; one can explain the double spin asymmetry $ A_{LT} $ in SIDIS~\cite{Kotzinian:2006dw} using $ g_{1T}(x,k^\perp) $, which encodes the information of a longitudinally polarized parton in a transversely polarized hadron; one can employ the distribution corresponding to a transversely polarized parton in a transversely polarized hadron, $ h_{1T}^{\LCperp}(x,k^\perp) $, also known as pretzelosity TMD, to explain the $ A_{UT}^{\sin(3\phi_h-\phi_S)} $single spin asymmetry~\cite{Lefky:2014eia}.

TMDs being non-perturbative in nature are always difficult to obtain from the first principles of QCD and thus, they have been investigated in several QCD inspired models. Examples include the diquark spectator model~\cite{Jakob:1997wg,Bacchetta:2008af}, MIT bag model~\cite{Avakian:2010br,Cherednikov:2006zn,Yuan:2003wk}, covariant parton model~\cite{Efremov:2009ze}, light-cone constituent quark model~\cite{Pasquini:2008ax} and light-front quark-diquark model with AdS/QCD predicted wave functions~\cite{Maji:2015vsa,Maji:2017bcz}. Meanwhile, promising theoretical frameworks for illuminating TMDs also include the Dyson–Schwinger equations approach~\cite{Shi:2018zqd,Shi:2020pqe} and the discretized space-time euclidean lattice~\cite{Musch:2010ka,Musch:2011er,Ji:2014hxa,Yoon:2017qzo,Lin:2020rut}. However, both the Dyson-Schwinger equations and the lattice QCD approaches are mostly working in the euclidean spacetime and therefore encounter challenges in accessing TMDs directly.

In recent years, Basis Light-front Quantization (BLFQ)~\cite{Vary:2009gt,Zhao:2014hpa} has been developed as a non-perturbative approach which is very useful for studying the structure of relativistic bound states~\cite{zhao2014electrong2,Wiecki:2014ola,LY2016,Jia:2018ary,Xu:2019xhk,Lan:2019vui}. Starting with the light-front Hamiltonian obtained from the Lagrangian via the Legendre transformation, we solve for the light-front wave functions (LFWFs). Once we have the LFWFs, we are able to investigate several properties of the physical system, including the anomalous magnetic moment and GPDs of electrons~\cite{zhao2014electrong2,Chakrabarti2014GPDinBLFQ}, internal structure of baryons~\cite{Xu:2019xhk} and mesons~\cite{Lan:2019img,Lan:2019vui}, as well as scattering phenomena~\cite{xbzhao2013tBLFQ,Zhao:2013jia,Hu:2019hjx}.

In Quantum Electrodynamics (QED), while the electron is treated as an elementary field, it can fluctuate into an electron-photon pair, \eg{}, $e\to e\gamma \to e$, with the same quantum number. The virtual photon can further break up into virtual electron and positron pairs with all possible combinations. Thus, an electron is no longer an isolated particle but is surrounded by virtual clouds of electrons, positrons and photons.  In analogy to the partonic structure of hadrons, one can interpret them as partons contained in the physical electron. TMDs of the physical electron, as a system composed of a bare electron and a bare photon, have been investigated in Ref.~\cite{bacchetta2016electron} using the lowest-order perturbation theory. Many other observables of the electron, \eg{}, the electromagnetic and gravitational form factors as well as spin and orbital angular momentum~\cite{Brodsky2000lightcone}, generalized parton distributions (GPDs)~\cite{Brodsky:2000xy,Brodsky:2006ku,Dahiya:2007is,Kumar:2015fta},Wigner distributions~\cite{Kumar:2017xcm}, \etc{} have been investigated in a model based on the quantum fluctuations of the electron in QED.

In this paper, we employ BLFQ to obtain TMDs of the physical spin-$1/2$ particle in QED, \ie{} electron. Only the leading two Fock sectors ($\ket{e}$ and $\ket{e\gamma}$) are considered. During the diagonalization of the Hamiltonian, we use a mass counter-term in the first Fock sector to renormalize the mass of the system to that of the physical electron and we obtain the LFWFs of the ground state for the physical electron. It is reasonable to treat the perturbative results as equivalent to experimental data in QED. We then compare the BLFQ results with the perturbative results reported in \paper{\cite{bacchetta2016electron}}. It is our aim to demonstrate that the BLFQ approach, which is inherently non-perturbative, provides results, up to finite basis artifacts, that agree with the perturbative results. We also aim to show how we may mitigate the finite basis artifacts by achieving convergence of suitably smoothed BLFQ results. In achieving these aims, we establish the methods to be used in applications that require a non-perturbative approach.

For the purposes of this work, we set the gauge link generating from infinite many photon exchanges in the final-state interaction~\cite{jxd2002where,Belitsky:2002sm} to unity. This leaves us five non-zero TMDs out of the eight leading-twist TMDs.  We also present spin-spin corrections between the bare electron and the physical electron for different polarization configurations. A detailed analysis of the TMDs with a nontrivial gauge link in the BLFQ approach will be reported in future studies.

The paper is organized as follows. We introduce the basics of the BLFQ approach in section~\ref{sec_BLFQ}. A discussion of mass renormalization is also given in this section. In section~\ref{sec_TMD}, the TMDs of electrons are evaluated. The numerical results of the TMDs and the spin-densities in the transverse momentum space for different spin configurations of the bare and the physical electrons are presented and discussed in section~\ref{sec_numerical}. The summary is given in section~\ref{sec_summary}.

\section{BASIS LIGHT-FRONT QUANTIZATION}
\label{sec_BLFQ}
BLFQ is based on the Hamiltonian formalism and adopts light-front quantization~\cite{brodsky1998treatise}. The mass spectrum and light-front Fock state wave functions are obtained from the solution of the following eigenvalue equation:
\begin{gather}
    \hat{H}|P,\Lambda\rangle= M^2|P,\Lambda\rangle\, , \label{eq:Heff_eigen_value} 
\end{gather}
where the Fock space representation of the system state with momentum $P$ and light-front helicity $ \Lambda $, $|P,\Lambda\rangle$ is expanded in terms of multi-particle light-front basis states as~\cite{Pasquini:2012jm},
\begin{widetext}
\begin{align}
  \label{eq:physical_lfwf}
| P,\Lambda \rangle = \sum_{n} \sum_{\lambda_1\dots \lambda_n}\int \prod_{i=1}^n \left[{dx_i d^2 k^{\LCperp}_{i} \over
\sqrt{x_i} 16 \pi^3} \right] 16 \pi^3 \delta\left(1- \sum_{i=1}^n x_i\right) \delta^2
\left(\sum_{i=1}^n k^{\LCperp}_{i}\right)\psi^{\Lambda}_{\lambda_1\dots \lambda_n}(\{x_i, k^{\LCperp}_{i}\})
\mid n, x_i P^+, x_i P^{\LCperp} + k^{\LCperp}_{i}, \lambda_i \rangle.
\end{align}
\end{widetext}
Here $x_i=k_i^+/P^+$ is the longitudinal fraction, $k^{\LCperp}_{i}$ is the relative transverse momentum and $\lambda_i$ is the light-front helicity, both of the $i$-th constituent. $n$ represents the number of particles in a Fock state. The physical transverse momenta of the constituents are $p^{\LCperp}_{i} = x_i P^{\LCperp} + k^{\LCperp}_{i}$ and the physical longitudinal momenta of the constituents $ p_i^+=k_i^+=x_i P^+ $. The boost invariant light-front wave functions $\psi^{\Lambda}_{\lambda_1\dots \lambda_n}$ are independent of the total momentum of the state, $P^+$ and $P^\LCperp$ and depend only on $x_i$ and  $k^{\LCperp}_{i}$. 

Like in QCD, we treat the physical electron as a composite particle, where we could find the bare electrons, positrons and photons as its partons. In this work, we restrict ourselves to only the first two Fock sectors:
\begin{align}
    \label{ephysical}
    \ket{e_{\mathrm{phy}}}=\ket{e}+\ket{e\gamma}\, .
\end{align}

The full Hamiltonian we diagonalize is
\begin{align}
    \label{htotal}
    \hat H= \hat H_{\mathrm{QED}}+\hat H^{\prime}\, ,
\end{align}
where the light-front\footnote{Here, we follow the convention for light-front four-vector $ v=(v^{\LCp},v^{\LCm},v^{\LCperp}) $, where $ v^{\LCp}=v^0+v^3 $, $ v^{\LCm}=v^0-v^3 $ and $ v^{\LCperp}=(v^1, v^2) $.} QED Hamiltonian $ \hat H_{\mathrm{QED}}=P^{\LCp}\hat P_{\mathrm{QED}}^{\LCm}- (\hat P^{\LCperp})^2 $. We obtain $ \hat P_{\mathrm{QED}}^{\LCm} $ from the QED Lagrangian~\cite{kogut&soper1970QED_infi_mom, brodsky1998treatise} through the Legendre transform and adopting the light-cone gauge $ A^{\LCp}=0 $. Restricted to the leading two Fock sectors, $ \hat P_{\mathrm{QED}}^{\LCm} $ is given by~\cite{xbzhao2013tBLFQ,zhao2014electrong2}
\begin{align}
    \label{QED_Pminus}
    P_{\mathrm{QED}}^{\LCm} &= \int\rmd^2x^{\LCperp}\rmd x^\LCm \; \left[ \frac{1}{2}\bar{{\Psi}} \gamma^\LCp \frac{m_e^2+(i\partial^{\LCperp})^2}{i\partial^\LCp}\Psi \right.\\ \nonumber
    &\left. + \frac{1}{2} { A}^j (i\partial^\LCperp)^2 { A}^j  + e{j}^\mu {A}_\mu  \right] \,,
\end{align}
with the electron mass $m_e$ and the physical electromagnetic coupling constant $e$. $\Psi$ and $A_\mu$ are the fermion and the gauge boson fields, respectively.

In the single-particle coordinate, $ \hat H_{\mathrm{QED}}$ incorporates the transverse center-of-mass (CM) motion, which is not necessarily zero. The advantage of using the single-particle coordinate is that each particle in the Fock space can be treated on equal footing, and this facilitates dealing with (anti-) symmetry among identical particles when higher Fock sectors are taken into account~\cite{zhao2014electrong2}. The disadvantage is that the CM motion of the system is mixed up with intrinsic motion.

In order to overcome this disadvantage, we add a constraint term 
\begin{align}
    \label{Hprime}
    \hat H^{\prime}=\lambda_{L} (\hat H_{CM}-2b^2\hat I)\, ,
\end{align}
to the light-front QED Hamiltonian to factorize out the transverse CM motion from the intrinsic motion, where the CM motion is then governed by a harmonic oscillator
\begin{align}
    \hat H_{CM}=\left(\sum_i \hat p^{\LCperp}_i\right)^2+b^4\left(\sum_i x_i \hat r^{\LCperp}_i\right)^2\,.
\end{align}
Moreover, by subtracting the zero-point energy $ 2b^2 $ and multiplying by a Lagrange multiplier $ \lambda_{L} $, we are able to shift the excited states of CM motion to higher energy and retain low-lying states that contain only the simplest (gaussian) state of CM motion times the intrinsic motion. A detailed discussion of the CM factorization within the BLFQ framework can be found in \papers{\cite{maris2013bound,Wiecki:2014ola}}.

The basis state used in BLFQ for a single particle is the direct product of the momentum eigenstate in the longitudinal direction, the two-dimensional harmonic oscillator basis state in the transverse direction and the helicity eigenstate. This choice is motivated by the success of \textit{ab initio} calculation of nuclei structure~\cite{vary2000NCFC, vary2009NCFC} and also holographic QCD~\cite{LY2016,teramond2009holography}. 

In the longitudinal direction, we impose (anti) periodic boundary condition on (fermions) bosons inside a box of length $ 2L $, resulting in
\begin{align}
    p^{\LCp}_i=\frac{2\pi}{L}k_i\,,
\end{align}
where $ k_i $ is an integer (boson) or half-integer (fermion)\footnote {We neglect the zero mode for bosons. For convenience we subtract $ \frac{1}{2} $ from state labels $ k_i $ for fermions so that the longitudinal quantum number of all particles and their sum $ K $ will be designated as integers.}. In the transverse direction, the basis states are the eigenstates of the two dimensional harmonic oscillator (2D-HO) Hamiltonian with two quantum numbers $ n_i,m_i $ and corresponding eigenenergies $ 2(2n_i+|m_i|+1)b^2 $~\cite{maris2013bound}, where $b$ is the HO scale parameter. Here $n_i$ represents the radial quantum number and $m_i$ the quantum number of orbital motion in the transverse plane. In the momentum space, the transverse basis states are given by
\begin{align}
    \phi_{n m}(p^{\LCperp})=\frac{1}{b}\sqrt{\frac{4\pi \times n!}{(n+|m|)!}}e^{im \theta}L_{n}^{|m|}(\rho^2)\rho^{|m|}e^{-\rho^2/2}\, .
\end{align}
Here, $ \theta=\mathrm{arg}(p^{\LCperp}) $, $ \rho=|p^{\LCperp}|/b $ and $ L_{n}^{|m|} $ is the associated Laguerre function. Each single-particle basis state is identified using four quantum numbers:
\begin{align}
    \alpha_i=( k_i,n_i,m_i,\lambda_i )\,,
\end{align}
where $ \lambda_i $ represents the light-front helicity~\cite{soper1972infi_mom_helicity} of the particle.

The basis states of a Fock sector with multiple particles are then expressed as the direct product of each single-particle basis state in the Fock sector
\begin{align}
    \ket{\alpha}=\otimes_i \ket{\alpha_i}\, .
\end{align}
All the basis states have a well defined value of the total angular momentum projection
\begin{align}
    \label{spin_constraint}
    \sum_i \; \left(\lambda_i+m_i\right)=\Lambda.
\end{align}
This follows from the commutation between $ \hat{J_Z} $ and $ \hat P_{QED}^{\LCm} $\footnote{ See \paper{\cite{kogut&soper1970QED_infi_mom}} for detailed discussion about the symmetry in the light-front dynamics.}. In addition, they all have the same total longitudinal momentum $P^+=\sum_ip_i^+$. One then parameterizes $P^+$ using a dimensionless variable $K=\sum_i k_i$ such that $P^+=\frac{2\pi}{L}K$. For a given particle $i$, the longitudinal momentum fraction is then represented as $x_i=p_i^+/P^+=k_i/K$.

Another element in reducing the infinite-dimensional basis to a finite-dimensional basis is the truncation of quantum numbers in the transverse direction. We retain the basis states satisfying:
\begin{align}
    \label{transverse_truncation}
    \sum_i(2n_i+|m_i|+1)\leq N_{\mathrm{max}}\,.
\end{align}
Note that $N_{\mathrm{max}}$ not only restricts the HO quanta, but also introduces an IR ($ \sim b/\sqrt{N_{\mathrm{max}}} $) and UV ($ \sim b\sqrt{N_{\mathrm{max}}} $) cutoff in the transverse direction.~\cite{xbzhao2013tBLFQ}

Overall, there are three basis parameters in our calculation, the scale of the 2D-HO $ b $, the transverse truncation $ N_{\mathrm{max}} $, and the longitudinal resolution $ K $. We will discuss the dependence of our results upon these parameters while discussing the numerical results in section \ref{sec_numerical}.

After solving \eq{\ref{eq:Heff_eigen_value}} in the BLFQ basis under the Fock sector truncation, \eq{\ref{ephysical}}, we get the single-particle coordinate LFWF in the momentum space for the second Fock sector as
\begin{gather}
    \begin{aligned}
    \label{lfwf_momentum}
    \psi^{\Lambda}_{\lambda_e, \lambda_{\gamma}}(x_e,p_e^{\LCperp},x_\gamma,p_\gamma^{\LCperp})&=\sum_{\substack{n_e,m_e\\n_\gamma,m_\gamma}}\left[\psi(\alpha_e,\alpha_\gamma)\times\right. \\
    &\left.\phi_{n_e m_e}(p_e^{\LCperp})\,\phi_{n_\gamma m_\gamma}(p_\gamma^{\LCperp})\right] \, .
    \end{aligned}
\end{gather}
where $ \psi(\alpha_e,\alpha_\gamma)=\braket{\alpha_e,\alpha_\gamma}{P,\Lambda} $ are the eigenvectors obtained by diagonalizing the Hamiltonian \eq{\ref{htotal}}. We then convert the non-perturbative solutions in the single-particle coordinates, as in \eq{\ref{lfwf_momentum}}, to that in the relative coordinates, $ \psi^{\Lambda}_{\lambda_e, \lambda_{\gamma}}(x_e,k_e^{\LCperp},x_\gamma,k_\gamma^{\LCperp}) $ in \eq{\ref{eq:physical_lfwf}}, by factorizing out the CM motion from \eq{\ref{lfwf_momentum}}~\cite{maris2013bound,Wiecki:2014ola,Moshinsky:1959qbh}.

\subsection{$ \Delta m_e $ \lowercase{and} $ Z_2 $}
As mentioned before, we need to perform mass renormalization during the diagonalization and we employ a Fock sector dependent renormalization procedure~\cite{Karmanov2008systematic, Karmanov2012abinitio}. The electron mass in the first Fock sector equals the bare electron mass, $m_e^{e}=m_0$, and it depends on the regulators (cut-off) in the renormalization process. We numerically diagonalize the Hamiltonian matrix in an iterative scheme, where we adjust $ m_e^{e} $ in the first Fock sector $\ket{e}$ to obtain the mass of the physical electron $M_e=0.511\, \mathrm{MeV}$ as the mass of the entire system. Since in our Fock space truncation, Eq.~(\ref{ephysical}), the photon cannot fluctuate into electron-positron pairs, for basis states in $|e\gamma\rangle$ sector the electron mass remains the same as the physical value, $ m_e^{e\gamma}=M_e $.

\begin{figure}[!h]
    \includegraphics[width=0.45\textwidth]{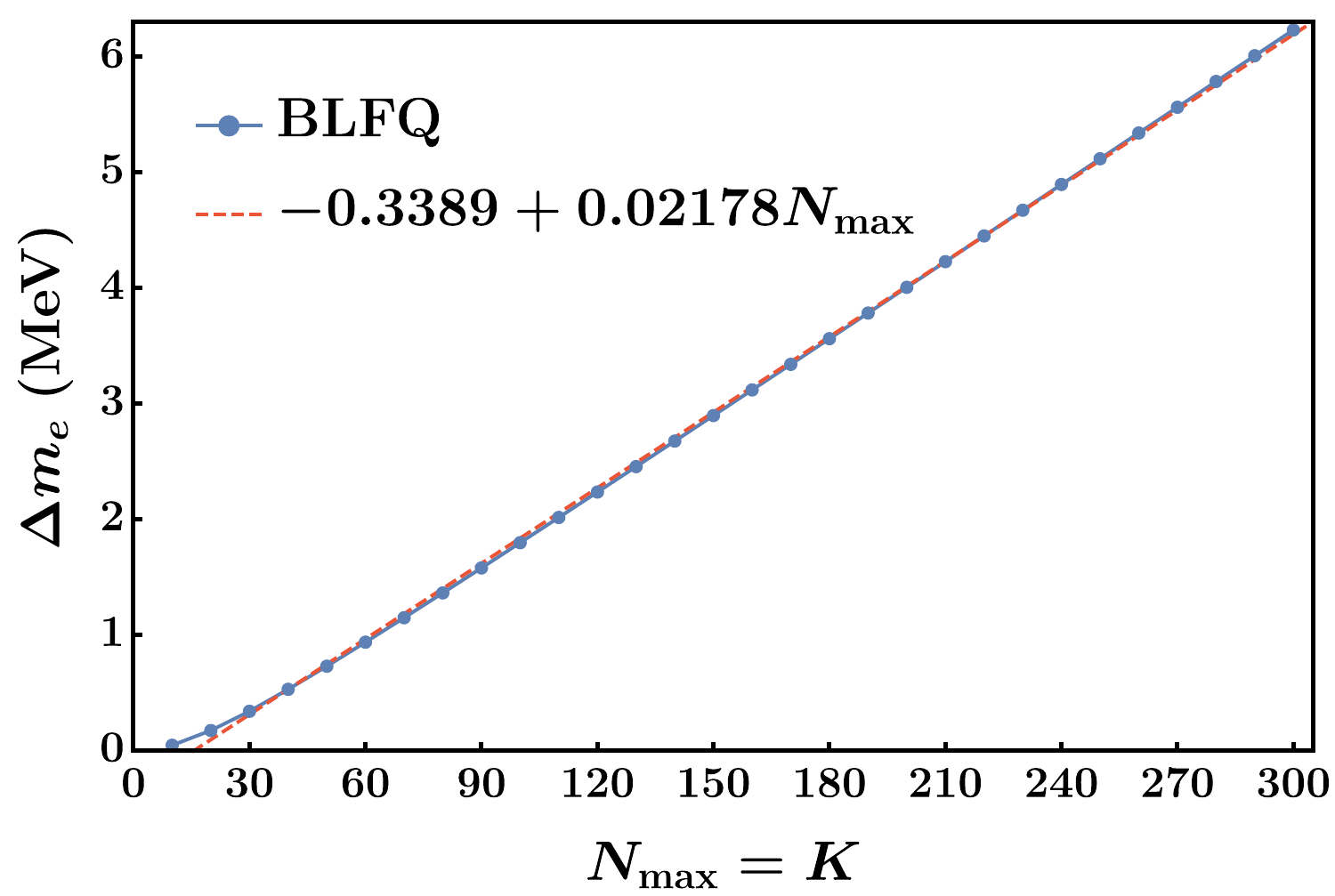}
    \caption{\label{plotdeltame}(color online) The mass counter term $ \Delta m_e=m_e-M_e$ as a function of the basis truncation parameter in BLFQ. All the results are evaluated at $ K=N_{\textrm{max}} $ and $ b=M_e $. The solid (blue) line with dots represents the BLFQ results, while the dashed (red) line corresponds to the linear fitting of the results: $\Delta m_e=-0.339+0.0218~ N_{\rm max}$.}
\end{figure}

We introduce the mass counter-term $ \Delta m_e=m_0-M_e $ that represents the mass correction due to the quantum fluctuations to higher Fock sectors. According to the light-front perturbation calculation~\cite{brodsky1998treatise}, $ \Delta m_e$ should increase with increasing basis truncation parameters, which leads to larger UV cutoff. This is confirmed in \fig{\ref{plotdeltame}}.

\begin{figure}[!h]
    \includegraphics[width=0.45\textwidth]{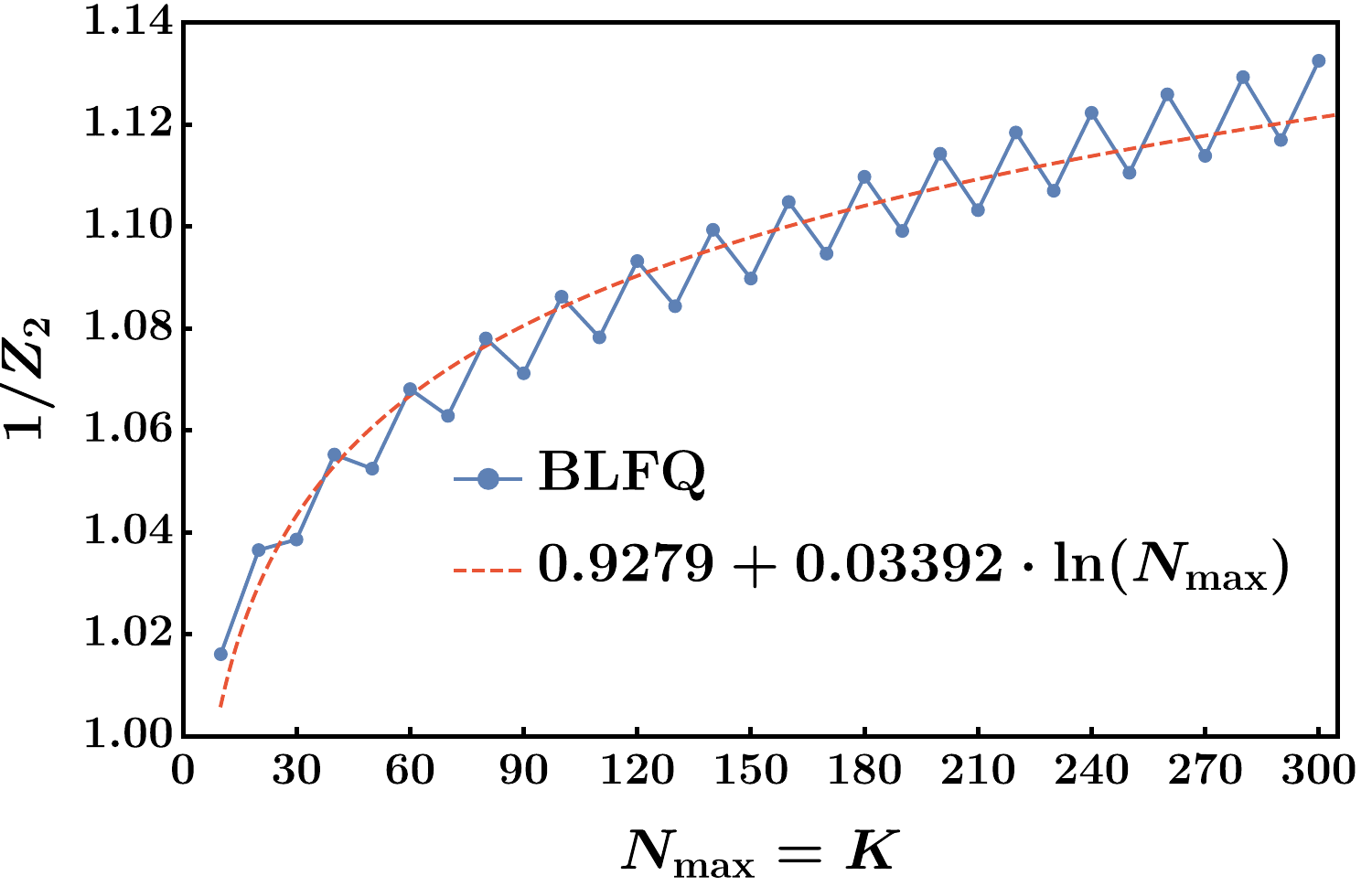}
    \caption{\label{plotz2}(color online) The electron wave function renormalization factor $ Z_2$, as a function of the basis truncation parameters in BLFQ. All the results are evaluated at $ K=N_{\textrm{max}} $ and $ b=M_e $. The dots connected by solid (blue) line segments represent the BLFQ results, while the dashed (red) line corresponds to the logarithm fitting of the results: $1/Z_2=0.9279+0.03392\ln(N_{\rm max})$.}
\end{figure}

Furthermore, following the discussions in~\papers{\cite{Zhao:2014hpa}, \cite{zhao2014electrong2}, \cite{Chakrabarti2014GPDinBLFQ}, \cite{Brodsky2004nonperturbative}}, we rescale the naive TMDs by $ Z_2 $
\begin{gather}
    \langle\,\mathrm{TMD}\,\rangle=\frac{\langle\,\mathrm{TMD}\,\rangle^{\prime}}{Z_2} \, ,
\end{gather}
to correct an artifact on the normalization of the LFWF in the $ \ket{e\gamma} $ Fock sector, which is introduced by our Fock sector truncation. Here, prime designates that the results are calculated directly by using the LFWF from \eq{\ref{lfwf_momentum}}. $ Z_2 $ is the electron wave function renormalization factor which in our truncation can be interpreted as the probability of finding a bare electron within a physical electron:
\begin{align}
    Z_2=\sum_{e} |\braket{e}{e_{\mathrm{phy}}}|^2\, .
\end{align}
The summation runs over all the basis states in the $|e\rangle$ sector. $ Z_2 $ incorporates the contribution from the quantum fluctuation between the $|e\rangle$ and $|e\gamma\rangle$ sectors and goes to zero in the infinite basis limit. We show $ 1/Z_2 $ as a function of basis truncation parameters in \fig{\ref{plotz2}}. We comment that such sawtooth patterns are a familiar odd-even effect in the BLFQ formalism~\cite{zhao2014electrong2}.

\section{TMD in QED}
\label{sec_TMD}
We treat the physical electron as a spin-$1/2$ composite particle, with a bare electron and a photon emerging from quantum fluctuation as its partons. The TMDs of a fermion inside the physical electron are defined through the fermion-fermion correlator function as\footnote{Our convention is different from that of \paper{\cite{bacchetta2016electron}} by a factor of $1/2$. This comes from the different convention for the light-front four vector.}
\begin{align}
    \label{TMD_cor_func}
    \Phi^{[\Gamma]}(x,k^{\LCperp};P,S)&=\frac{1}{2} \int \frac{\rmd z^{\LCm} \rmd^2 z^{\LCperp} }{2(2\pi)^3} \, e^{i k \cdot z}\\ \nonumber 
    &\bra{P,S} \bar{\Psi}(0)\mathcal{U}(0,z)\Gamma\Psi(z) \ket{P,S} |_{z^{\LCp}=0} \, ,
\end{align}
where $ x=\frac{k^{\LCp}}{P^{\LCp}} $, and $ k^{\LCperp} $ stands for the relative transverse momentum of the bare electron. $\Gamma$ is a Dirac $\gamma$-matrix. For the leading-twist TMDs, $\Gamma=\gamma^+$, $\gamma^+\gamma_5$ or $i\sigma^{j+}\gamma_5$, corresponding to unpolarized, longitudinally polarized or transversely polarized fermion inside the physical electron respectively. The gauge link $\mathcal{U}(0,z)$ guarantees the gauge invariance of this non-local operator. $\ket{P,S}$ ($\bra{P,S}$) stands for the initial (final) state of the physical electron with momentum $ P $ and spin four-vector $ S $.

We choose the light-cone gauge $A^+=0$. We work in a frame where the momentum and spin four-vector of the physical electron are represented as $ P=(P^{\LCp},\frac{M_e^2}{P^{\LCp}}, 0^{\LCperp}) $ and $ S=(S^3\frac{P^{\LCp}}{M_e},-S^3\frac{M_e}{P^{\LCp}}, S^{\LCperp})$. It is easy to verify that this covariant spin vector satisfies $ P\cdot S=0 $ and also $ S^2=-1 $ if we choose the three-dimensional spin vector to be
\begin{align}
    \vec{S}=(S^1, S^2, S^3)=(\sin\theta \cos\varphi, \sin\theta \sin\varphi, \cos\theta)\, .
\end{align}
The spin vector state can then be transformed from the light-front helicity state~\cite{soper1972infi_mom_helicity} using the following formulae~\cite{Pasquini:2012jm}
\begin{align}
    \label{veccortohelicor_matrix}
    \begin{pmatrix}
        \ket{P,S}  \\
        \ket{P,-S} \\
    \end{pmatrix}
    =
    \begin{pmatrix}
        \cos\frac{\theta}{2}e^{-i\frac{\varphi}{2}}  &  \sin\frac{\theta}{2}e^{i\frac{\varphi}{2}}  \\
        -\sin\frac{\theta}{2}e^{-i\frac{\varphi}{2}}  &  \cos\frac{\theta}{2}e^{i\frac{\varphi}{2}}  \\
    \end{pmatrix}
    \begin{pmatrix}
        \ket{P,+}  \\
        \ket{P,-} \\
    \end{pmatrix}\,.
\end{align}
This transformation is crucial because the eigenstates we obtain in BLFQ are the states associated with the light-front helicities $ \Lambda=\pm(1/2) $ rather than the spin vectors $ S $.

There exist eight leading-twist TMDs which are defined as
\begin{align}
    &\Phi^{[\gamma^{\LCp}]}(x,k^{\LCperp};P,S)=f_{1}^{e}-\frac{\epsilon^{ij}k^iS^j}{M_e}f_{1T}^{\LCperp e}\, , \label{veccortotmdplus}\\
    &\Phi^{[\gamma^{\LCp}\gamma^{5}]}(x,k^{\LCperp};P,S)=S^{3}g_{1L}^{e}+\frac{k^{\LCperp}\cdot S^{\LCperp}}{M_e}g_{1T}^{e} \, , \label{veccortotmdplus5} \\
    &\begin{aligned} \label{veccortotmdsigmaj}
    &\Phi^{[i\sigma^{j\LCp}\gamma^{5}]}(x,k^{\LCperp};P,S)=S^{j}h_{1}^{e}+S^{3}\frac{k^{j}}{M_e}h_{1L}^{\LCperp e}  \\
    &+ S^i \frac{2k^i k^j-(k^{\LCperp})^2\delta^{ij}}{2M_e^2}h_{1T}^{\LCperp e} + \frac{\epsilon^{ij}k^i}{M_e}h_1^{\LCperp e}\, .
    \end{aligned}
\end{align}
Here, $ i,j $ run in the transverse plane, and anti-symmetric tensor $ \epsilon^{12}=-\epsilon^{21}=1 $. However, if we set the gauge link to unity, two of them, the Boer-Mulders function $ h_1^{\LCperp e} $~\cite{boer1998todd} and the Sivers function $ f_{1T}^{\LCperp e} $~\cite{sivers1990ssa} reduce to zero because of their T-odd property. Furthermore, we find that $ h_{1T}^{\LCperp e} $ mainly comes from the D-wave component and the D-wave component is zero in our current Fock sector truncation. Thus, $ h_{1T}^{\LCperp e}$ is also zero. This is consistent with the result of the perturbative light-front calculation~\cite{bacchetta2016electron}.

Therefore, under the approximation that the gauge link is the identity operator and our current Fock sector truncation, there exist five non-zero TMDs at leading-twist. We are now able to evaluate those TMDs using the LFWF in the $ \ket{e\gamma} $ Fock sector ($ \psi^{\Lambda}_{\lambda_e, \lambda_{\gamma}}(x_e,k_e^{\LCperp},x_\gamma,k_\gamma^{\LCperp}) $ in \eq{\ref{eq:physical_lfwf}}) as~\cite{Pasquini:2012jm,bacchetta2016electron}
\begin{align}
    f_1^e&=\int [\rmd e\gamma]~\sum_{\lambda_{\gamma}} \left[ |\psi^{+}_{+,\lambda_{\gamma}}|^2 + |\psi^{+}_{-,\lambda_{\gamma}}|^2 \right] \, ,\label{f1e} \\
    g_{1L}^e&=\int [\rmd e\gamma]~\sum_{\lambda_{\gamma}} \left[ |\psi^{+}_{+,\lambda_{\gamma}}|^2 - |\psi^{+}_{-,\lambda_{\gamma}}|^2 \right] \, , \label{g1T}\\
    g_{1T}^e&=\frac{M_e}{(k^{\LCperp})^2}\int [\rmd e\gamma]~\sum_{\lambda_{\gamma}}  2\,{\rm Re} \left[ k^{\LCperp}_R~ \psi^{+\, *}_{+,\lambda_{\gamma}} \psi^{-}_{+,\lambda_{\gamma}} \right] , \label{g1L}\\
    h_1^e&=\int [\rmd e\gamma] \sum_{\lambda_{\gamma}} \left[ \psi^{+\, *}_{+,\lambda_{\gamma}} \psi^{-}_{-,\lambda_{\gamma}} \right] \, , \label{h1e}\\
    h_{1L}^{\LCperp e}&=\frac{M_e}{(k^{\LCperp})^2} \int [\rmd e\gamma] ~\sum_{\lambda_{\gamma}} 2\,{\rm Re} \left[ k^{\LCperp}_R ~\psi^{+\, *}_{-,\lambda{\gamma}} \psi^{+}_{+,\lambda{\gamma}} \right] \, , \label{h1L}
\end{align}
where $k^{\LCperp}_R=k^1+ik^2$ and we use the abbreviation
\begin{align}
    [\rmd e\gamma]&=\frac{\rmd x_e \rmd x_{\gamma} \rmd^2 k_e^{\LCperp} \rmd^2 k_{\gamma}^{\LCperp}}{16\pi^3} \delta(x_e+x_{\gamma}-1)\\ \nonumber
    &\times \delta^2(k_e^{\LCperp}+k_{\gamma}^{\LCperp})\delta(x-x_e)\delta^2(k^{\LCperp}-k_e^{\LCperp})\, .
\end{align}
We omit the argument of LFWFs ($\{x_i,k_i^{\LCperp}\}$) and TMDs ($x,k^{\LCperp}$) for the sake of conciseness.

The unpolarized TMD $f_1^e$ describes the distribution of an unpolarized bare electron in an unpolarized physical electron, while the helicity TMD $g_{1L}^e$ gives the information of the momentum distribution of a longitudinally polarized bare electron when the physical electron is also longitudinally polarized. $f_1^e$ and $g_{1L}^e$ are obtained by the sum of and the difference between the squares of S- and P-wave components under our Fock sector truncation, respectively. On the other hand, the transversity TMD $h_1^e$ describes the correlation between the transversely polarized bare electron and the transversely polarized physical electron. Being chiral-odd in nature, $h_1^e$ involves a helicity-flip of the bare electron from initial to final state, which is accompanied by a helicity flip of the physical electron in the same direction. This TMD receives contributions from the partial waves with $L_z=\pm1$. The TMD $g_{1T}^e$ is the distribution of a longitudinally polarized bare electron in a transversely polarized physical electron, whereas $h_{1L}^{\LCperp e}$ provides the momentum distribution for the transversely polarized bare electron in a longitudinally polarized physical electron.

Meanwhile, as a spin-$ 1/2 $ composite system, the three leading-twist PDFs of the physical electron can be retrieved by integrating $f_1^e(x,k^\LCperp)$, $g_{1L}^e(x,k^\LCperp)$, and $h_1^e(x,k^\LCperp)$ over $k^\LCperp$,
\begin{align}
    f_1^e(x)&=\int \rmd^2  k^\LCperp f_1^e(x,k^\LCperp)\, ,\label{f1} \\
    g_1^e(x)&=\int \rmd^2  k^\LCperp g_{1L}^e(x,k^\LCperp) \, , \label{g1}\\
    h_1^e(x)&=\int \rmd^2  k^\LCperp h_1^e(x,k^\LCperp) \, . \label{h1}
\end{align}

\section{Numerical Results}
\label{sec_numerical}
\subsection{Averaging Method and Convergence Behaviour}
\begin{figure*}
    \includegraphics[width=0.49\textwidth]{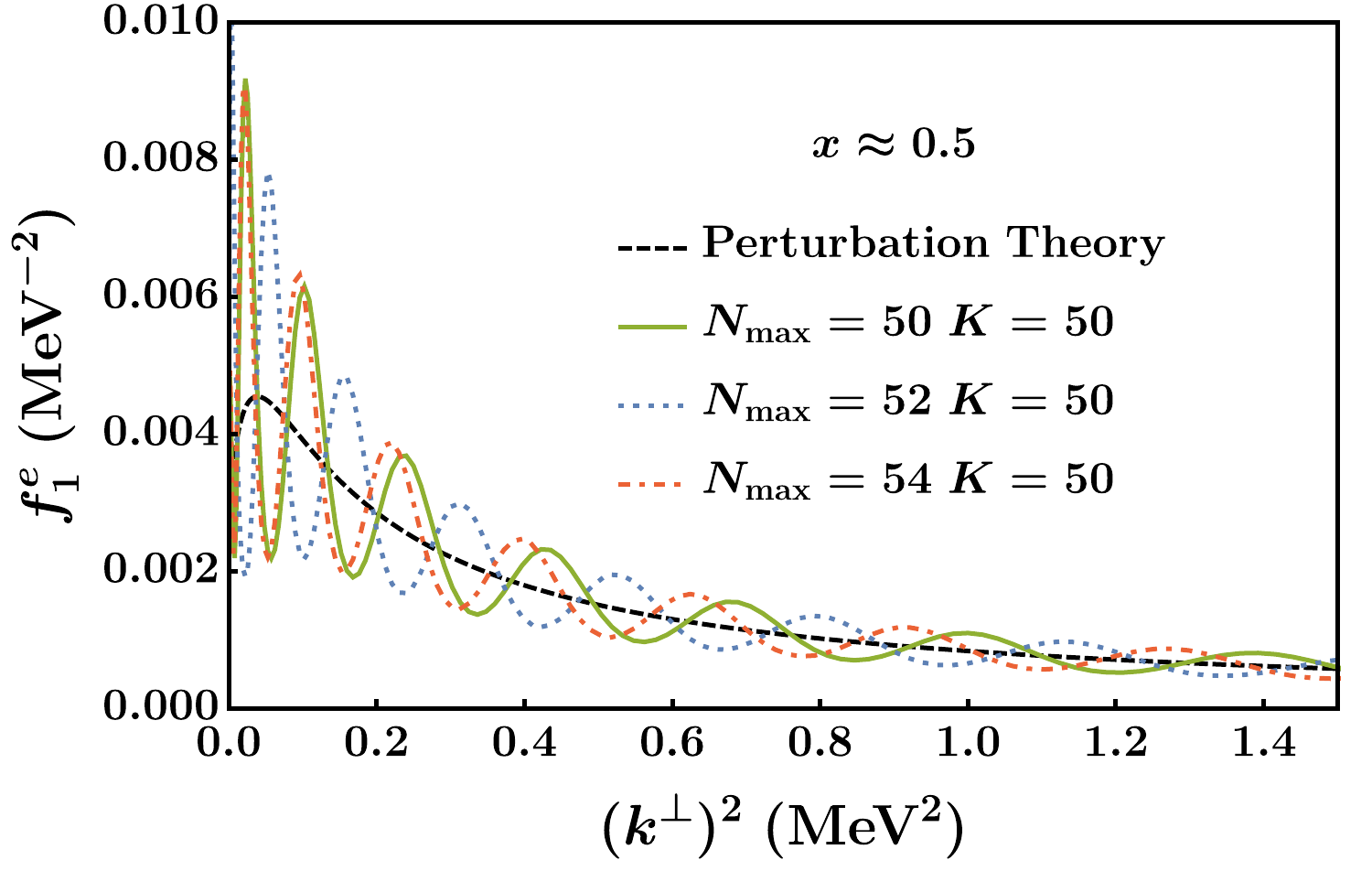}
    \includegraphics[width=0.49\textwidth]{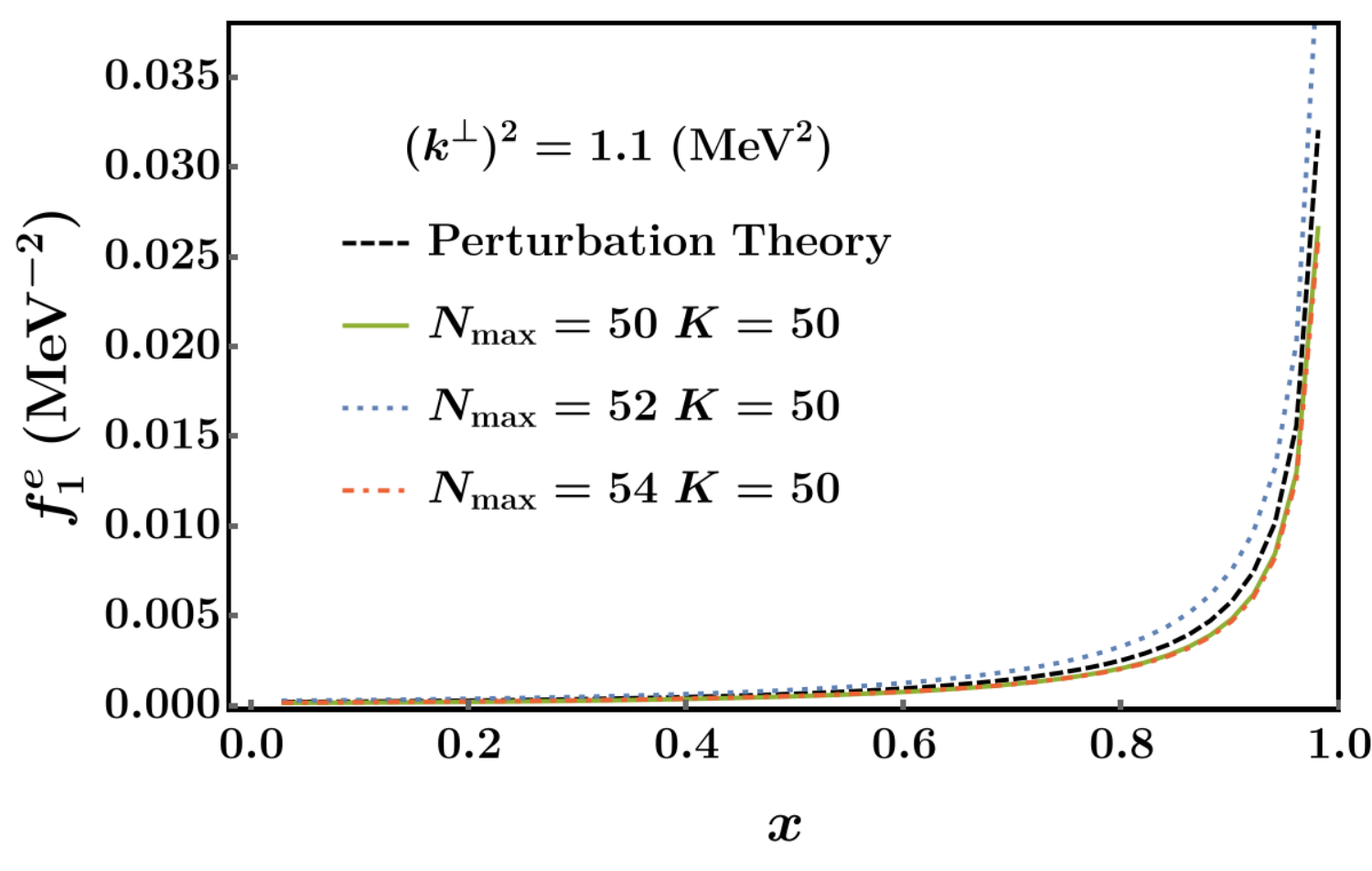}
    \caption{\label{plotN}(color online) The BLFQ computations at different truncation parameters and the perturbative results (black dashed lines)~\cite{bacchetta2016electron} for $ f_1^e $ in transverse direction (left) and longitudinal direction (right). We use $ f_1^e $ at $ x\approx 0.5 $ and at $ (k^{\LCperp})^2=1.1\,\mathrm{(MeV^2)} $ as examples. Curves with different dash styles (colors) are the BLFQ results at different $ N_{\textrm{max}} $ with fixed $ K=50 $ and $ b=M_e $.}
\end{figure*}

In \fig{\ref{plotN}}, we show the  TMD $ f_1^e(x,k^\LCperp) $ in $x$ and $ k^\LCperp $ directions for fixed values of  $ k^\LCperp $ and $x$ respectively. In these plots, the BLFQ computations are compared with those from the perturbative calculations in \paper{\cite{bacchetta2016electron}}. It can be noticed that in the transverse direction, the BLFQ results oscillate around the perturbative results, which can be viewed as a proxy for the experiment data in QED.

In this section we discuss the method we adopt to reduce the oscillation in the transverse direction \ie{} with respect to $ k^\LCperp $ and also study the convergence behaviour of our results as a function of the basis truncation parameters.

The oscillation in the TMD computed by the BLFQ method finds its root in the oscillating behaviour of the HO basis states that we use in the transverse plane. As mentioned in section~\ref{sec_BLFQ}, $ N_{\textrm{max}} $, the truncation parameter in the transverse direction, together with $ b $ (the scale of the transverse 2D-HO basis), determines our transverse basis states. This motivates us to average over the BLFQ results at different $ N_{\textrm{max}} $ and $ b $ to reduce these finite basis artifacts. We also have another truncation parameter $ K $ in the longitudinal direction. $ N_{\textrm{max}} $ and $ K $ are truncation parameters introduced by our calculation method, so their effects must be studied to differentiate between truncation artifacts and the underlying physics extracted from the BLFQ results.

\begin{figure*}
    \includegraphics[width=0.49\textwidth]{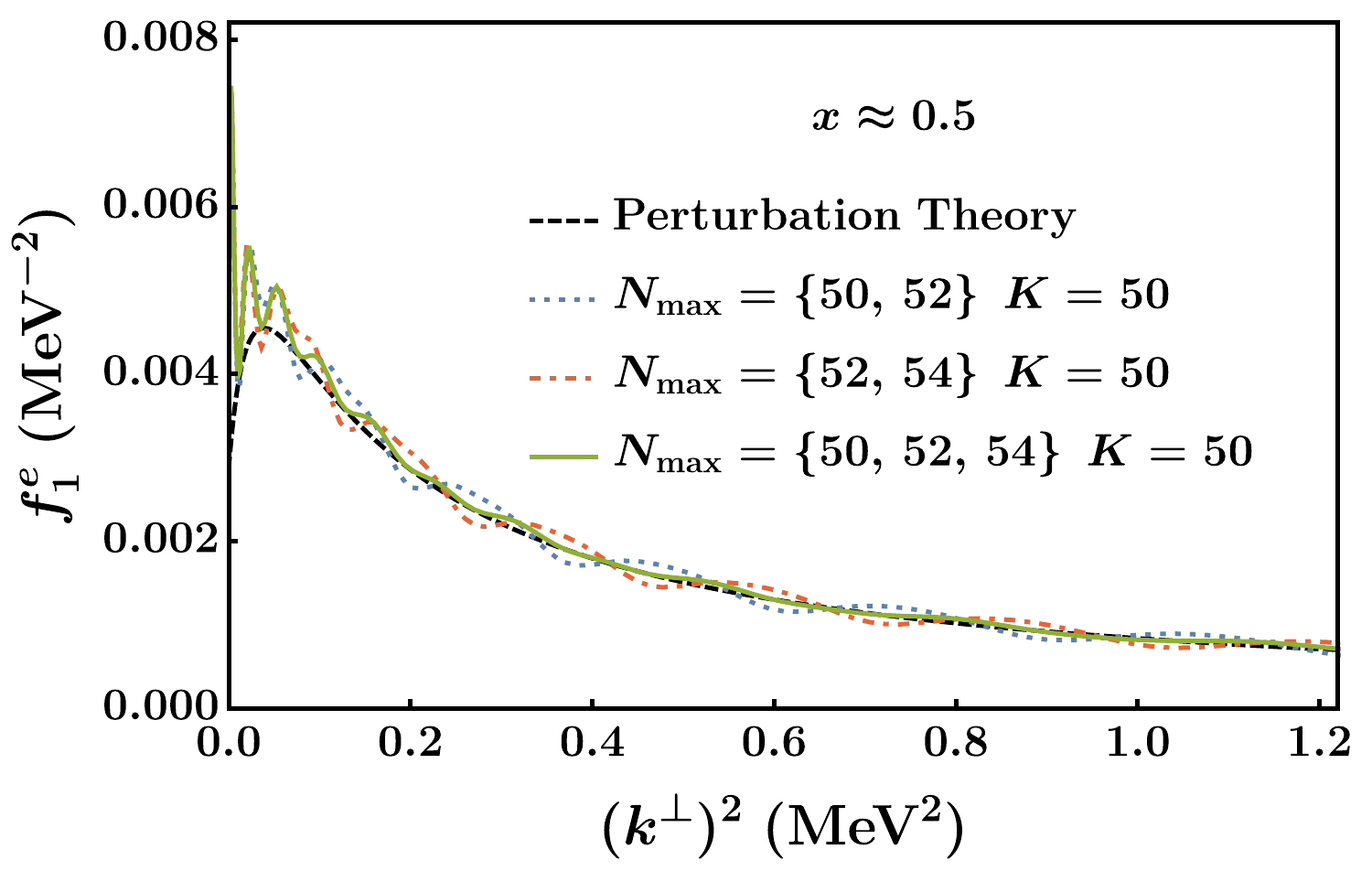}
    \includegraphics[width=0.49\textwidth]{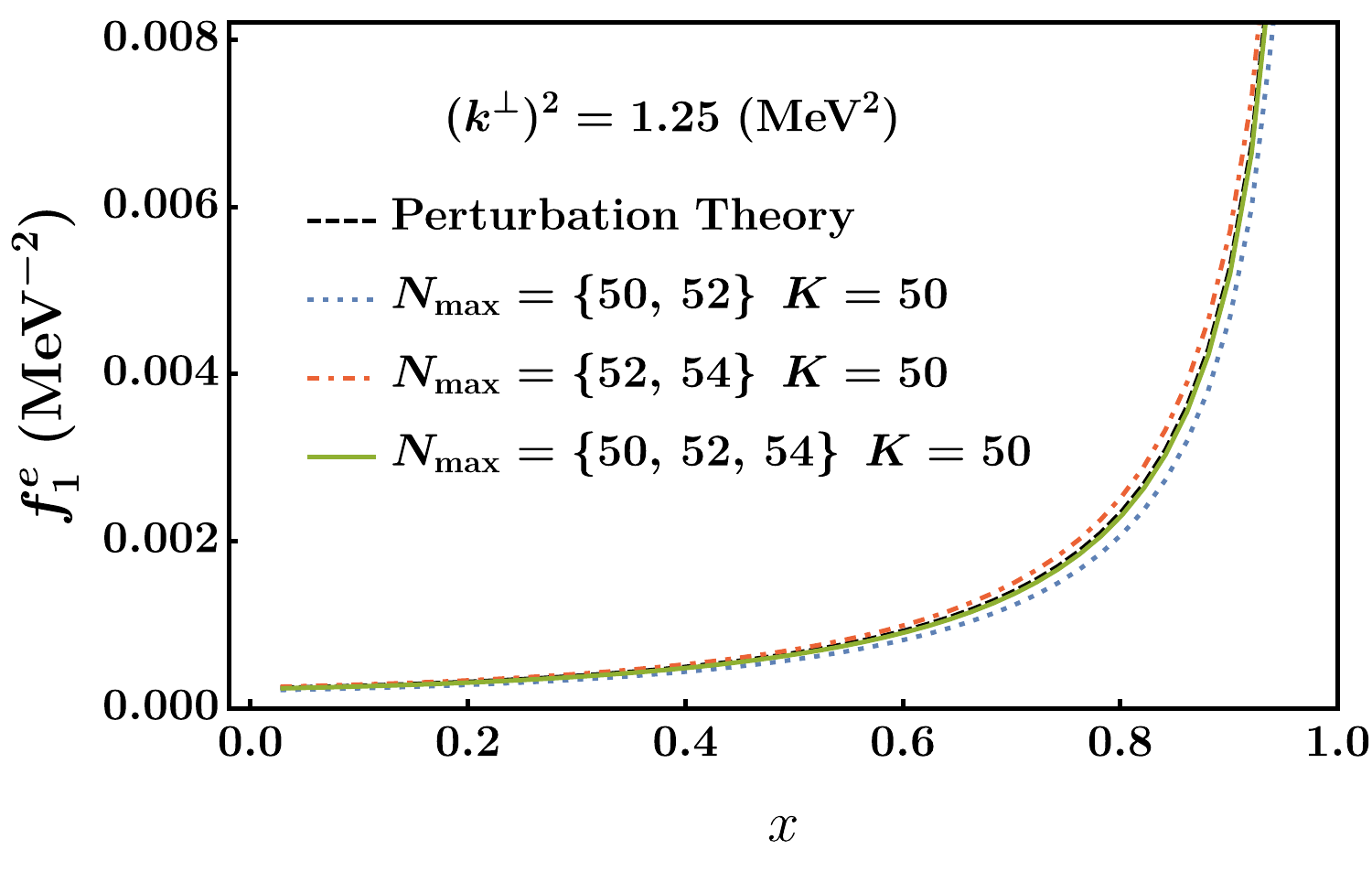}
    \caption{\label{plotave}(color online) The BLFQ results after averaging and the perturbative results (black dashed lines)~\cite{bacchetta2016electron} for $ f_1^e $ in the transverse direction (left) and longitudinal direction (right). Calculations are performed at $ x_e\approx 0.5$ (left) and at $ k_e^{\LCperp}=1.25\,(\mathrm{MeV^2}) $ (right). Curves with different dash styles (colors) are results averaged from the BLFQ computations at different $ N_{\textrm{max}} $ sets with fixed $ K=50 $ and $ b=M_e $.}
\end{figure*}

In our calculations, we set the 2D-HO scale $b$ the same as the mass of physical electron \ie{} $b=M_e$. Here, we only discuss the dependence of our results on $ N_{\rm max} $ and $ K $. In \fig{\ref{plotN}}, we illustrate the BLFQ computations at different $N_{\rm max}$ with $ K=50 $ and $ b=M_e $. We see that if we only change $ N_{\textrm{max}} $, the phase of oscillation changes. The phase shift between computations at $ N_{\textrm{max}}=n+2 $ and at $ N_{\textrm{max}}=n $ is nearly a half period. This property is very useful for approximately removing these truncation artifacts. In \fig{\ref{plotave}} we illustrate averages between the BLFQ computations at $ N_{\textrm{max}}=n $ and $ N_{\textrm{max}}=n+2 $, with $ K=n $. We further observe that the results obtained from averaging between $\{ N_{\textrm{max}}=n,\,n+2 \}$ and between $\{ N_{\textrm{max}}=n+2,\,n+4\} $ also shift from each other by nearly a phase of $ \pi $. Based on these observations, we propose a two-step averaging method involving the BLFQ computations at $\{ N_{\textrm{max}}=n,\,n+2,\,n+4 \}$, with $ K=n $:
\begin{enumerate}
    \item First take averages between the BLFQ computations at $N_{\textrm{max}}=n $ and $ N_{\textrm{max}}=n+2 $, with $ K=n $. Then take averages between the BLFQ computations at $ N_{\textrm{max}}=n+2 $ and $ N_{\textrm{max}}=n+4 $, with also $ K=n $.
    \item The final results are obtained by averaging between those two averages produced in the first step.
\end{enumerate}
The final averages for $ n=50 $ are also shown in \fig{\ref{plotave}} and we see that they are very close to the perturbative results.

\begin{figure*}
    \includegraphics[width=0.49\textwidth]{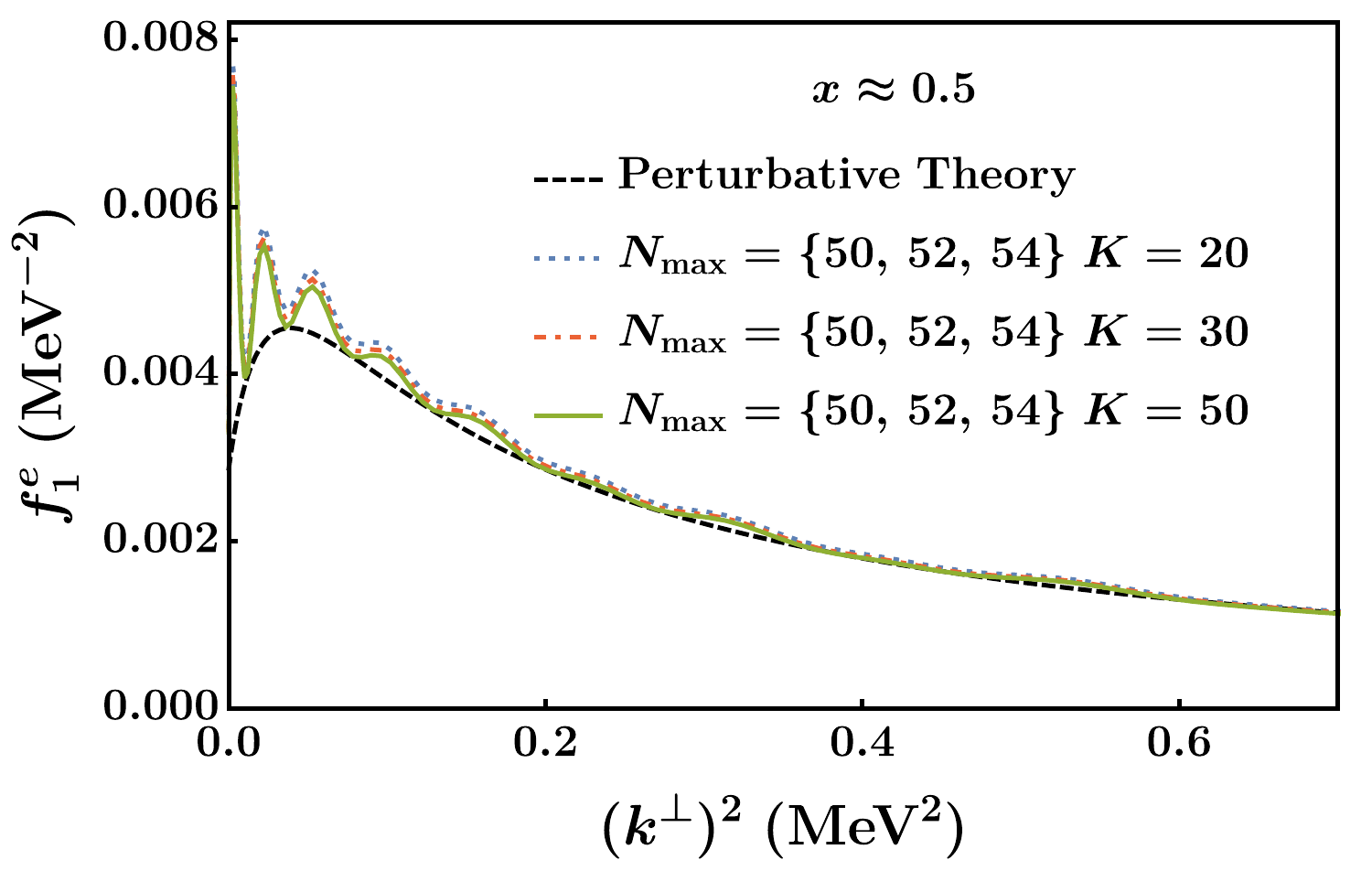}
    \includegraphics[width=0.49\textwidth]{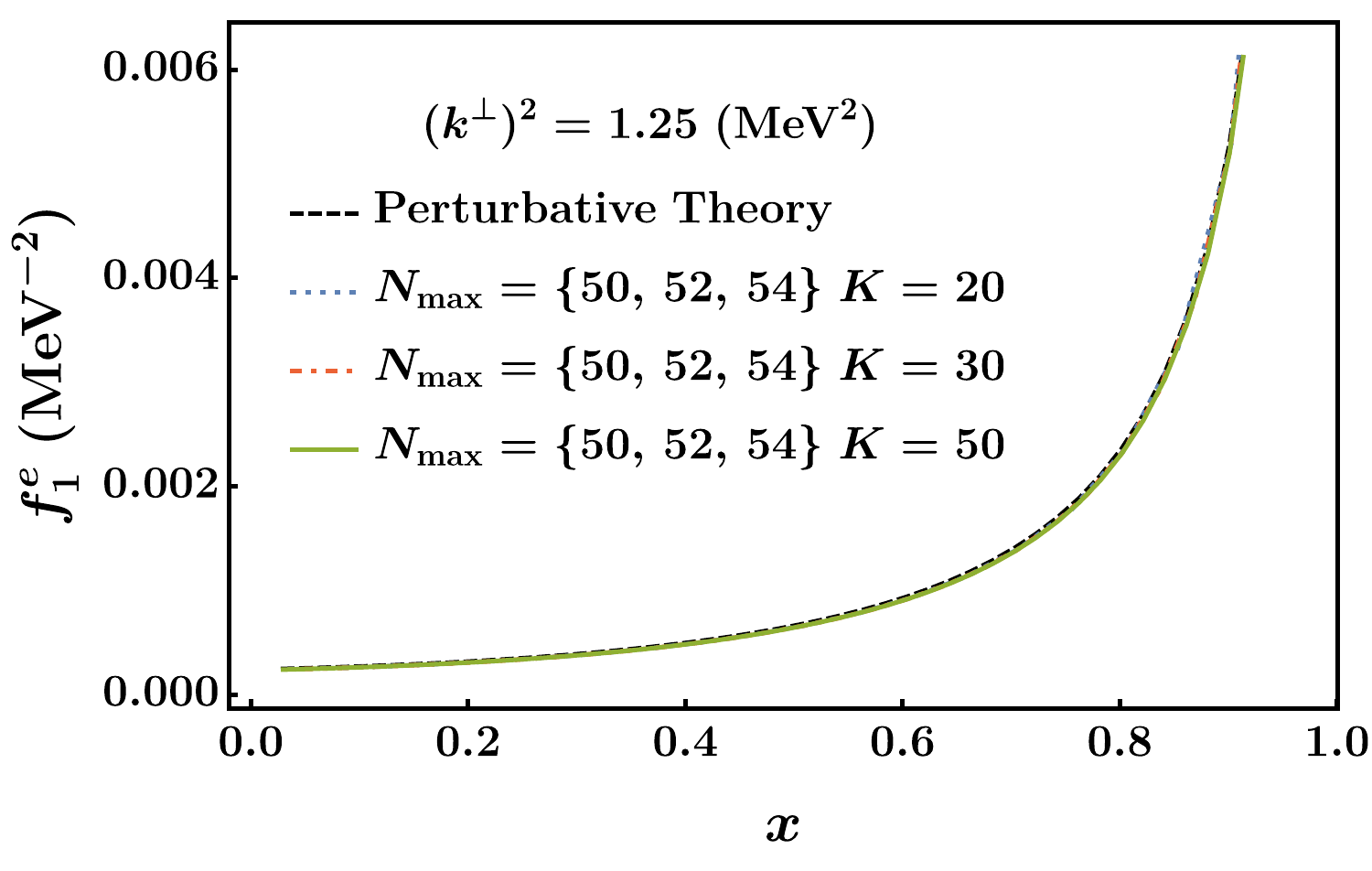}
    \caption{\label{aveK}(color online) The BLFQ results after averaging and the perturbative results (black dashed lines)~\cite{bacchetta2016electron} for $ f_1^e $ in the transverse direction (left) and longitudinal direction (right). Calculations are performed at $ x_e\approx 0.5$ (left) and at $ k_e^{\LCperp}=1.25\,(\mathrm{MeV^2}) $ (right). Curves with different dash styles (colors) are results averaged from the BLFQ computations at $ N_{\textrm{max}}=\{50,\,52,\,54\} $, $ b=M_e $ but different $ K $ as specified in the legends.}
\end{figure*}

In \fig{\ref{aveK}} we plot the BLFQ results after implementing the two-step averaging method using the same $ N_{\mathrm{max}} $ set but different $ K $. We observe that the BLFQ results change minimally when $ K $ changes significantly since it is difficult to distinguish the difference between those curves at different $ K $ in \fig{\ref{aveK}}. This suggests that our TMD results converge very quickly with longitudinal truncation parameter $ K $. For simplicity and resolution in the longitudinal direction we from now on set $ K=n $ while implementing our averaging method.

\begin{figure*}
    \includegraphics[width=0.49\textwidth]{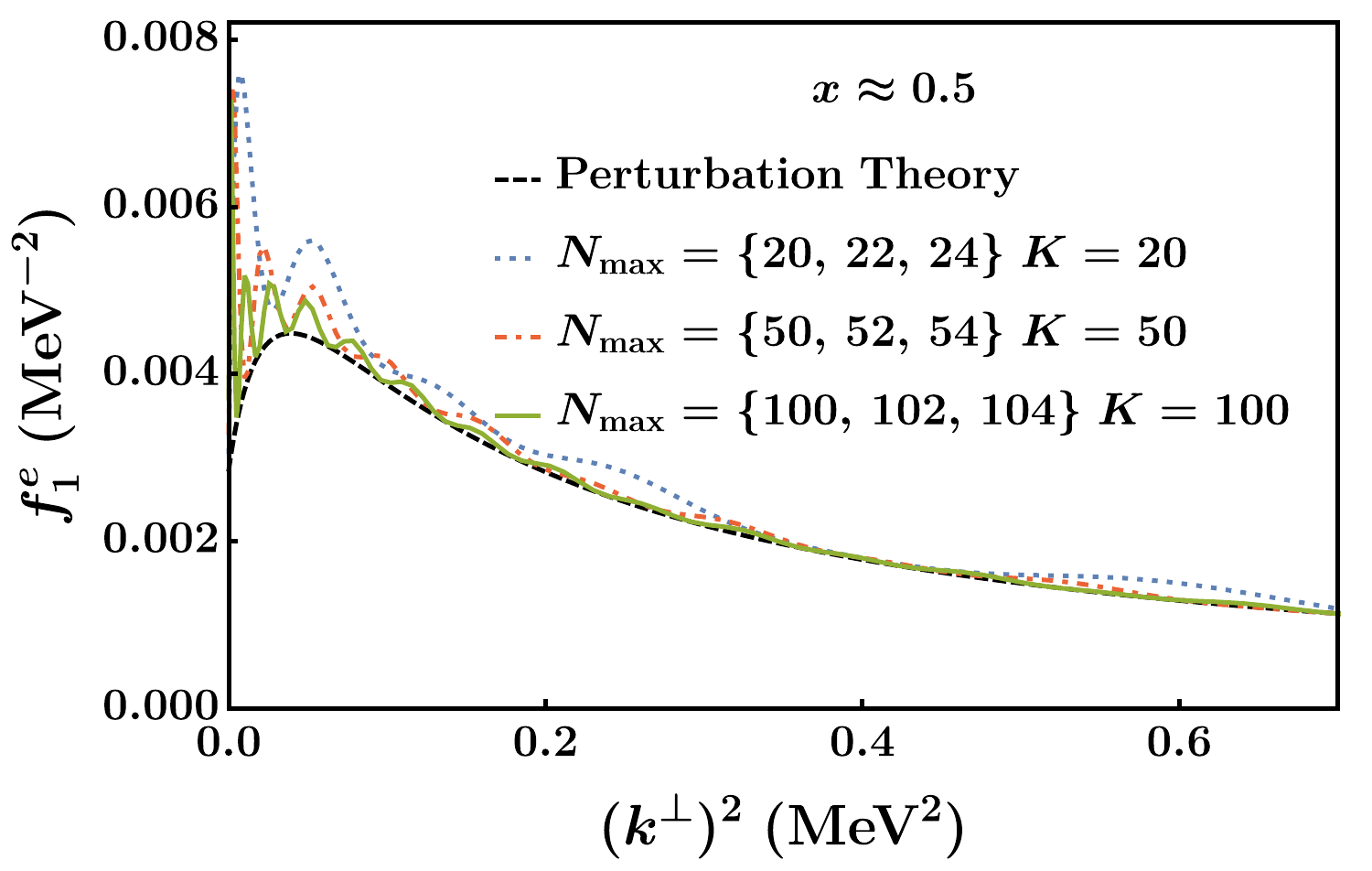}
    \includegraphics[width=0.49\textwidth]{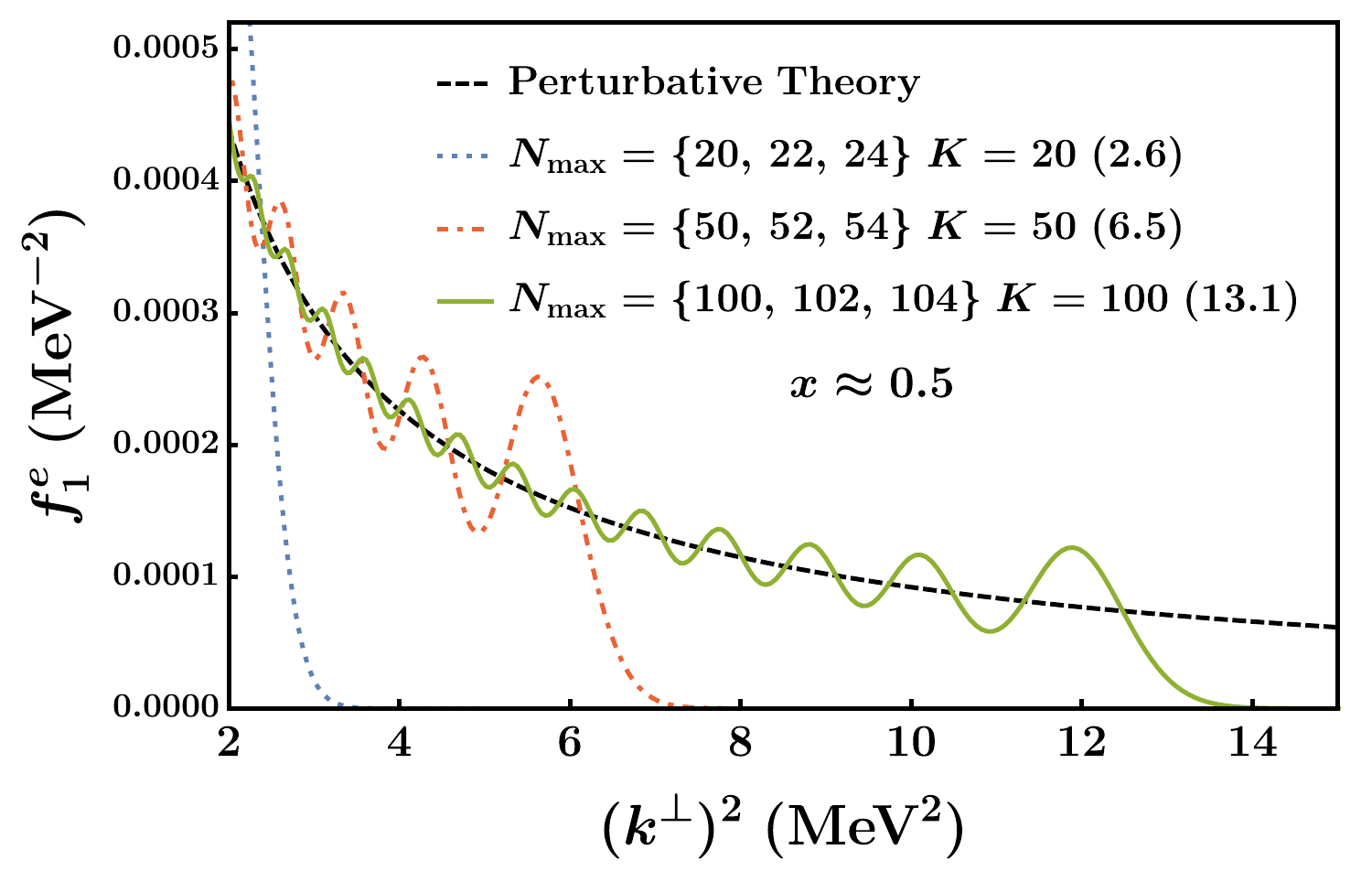}
    \caption{\label{aveN}(color online) $ f_1^e $ after averaging over the BLFQ computations at $ N_{\textrm{max}}=\{n,\,n+2,\,n+4\} $, $ K=n $ and $ b=M_e $. (Here $ n $ is the lowest $ N_{\textrm{max}}$ we use for averaging.) We also plot the perturbative results (black dashed lines)~\cite{bacchetta2016electron} for comparison. In the left panel we plot in the IR region. In the right panel we show the UV region. In the legend of the right panel, we list the square of the UV cutoff at $ x=0.5 $ introduced by the $ N_{\mathrm{max}} $ truncation, \ie{} $ (b\cdot \sqrt{N_{\textrm{max}}/2})^2 $. Curves with different dash styles (colors) are results averaged from the BLFQ computations at different $ N_{\textrm{max}} $ and $ K $, with fixed $ b=M_e $.}
\end{figure*}

In \fig{\ref{aveN}} we plot the BLFQ results averaged over $ N_{\mathrm{max}}=\{n,\,n+2,\,n+4\} $, $ K=n $ for different $ n $ in the IR region (left panel) and also in the UV region (right panel) in the transverse direction. It is clear from the left panel of \fig{\ref{aveN}} that, after implementing our averaging method, the BLFQ results approach the perturbative result as $ N_{\textrm{max}} $ increases. When we plot the BLFQ results in the UV region (right panel of \fig{\ref{aveN}}), we observe that our result drops to zero abruptly after the UV cutoff ($\sim b\sqrt{N_{\textrm{max}}} $). Since the UV cutoff increases as $ N_{\textrm{max}} $ increases the agreement extends to the relevant  higher momentum region.

\subsection{TMDs}
\begin{figure*}
    \centering
    \includegraphics[width=0.49\textwidth]{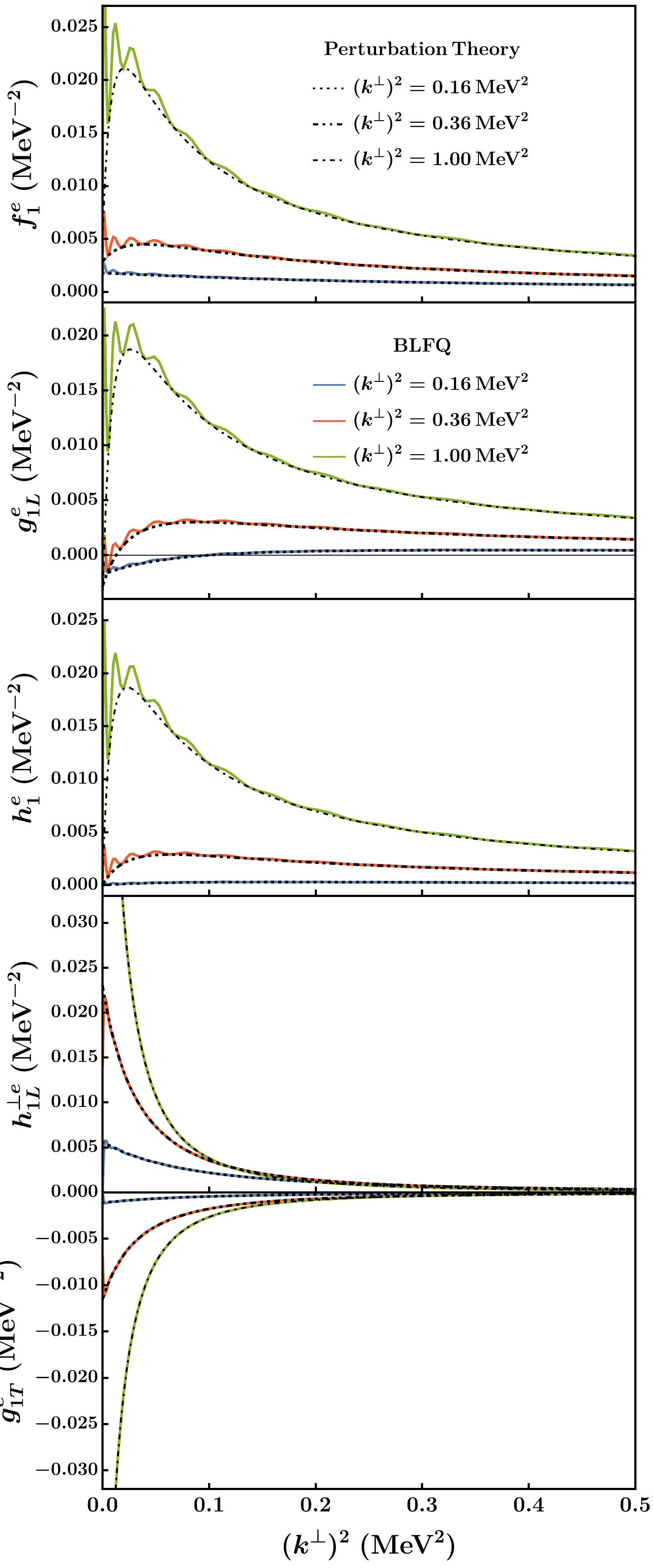}
    \includegraphics[width=0.49\textwidth]{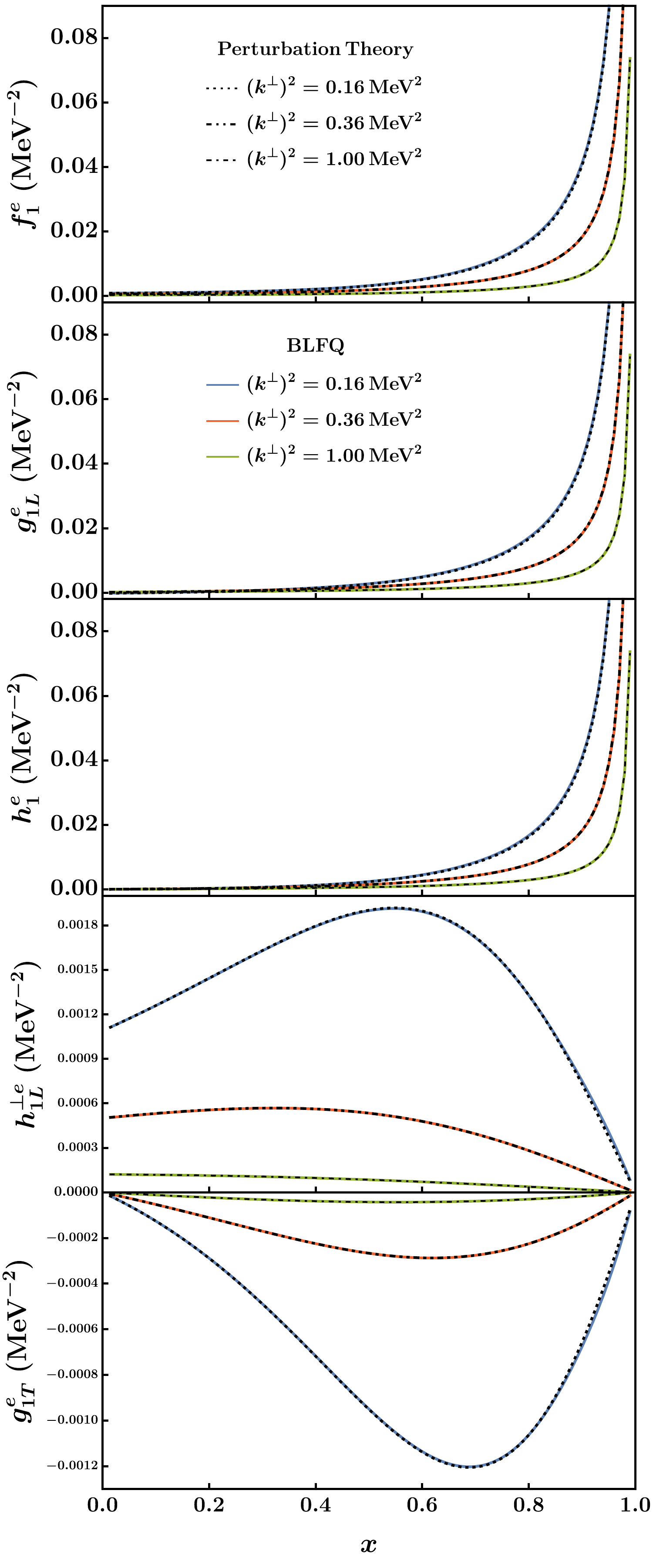}
    \caption{\label{plot2D}(color online) The BLFQ results and the perturbative results for five non-zero leading-twist TMDs neglecting the gauge link. The BLFQ results are obtained by averaging over the BLFQ computations at $ N_{\textrm{max}}=\{100,\,102,\,104 \}$, $ K=100 $ and $ b=M_e $. The left panels show $ k^{\LCperp} $-dependence at selected $ x $ and the right ones show $ x $-dependence at selected $ k^{\LCperp} $. Curves with different dash styles or colors are results at different $ x $ ($ k^{\LCperp} $) for left (right) panels.}
\end{figure*}

We evaluate the TMDs following \eqs{\ref{f1e})-(\ref{h1L}} and then obtain the BLFQ results by averaging over the BLFQ computations at $ N_{\textrm{max}}=\{100,~102,~104 \}$, $ K=100$ and $ b=M_e $. If not stated otherwise, the BLFQ results that we show in the following are all obtained in the same way. In the left panel of \fig{\ref{plot2D}}, we show our five leading-twist TMDs as functions of $k^{\LCperp}$ for different values of $x$, and in the right panel of \fig{\ref{plot2D}}, we show them as functions of $x$ for different values of $k^{\LCperp}$. In these figures, we compare the BLFQ results with the perturbative results evaluated in Ref.~\cite{bacchetta2016electron}. We find that all the five non-zero TMDs obtained by the BLFQ approach are in excellent agreement with the perturbative results over the entire region of $x$ and with moderately large $ |k^{\LCperp}| $. We argue that the discrepancy in the small $ |k^{\LCperp}| $ region is due to relatively small truncation parameters leading to larger IR cutoffs, and the agreement improves when we relax the truncation parameters (see left panel of \fig{\ref{aveN}}).

\begin{figure*}
    \includegraphics[width=0.45\textwidth]{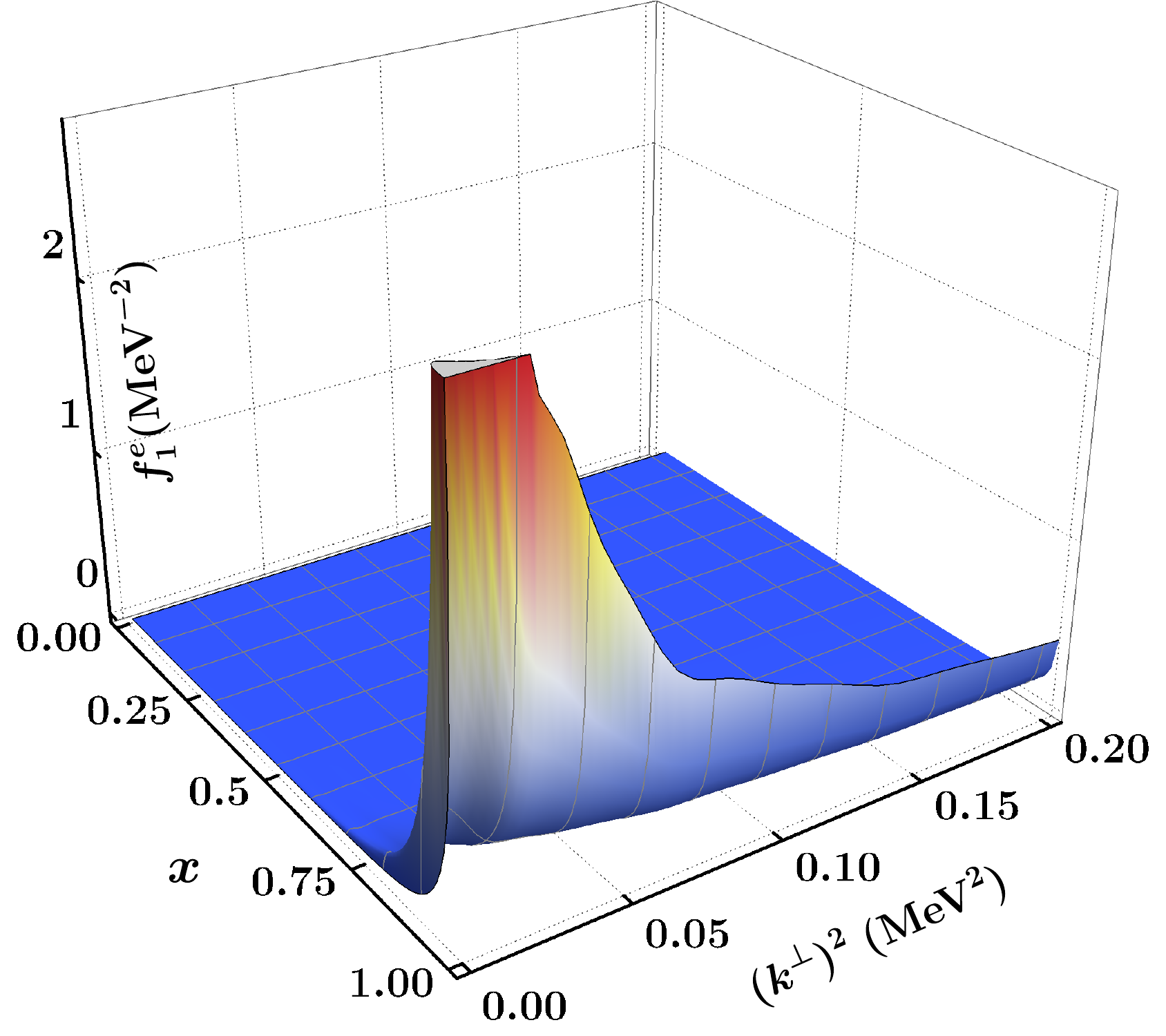}
    \includegraphics[width=0.45\textwidth]{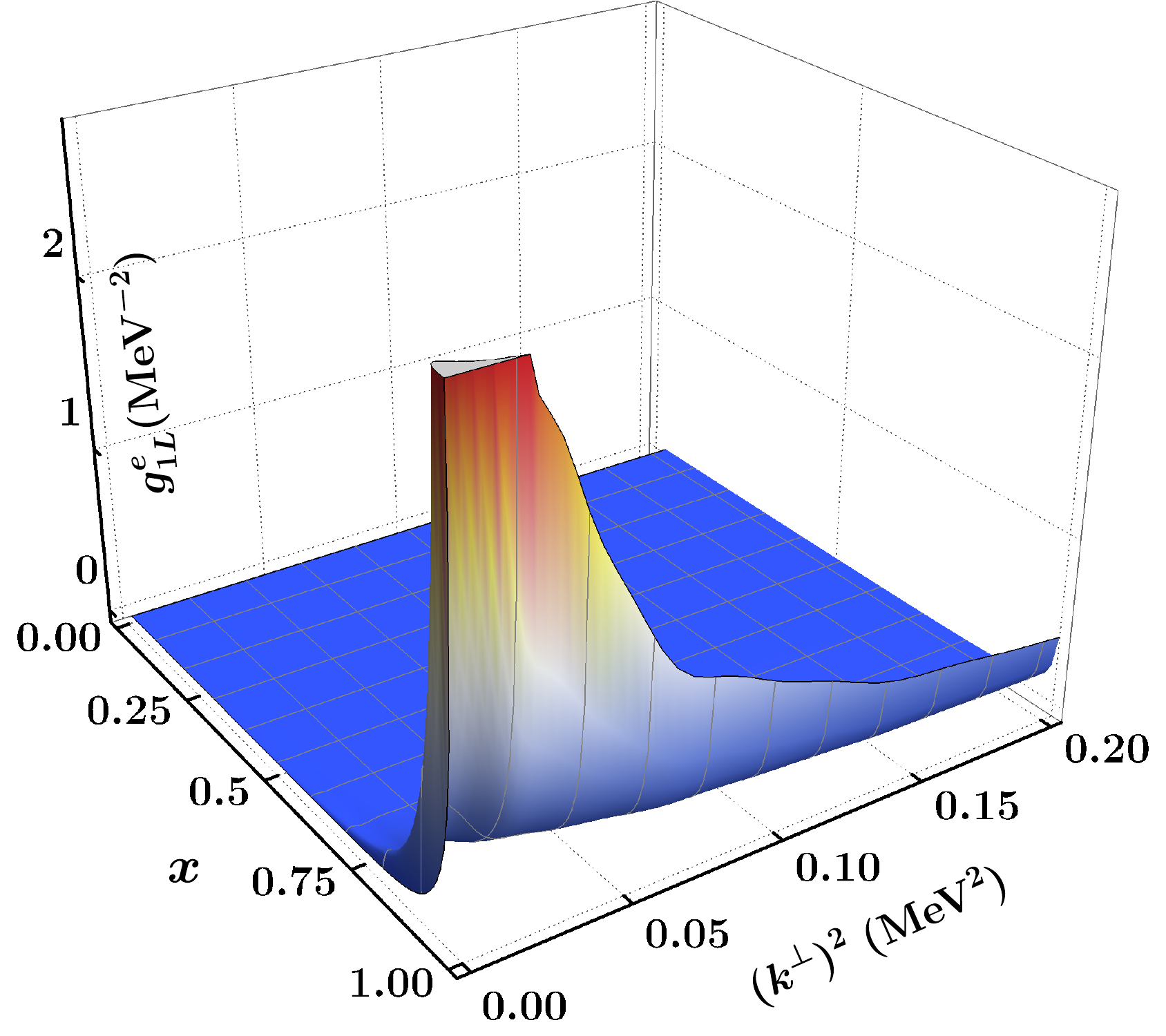}
    \includegraphics[width=0.45\textwidth]{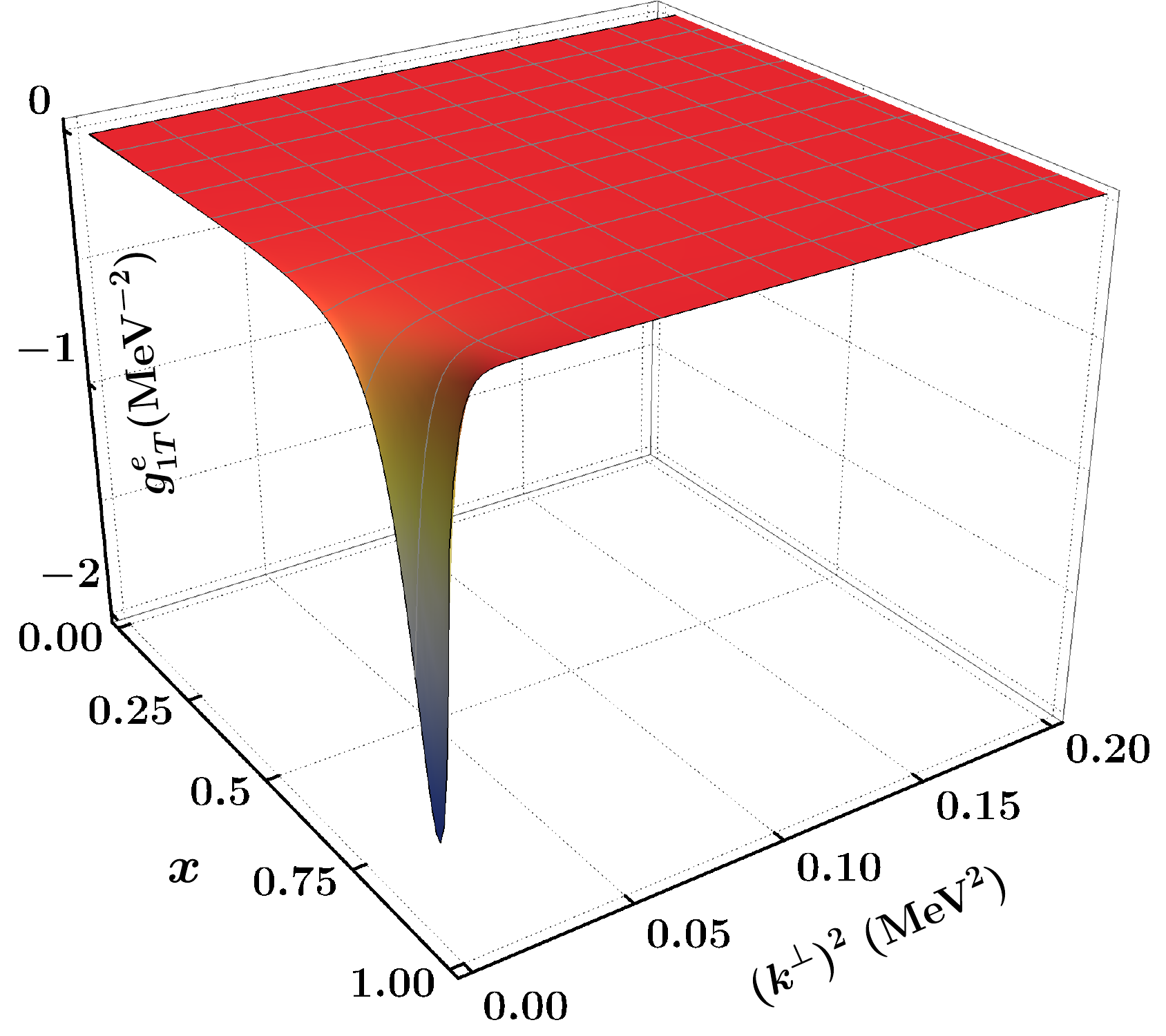}
    \includegraphics[width=0.45\textwidth]{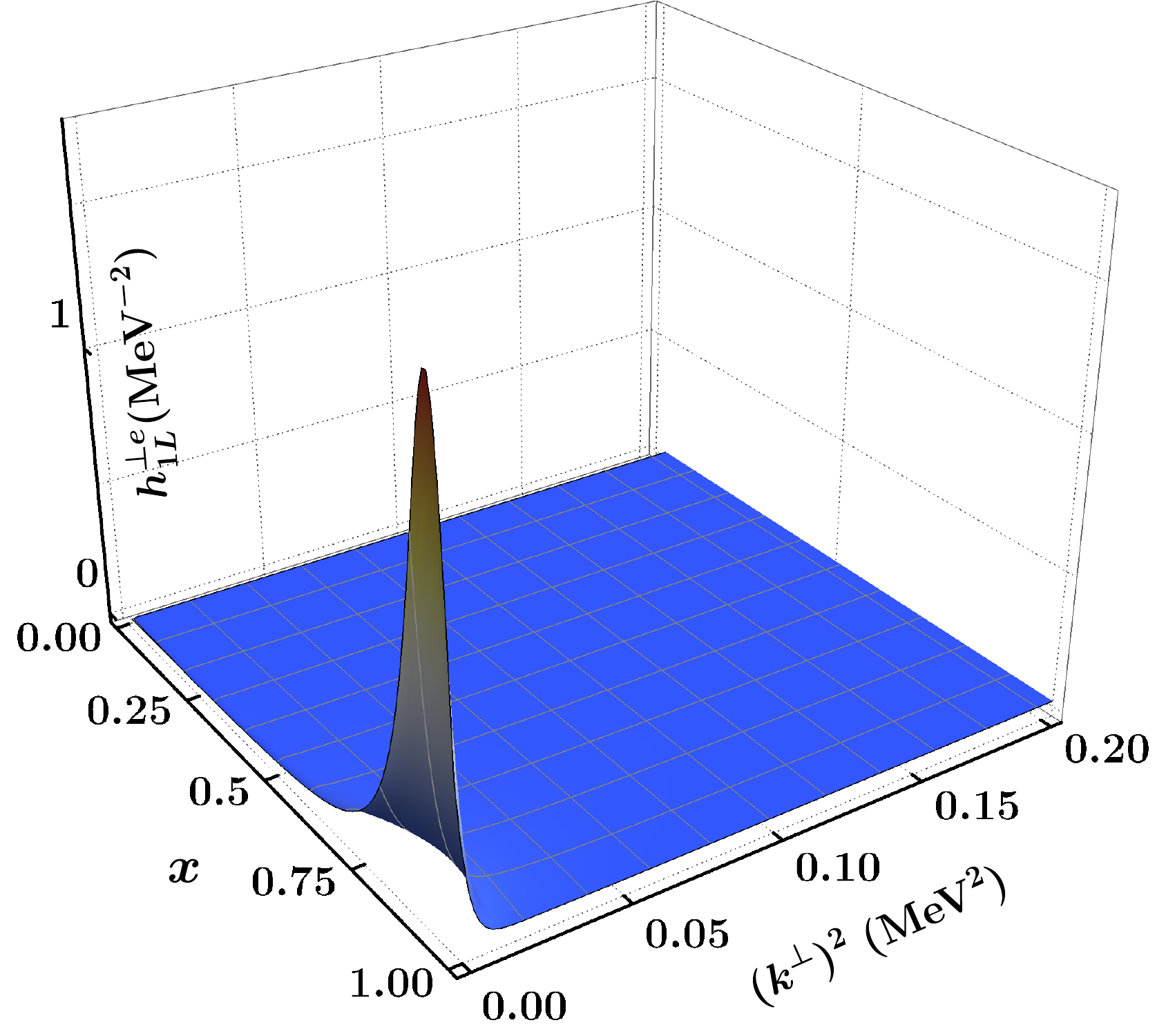}
    \includegraphics[width=0.45\textwidth]{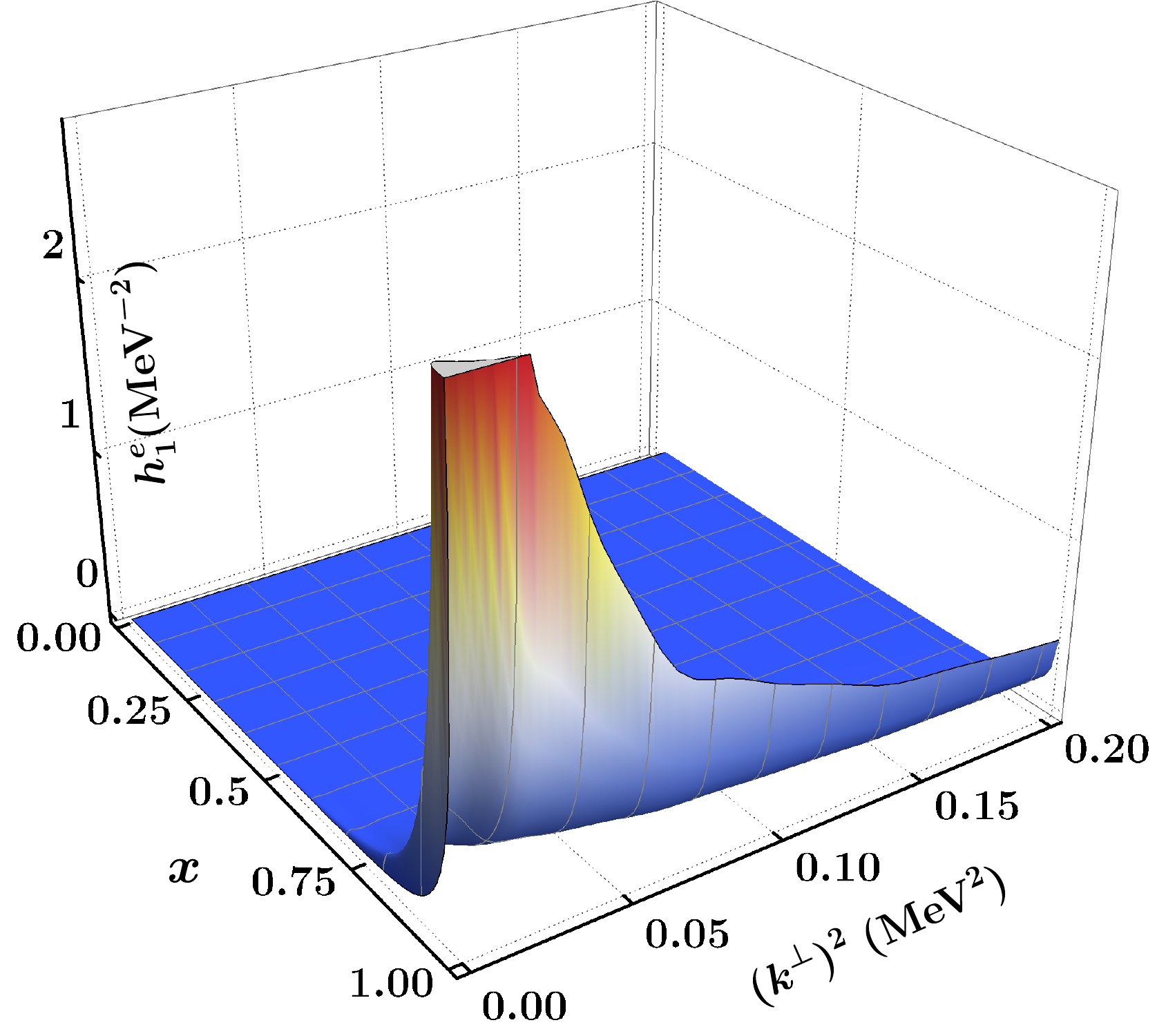}
    \caption{\label{plot3D} $ 3- $D plots of the BLFQ results for five non-zero TMDs neglecting the gauge link. Results are all obtained by averaging over the BLFQ computations at $ N_{\textrm{max}}=\{100,\,102,\,104\} $, $ K=100 $ and $ b=M_e $.}
\end{figure*}

The three-dimensional structures of our BLFQ results after averaging are shown in \fig{\ref{plot3D}}. It is clearly seen there that all five non-zero TMDs have peaks near $(x,k^{\LCperp})=(1,0)$. These peaks indicate the dominant probability is to find a bare electron in the physical electron in each of the different polarization configurations. For $ f_1^e,g_{1L}^e\,\mathrm{and}\,h_1^e $, which appear very similar by eye, the peak in the transverse direction runs higher with increasing $ x $ but falls rapidly to zero or even negative region (for $ g_{1L}^e $) after $ x $ decreases below $ 0.7 $. For $ g_{1T}^e\,\mathrm{and}\,h_{1L}^{\LCperp e} $, the peak occurs very close to $ k^{\LCperp}=0 $ and falls rapidly towards zero as $ (k^{\LCperp})^2 $ increases towards $ 0.01\,(\mathrm{MeV^2}) $.

\begin{figure}
    \includegraphics[width=0.48\textwidth]{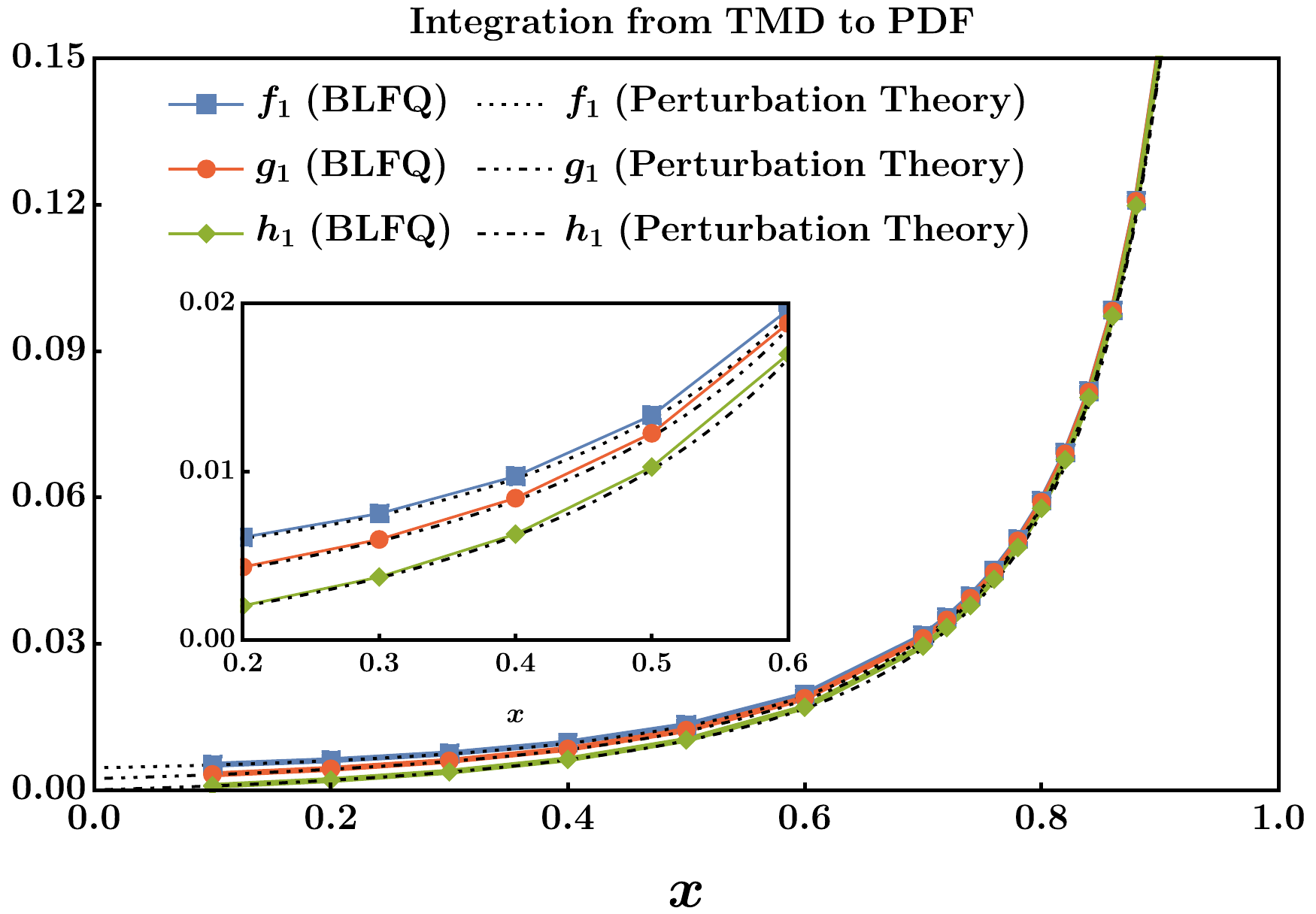}
    \caption{\label{pdf}(color online) Plots for the electron PDFs $ f_1^e(x) $, $ g_1^e(x) $ and $ h_1^e(x) $. The BLFQ results (solid lines with different markers) are compared with the perturbative results (black lines with different dashing styles). The BLFQ results are obtained by averaging over the BLFQ computations at $ N_{\textrm{max}}=\{100,\,102,\,104\} $, $ K=100 $ and $ b=M_e $. We also show an extended version of curves in the range $ 0.2\le x\le 0.6 $ for clearer comparison}
\end{figure}

PDFs of the bare electron within the physical electron are illustrated in \fig{\ref{pdf}}, where again we find a good agreement between our BLFQ results after averaging and the perturbative results. We remark that in \fig{\ref{pdf}}, those three PDFs are very close and in the limit $ x\rightarrow 1 $ they basically become the same. This feature is also clearly seen if you take the same limit for the analytic perturbative equations, \ie{} Eqs. (38),(39),(43) in \paper{\cite{bacchetta2016electron}}. In that limit, those three TMDs actually equal each other and thus their integration in the transverse direction also become the same.

\subsection{Spin Densities in Momentum Space}
\label{densities}

By designating the three-dimensional spin vectors of the bare electron and physical electron by $ \vec s=(s^1, s^2, s^3) $ and $ \vec S=(S^1,S^2,S^3) $ respectively, we can define the bare electron's momentum distribution inside a physical electron with different polarization combinations via those five non-zero leading-twist TMDs as~\cite{bacchetta2016electron}
\begin{align}
    \rho(x,k^{\LCperp},s,S)&=\frac{1}{2}\left[ f_1^e+s^3 S^3 g_{1L}^e + s^3  \frac{S^{\LCperp}\cdot k^{\LCperp}}{M_e} g_{1T}^e  \right. \\ \nonumber
    & \left. +S^3  \frac{s^{\LCperp}\cdot k^{\LCperp}}{M_e} h_{1L}^{\LCperp e} +s^{\LCperp}\cdot S^{\LCperp} h_1^e\right]\,.
\end{align}
Using this formula, we find that the momentum distribution for both the bare and the physical electron polarized in the same transverse direction can be computed by $ (f_1^e+h_1^e)/2 $. Other polarization combinations are $ (f_1^e+g_{1L}^e)/2 $ for both polarized in the same longitudinal direction, $ (f_1^e+\frac{k^y}{M_e}g_{1T}^e)/2 $ for longitudinally polarized bare electron inside $ y- $axis polarized physical electron and $ (f_1^e+\frac{k^y}{M_e}h_{1L}^{\LCperp  e})/2 $ for the converse arrangement.

\begin{figure*}
    \includegraphics[width=0.98\textwidth]{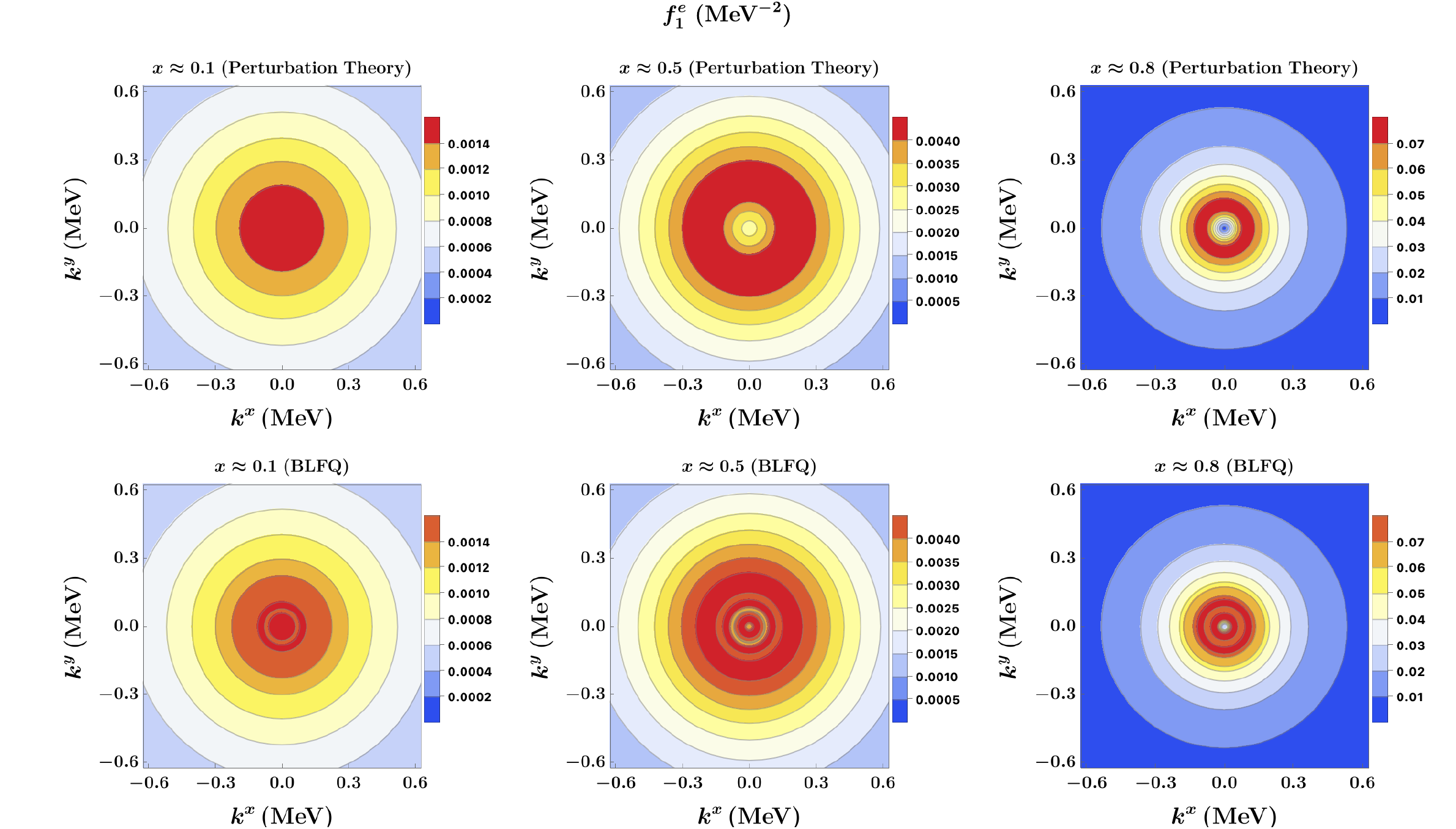}
    \caption{\label{plotdenxf1e}Density plots in the transverse-momentum plane at different $ x $ for the unpolarized distribution, \ie{} $ f_1^e $. The BLFQ results are shown in the lower row and the perturbative results are shown in the upper row. From left to right, the longitudinal momentum fraction of the bare electron is $ 0.1 $, $ 0.5 $ and $ 0.8 $ respectively. The BLFQ results are obtained by averaging over the BLFQ computations at $ N_{\textrm{max}}=\{100,\,102,\,104\} $, $ K=100 $ and $ b=M_e $.}
\end{figure*}

\begin{figure*}
    \includegraphics[width=0.98\textwidth]{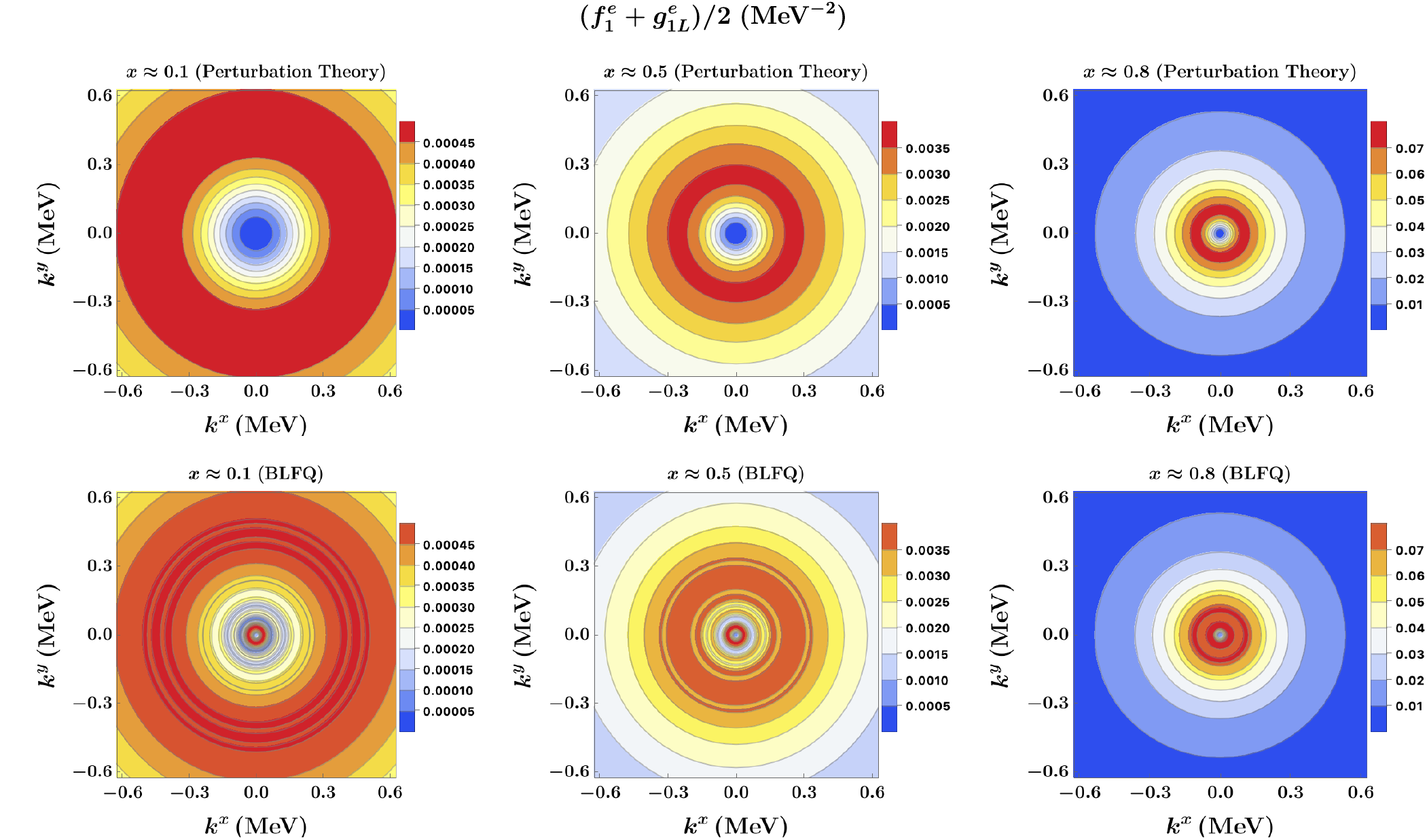}
    \caption{\label{plotdenxg1L}Density plots in the transverse-momentum plane at different $ x $ for distribution with both physical and bare electron polarized in the same longitudinal direction, \ie{} $ (f_i^e+g_{1L}^e)/2 $. The BLFQ results are shown in the lower row and the perturbative results are shown in the upper row. From left to right, the longitudinal momentum fraction of the bare electron is $ 0.1 $, $ 0.5 $ and $ 0.8 $ respectively. The BLFQ results are obtained by averaging over the BLFQ computations at $ N_{\textrm{max}}=\{100,\,102,\,104\} $, $ K=100 $ and $ b=M_e $.}
\end{figure*}

\begin{figure*}
    \includegraphics[width=0.98\textwidth]{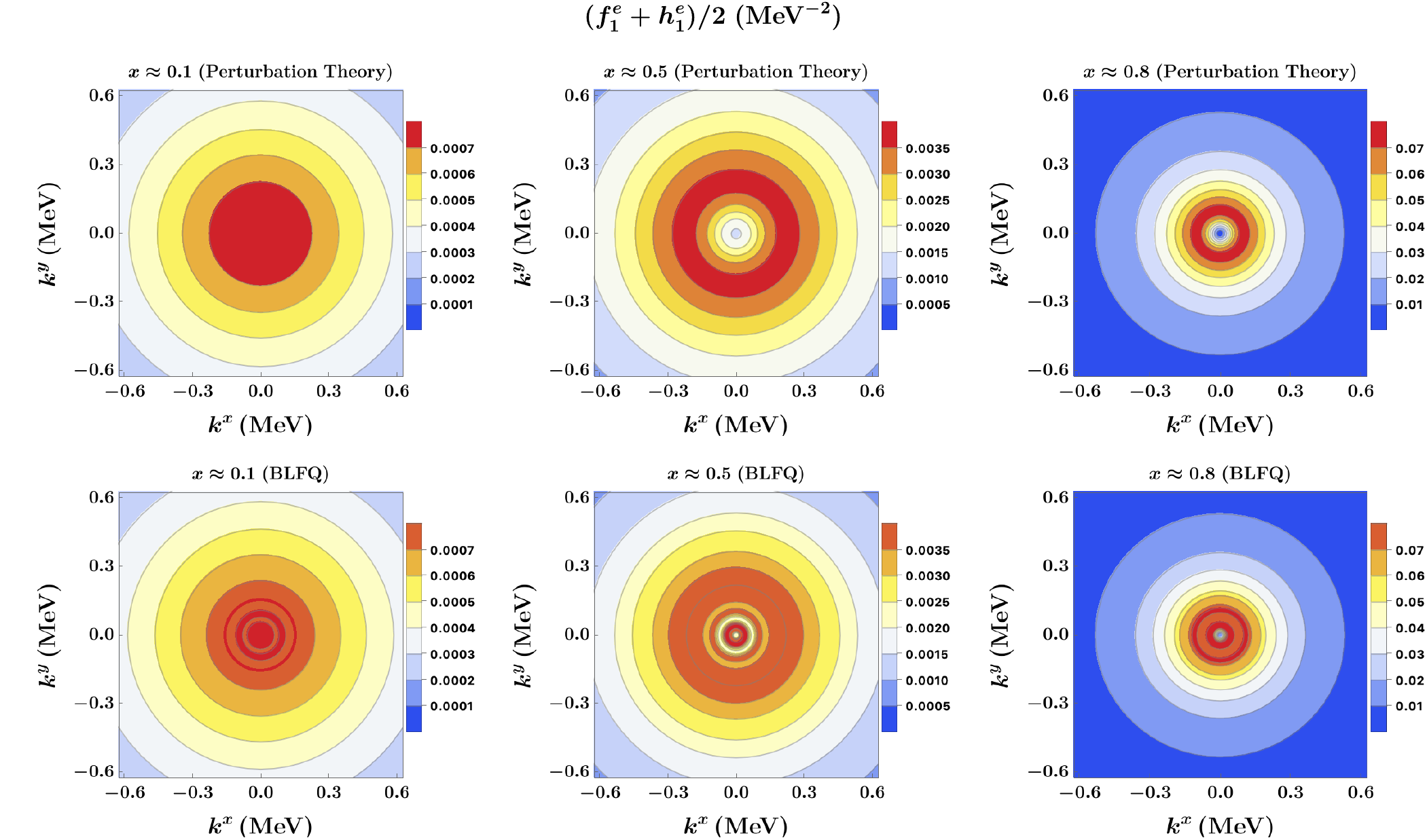}
    \caption{\label{plotdenxh1e}Density plots in the transverse-momentum plane at different $ x $ for distribution with both physical and bare electron polarized in the same transverse direction, \ie{} $ (f_i^e+h_1^e)/2 $. The BLFQ results are shown in the lower row and the perturbative results are shown in the upper row. From left to right, the longitudinal momentum fraction of the bare electron is $ 0.1 $, $ 0.5 $ and $ 0.8 $ respectively. The BLFQ results are obtained by averaging over the BLFQ computations at $ N_{\textrm{max}}=\{100,\,102,\,104\} $, $ K=100 $ and $ b=M_e $.}
\end{figure*}

In \figs{\ref{plotdenxf1e}, \ref{plotdenxg1L}, and \ref{plotdenxh1e}}, we present the bare electron densities in the transverse momentum plane at different values of $ x $ when both the bare and the physical electrons are unpolarized, longitudinally, and transversely polarized, respectively. We find that those three distributions in momentum space are azimuthally symmetric, having the maximum at the centre and rapidly fall off towards the periphery of the transverse momentum plane. Note that $h_{1T}^{\LCperp e}$ and $h_{1}^{\LCperp e}$ vanish in our calculation which makes the transversely polarized density symmetric. In principle, with nontrivial gauge link they will not vanish and the transversely polarized density may not be symmetric. The densities grow larger and larger with increasing $ x $, which is consistent with the right panels of \fig{\ref{plot2D}}. It is also observed from the right panels of \fig{\ref{plot2D}} that $f_1^e$, $g_{1L}^e$ and $h_1^e$ follow similar patterns in the longitudinal direction, though they are not identical. Thus, changes in the longitudinal direction in \figs{\ref{plotdenxf1e}, \ref{plotdenxg1L}, and \ref{plotdenxh1e}} are also quite similar. However, longitudinally and transversely polarized distributions (\figs{\ref{plotdenxg1L}, \ref{plotdenxh1e}}) have smaller values in the center with respect to unpolarized distribution (\fig{\ref{plotdenxf1e}}), owing to averaging $ f_1^e $ with $g_{1L}^e$ and $ f_1^e $ with $h^e_1$ respectively.

\begin{figure*}
    \includegraphics[width=0.98\textwidth]{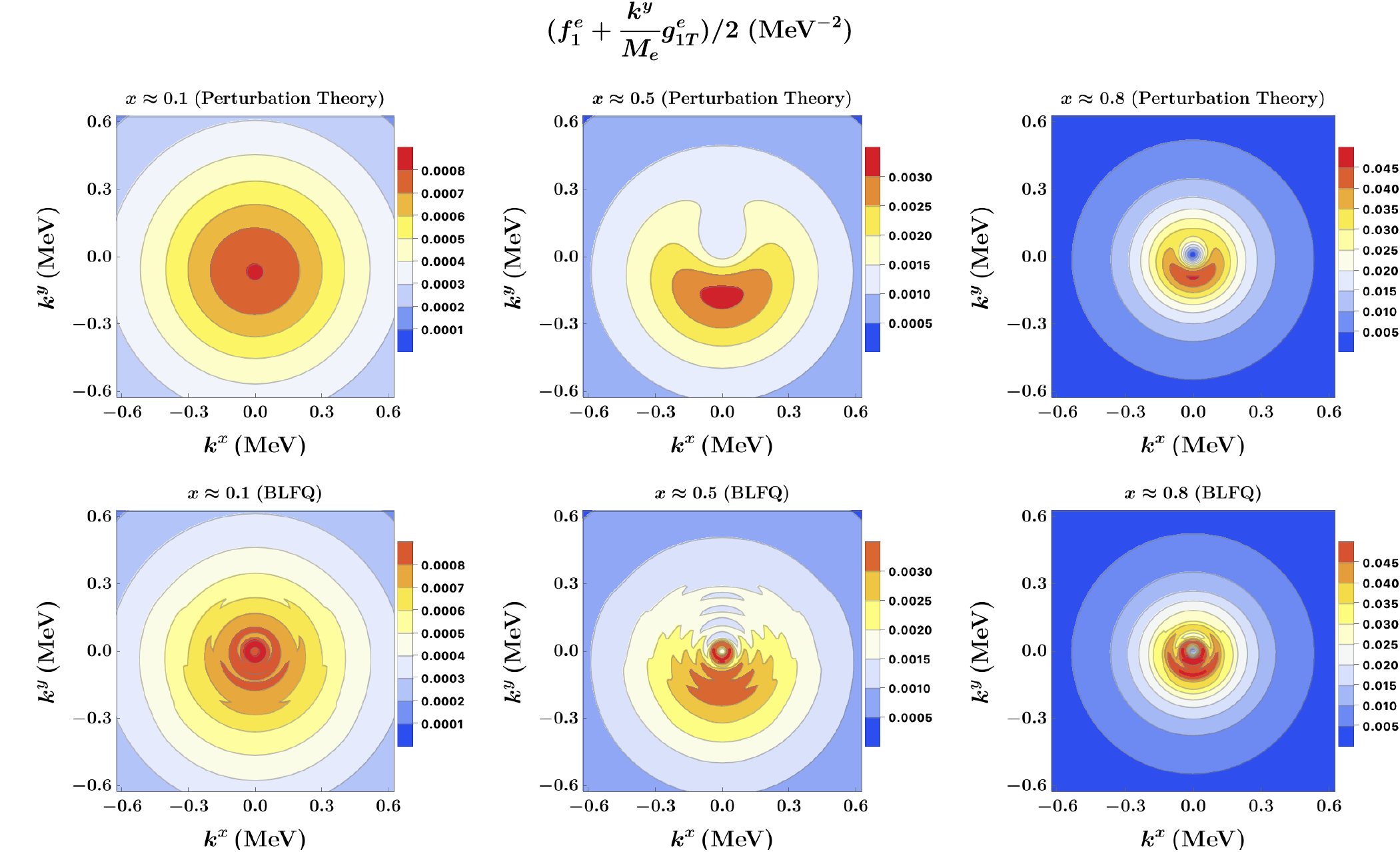}
    \caption{\label{plotdenxg1T}Density plots in the transverse-momentum plane at different $ x $ for distribution of a longitudinally polarized bare electron in a $ y- $axis polarized physical electron, \ie{} $ (f_i^e+\frac{k^y}{M_e}g_{1T}^e)/2 $. The BLFQ results are shown in the lower row and the perturbative results are shown in the upper row. From left to right, the longitudinal momentum fraction of the bare electron is $ 0.1 $, $ 0.5 $ and $ 0.8 $ respectively. The BLFQ results are obtained by averaging over the BLFQ computations at $ N_{\textrm{max}}=\{100,\,102,\,104\} $, $ K=100 $ and $ b=M_e $.}
\end{figure*}

\begin{figure*}
    \includegraphics[width=0.98\textwidth]{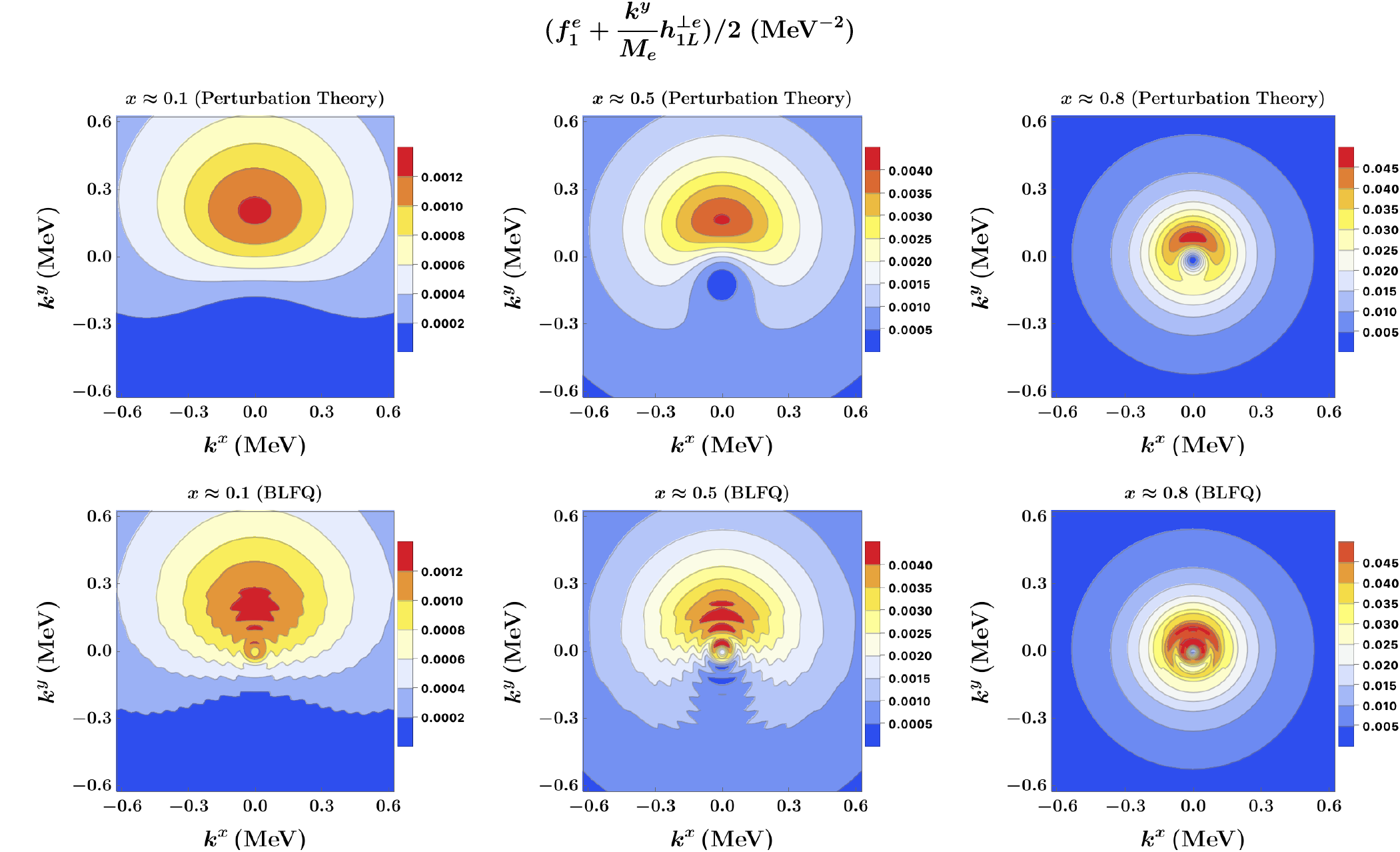}
    \caption{\label{plotdenxh1L}Density plots in the transverse-momentum plane at different $ x $ for distribution of a $ y- $axis polarized bare electron in a longitudinally polarized physical electron, \ie{} $ (f_i^e+\frac{k^y}{M_e}h_{1L}^{\LCperp e})/2 $. The BLFQ results are shown in the lower row and the perturbative results are shown in the upper row. From left to right, the longitudinal momentum fraction of the bare electron is $ 0.1 $, $ 0.5 $ and $ 0.8 $ respectively. The BLFQ results are obtained by averaging over the BLFQ computations at $ N_{\textrm{max}}=\{100,\,102,\,104\} $, $ K=100 $ and $ b=M_e $.}
\end{figure*}

The distributions $ (f_1^e+\frac{k^y}{M_e}g_{1T}^e)/2 $ and $ (f_1^e+\frac{k^y}{M_e}h_{1L}^{\LCperp e})/2 $ are shown in \figs{\ref{plotdenxg1T} and \ref{plotdenxh1L}}, respectively, where we clearly see the distortion introduced by different polarization configurations of the bare and the physical electron. Both densities feature a significant dipole deformation arising from the terms $\frac{k^y}{M_e}g_{1T}^e$ and $\frac{k^y}{M_e}h_{1L}^{e\LCperp}$ in $ y- $axis. It can also be noticed that the distortion of a longitudinally polarized bare electron in a transversely polarized parent electron is opposite to that of a transversely polarized bare electron in a longitudinally polarized physical electron. The reason is that $h_{1L}^{\LCperp e}$ is positive, while $g_{1T}^e$ is negative. Since both the TMDs have the peaks at large $x$, the distortions are stronger near $x\sim0.8$ compared to other values of $x$.

The oscillation of the BLFQ results after averaging (lower rows) around the perturbative results (upper rows) is also observed in \fig{\ref{plotdenxf1e}} to \fig{\ref{plotdenxh1L}}. Although we utilize BLFQ results after averaging to calculate those density plots, we could still see the oscillation clearly. For \figs{\ref{plotdenxf1e}, \ref{plotdenxg1L} and \ref{plotdenxh1e}}, the oscillation comes directly from the oscillation of $ f_1^e\;g_{1L}^e\;\mathrm{and}\;h_1^e $ in the small $ k^{\LCperp} $ region even after averaging, which can be seen from the left panels of \fig{\ref{plot2D}}. And for \figs{\ref{plotdenxg1T} and \ref{plotdenxh1L}}, the original oscillation shown in \fig{\ref{plot2D}} is small but are amplified by the transverse momentum $ k^y $ in the definition of those distributions. We anticipate that BLFQ calculations in larger basis spaces, accompanied by a larger averaging domain, would further reduce these finite basis artifacts.

\section{Summary}\label{sec_summary}
In this paper, we have investigated the leading-twist TMDs of the physical electron from its light-front wave functions in the framework of Basis Light-front Quantization. These wave functions were obtained from the eigenvectors of the light-front QED Hamiltonian in the light-cone gauge. Our non-perturbative results have been compared with leading-order perturbative calculations. Our current study of the TMDs is performed within the lowest nontrivial Fock sectors. With a proper renormalization procedure and a rescaling of the naive TMDs correcting the artifacts introduced by the Fock space truncation, the BLFQ results are consistent with the perturbative calculations. During the calculation, we have also introduced an averaging method for reducing finite basis artifacts of our BLFQ results. This method may be used in the following non-perturbative investigation. In this study, the gauge link has been set to unity which leaves us five non-zero TMDs out of the eight leading-twist TMDs. Employing those non-zero TMDs, we have also studied the spin densities in the transverse momentum plane of a bare electron inside a physical electron with different polarization configurations. The quantitative results of the densities are also consistent with the perturbative results. These calculations constitute a comprehensive and accurate test of the BLFQ approach. For further investigation, future developments will  focus on the inclusion of a nontrivial gauge link that will provide a prediction of the Boer-Mulders function $h_1^{\LCperp e}$  and the Sivers function $f_{1T}^{\LCperp e}$ in QED. The main purpose of this study is to establish the foundation for studying the TMDs for strongly interacting systems in QCD, like the baryon and meson which are highly non-perturbative.

\begin{acknowledgments}
C. M. is supported by the National Natural Science Foundation of China (NSFC) under the Grants No. 11850410436 and No. 11950410753. X. Z. is supported by new faculty startup funding by the Institute of Modern Physics, Chinese Academy of Sciences and by Key Research Program of Frontier Sciences, Chinese Academy of Sciences, Grant No. ZDB-SLY-7020. J. P. V. is supported by the Department of Energy under Grants No. DE-FG02-87ER40371, and No. de-sc0018223 (SciDAC4/NUCLEI). This work of H. Z., S. X., C. M., and X. Z. is also supported by the Strategic Priority Research Program of the Chinese Academy of Sciences, Grant No. XDB34000000. This research used resources of the National Energy Research Scientific Computing Center (NERSC), a U.S. Department of Energy Office of Science User Facility operated under Contract No. DE-AC02-05CH11231. A portion of the computational resources were also provided by Gansu Computing Center.
\end{acknowledgments}


\begin{thebibliography}{76}%
\makeatletter
\providecommand \@ifxundefined [1]{%
 \@ifx{#1\undefined}
}%
\providecommand \@ifnum [1]{%
 \ifnum #1\expandafter \@firstoftwo
 \else \expandafter \@secondoftwo
 \fi
}%
\providecommand \@ifx [1]{%
 \ifx #1\expandafter \@firstoftwo
 \else \expandafter \@secondoftwo
 \fi
}%
\providecommand \natexlab [1]{#1}%
\providecommand \enquote  [1]{``#1''}%
\providecommand \bibnamefont  [1]{#1}%
\providecommand \bibfnamefont [1]{#1}%
\providecommand \citenamefont [1]{#1}%
\providecommand \href@noop [0]{\@secondoftwo}%
\providecommand \href [0]{\begingroup \@sanitize@url \@href}%
\providecommand \@href[1]{\@@startlink{#1}\@@href}%
\providecommand \@@href[1]{\endgroup#1\@@endlink}%
\providecommand \@sanitize@url [0]{\catcode `\\12\catcode `\$12\catcode
  `\&12\catcode `\#12\catcode `\^12\catcode `\_12\catcode `\%12\relax}%
\providecommand \@@startlink[1]{}%
\providecommand \@@endlink[0]{}%
\providecommand \url  [0]{\begingroup\@sanitize@url \@url }%
\providecommand \@url [1]{\endgroup\@href {#1}{\urlprefix }}%
\providecommand \urlprefix  [0]{URL }%
\providecommand \Eprint [0]{\href }%
\providecommand \doibase [0]{http://dx.doi.org/}%
\providecommand \selectlanguage [0]{\@gobble}%
\providecommand \bibinfo  [0]{\@secondoftwo}%
\providecommand \bibfield  [0]{\@secondoftwo}%
\providecommand \translation [1]{[#1]}%
\providecommand \BibitemOpen [0]{}%
\providecommand \bibitemStop [0]{}%
\providecommand \bibitemNoStop [0]{.\EOS\space}%
\providecommand \EOS [0]{\spacefactor3000\relax}%
\providecommand \BibitemShut  [1]{\csname bibitem#1\endcsname}%
\let\auto@bib@innerbib\@empty
\bibitem [{\citenamefont {Collins}\ and\ \citenamefont
  {Soper}(1982)}]{Collins:1981uw}%
  \BibitemOpen
  \bibfield  {author} {\bibinfo {author} {\bibfnamefont {J.~C.}\ \bibnamefont
  {Collins}}\ and\ \bibinfo {author} {\bibfnamefont {D.~E.}\ \bibnamefont
  {Soper}},\ }\href {\doibase 10.1016/0550-3213(82)90021-9} {\bibfield
  {journal} {\bibinfo  {journal} {Nucl. Phys. B}\ }\textbf {\bibinfo {volume}
  {194}},\ \bibinfo {pages} {445} (\bibinfo {year} {1982})}\BibitemShut
  {NoStop}%
\bibitem [{\citenamefont {Martin}\ \emph {et~al.}(1998)\citenamefont {Martin},
  \citenamefont {Roberts}, \citenamefont {Stirling},\ and\ \citenamefont
  {Thorne}}]{Martin:1998sq}%
  \BibitemOpen
  \bibfield  {author} {\bibinfo {author} {\bibfnamefont {A.~D.}\ \bibnamefont
  {Martin}}, \bibinfo {author} {\bibfnamefont {R.}~\bibnamefont {Roberts}},
  \bibinfo {author} {\bibfnamefont {W.}~\bibnamefont {Stirling}}, \ and\
  \bibinfo {author} {\bibfnamefont {R.}~\bibnamefont {Thorne}},\ }\href
  {\doibase 10.1007/s100520050220} {\bibfield  {journal} {\bibinfo  {journal}
  {Eur. Phys. J. C}\ }\textbf {\bibinfo {volume} {4}},\ \bibinfo {pages} {463}
  (\bibinfo {year} {1998})},\ \Eprint {http://arxiv.org/abs/hep-ph/9803445}
  {arXiv:hep-ph/9803445} \BibitemShut {NoStop}%
\bibitem [{\citenamefont {Gluck}\ \emph {et~al.}(1995)\citenamefont {Gluck},
  \citenamefont {Reya},\ and\ \citenamefont {Vogt}}]{Gluck:1994uf}%
  \BibitemOpen
  \bibfield  {author} {\bibinfo {author} {\bibfnamefont {M.}~\bibnamefont
  {Gluck}}, \bibinfo {author} {\bibfnamefont {E.}~\bibnamefont {Reya}}, \ and\
  \bibinfo {author} {\bibfnamefont {A.}~\bibnamefont {Vogt}},\ }\href {\doibase
  10.1007/BF01624586} {\bibfield  {journal} {\bibinfo  {journal} {Z. Phys. C}\
  }\textbf {\bibinfo {volume} {67}},\ \bibinfo {pages} {433} (\bibinfo {year}
  {1995})}\BibitemShut {NoStop}%
\bibitem [{\citenamefont {Glück}\ \emph {et~al.}(1998)\citenamefont {Glück},
  \citenamefont {Reya},\ and\ \citenamefont {Vogt}}]{Gluck:1998xa}%
  \BibitemOpen
  \bibfield  {author} {\bibinfo {author} {\bibfnamefont {M.}~\bibnamefont
  {Glück}}, \bibinfo {author} {\bibfnamefont {E.}~\bibnamefont {Reya}}, \ and\
  \bibinfo {author} {\bibfnamefont {A.}~\bibnamefont {Vogt}},\ }\href {\doibase
  10.1007/s100520050289} {\bibfield  {journal} {\bibinfo  {journal} {Eur. Phys.
  J. C}\ }\textbf {\bibinfo {volume} {5}},\ \bibinfo {pages} {461} (\bibinfo
  {year} {1998})},\ \Eprint {http://arxiv.org/abs/hep-ph/9806404}
  {arXiv:hep-ph/9806404} \BibitemShut {NoStop}%
\bibitem [{\citenamefont {Angeles-Martinez}\ \emph {et~al.}(2015)\citenamefont
  {Angeles-Martinez} \emph {et~al.}}]{Angeles-Martinez:2015sea}%
  \BibitemOpen
  \bibfield  {author} {\bibinfo {author} {\bibfnamefont {R.}~\bibnamefont
  {Angeles-Martinez}} \emph {et~al.},\ }\href {\doibase
  10.5506/APhysPolB.46.2501} {\bibfield  {journal} {\bibinfo  {journal} {Acta
  Phys. Polon. B}\ }\textbf {\bibinfo {volume} {46}},\ \bibinfo {pages} {2501}
  (\bibinfo {year} {2015})},\ \Eprint {http://arxiv.org/abs/1507.05267}
  {arXiv:1507.05267 [hep-ph]} \BibitemShut {NoStop}%
\bibitem [{\citenamefont {Collins}(2011)}]{Collins:2011ca}%
  \BibitemOpen
  \bibfield  {author} {\bibinfo {author} {\bibfnamefont {J.}~\bibnamefont
  {Collins}},\ }\href {\doibase 10.1142/S2010194511001590} {\bibfield
  {journal} {\bibinfo  {journal} {Int. J. Mod. Phys. Conf. Ser.}\ }\textbf
  {\bibinfo {volume} {4}},\ \bibinfo {pages} {85} (\bibinfo {year} {2011})},\
  \Eprint {http://arxiv.org/abs/1107.4123} {arXiv:1107.4123 [hep-ph]}
  \BibitemShut {NoStop}%
\bibitem [{\citenamefont {Pasquini}\ and\ \citenamefont
  {Lorce'}(2012)}]{Pasquini:2012jm}%
  \BibitemOpen
  \bibfield  {author} {\bibinfo {author} {\bibfnamefont {B.}~\bibnamefont
  {Pasquini}}\ and\ \bibinfo {author} {\bibfnamefont {C.}~\bibnamefont
  {Lorce'}},\ }\href {\doibase 10.3254/978-1-61499-197-7-197} {\bibfield
  {journal} {\bibinfo  {journal} {Proc. Int. Sch. Phys. Fermi}\ }\textbf
  {\bibinfo {volume} {180}},\ \bibinfo {pages} {197} (\bibinfo {year}
  {2012})},\ \Eprint {http://arxiv.org/abs/1203.5006} {arXiv:1203.5006
  [hep-ph]} \BibitemShut {NoStop}%
\bibitem [{\citenamefont {Brodsky}\ \emph {et~al.}(2002)\citenamefont
  {Brodsky}, \citenamefont {Hwang},\ and\ \citenamefont
  {Schmidt}}]{Brodsky:2002cx}%
  \BibitemOpen
  \bibfield  {author} {\bibinfo {author} {\bibfnamefont {S.~J.}\ \bibnamefont
  {Brodsky}}, \bibinfo {author} {\bibfnamefont {D.~S.}\ \bibnamefont {Hwang}},
  \ and\ \bibinfo {author} {\bibfnamefont {I.}~\bibnamefont {Schmidt}},\ }\href
  {\doibase 10.1016/S0370-2693(02)01320-5} {\bibfield  {journal} {\bibinfo
  {journal} {Phys. Lett. B}\ }\textbf {\bibinfo {volume} {530}},\ \bibinfo
  {pages} {99} (\bibinfo {year} {2002})},\ \Eprint
  {http://arxiv.org/abs/hep-ph/0201296} {arXiv:hep-ph/0201296} \BibitemShut
  {NoStop}%
\bibitem [{\citenamefont {Bacchetta}\ \emph {et~al.}(2017)\citenamefont
  {Bacchetta}, \citenamefont {Delcarro}, \citenamefont {Pisano}, \citenamefont
  {Radici},\ and\ \citenamefont {Signori}}]{Bacchetta:2017gcc}%
  \BibitemOpen
  \bibfield  {author} {\bibinfo {author} {\bibfnamefont {A.}~\bibnamefont
  {Bacchetta}}, \bibinfo {author} {\bibfnamefont {F.}~\bibnamefont {Delcarro}},
  \bibinfo {author} {\bibfnamefont {C.}~\bibnamefont {Pisano}}, \bibinfo
  {author} {\bibfnamefont {M.}~\bibnamefont {Radici}}, \ and\ \bibinfo {author}
  {\bibfnamefont {A.}~\bibnamefont {Signori}},\ }\href {\doibase
  10.1007/JHEP06(2017)081} {\bibfield  {journal} {\bibinfo  {journal} {JHEP}\
  }\textbf {\bibinfo {volume} {06}},\ \bibinfo {pages} {081} (\bibinfo {year}
  {2017})},\ \bibinfo {note} {[Erratum: JHEP 06, 051 (2019)]},\ \Eprint
  {http://arxiv.org/abs/1703.10157} {arXiv:1703.10157 [hep-ph]} \BibitemShut
  {NoStop}%
\bibitem [{\citenamefont {Ralston}\ and\ \citenamefont
  {Soper}(1979)}]{Ralston:1979ys}%
  \BibitemOpen
  \bibfield  {author} {\bibinfo {author} {\bibfnamefont {J.~P.}\ \bibnamefont
  {Ralston}}\ and\ \bibinfo {author} {\bibfnamefont {D.~E.}\ \bibnamefont
  {Soper}},\ }\href {\doibase 10.1016/0550-3213(79)90082-8} {\bibfield
  {journal} {\bibinfo  {journal} {Nucl. Phys. B}\ }\textbf {\bibinfo {volume}
  {152}},\ \bibinfo {pages} {109} (\bibinfo {year} {1979})}\BibitemShut
  {NoStop}%
\bibitem [{\citenamefont {Donohue}\ and\ \citenamefont
  {Gottlieb}(1981)}]{Donohue:1980tn}%
  \BibitemOpen
  \bibfield  {author} {\bibinfo {author} {\bibfnamefont {J.}~\bibnamefont
  {Donohue}}\ and\ \bibinfo {author} {\bibfnamefont {S.~A.}\ \bibnamefont
  {Gottlieb}},\ }\href {\doibase 10.1103/PhysRevD.23.2577} {\bibfield
  {journal} {\bibinfo  {journal} {Phys. Rev. D}\ }\textbf {\bibinfo {volume}
  {23}},\ \bibinfo {pages} {2577} (\bibinfo {year} {1981})}\BibitemShut
  {NoStop}%
\bibitem [{\citenamefont {Tangerman}\ and\ \citenamefont
  {Mulders}(1995)}]{Tangerman:1994eh}%
  \BibitemOpen
  \bibfield  {author} {\bibinfo {author} {\bibfnamefont {R.}~\bibnamefont
  {Tangerman}}\ and\ \bibinfo {author} {\bibfnamefont {P.}~\bibnamefont
  {Mulders}},\ }\href {\doibase 10.1103/PhysRevD.51.3357} {\bibfield  {journal}
  {\bibinfo  {journal} {Phys. Rev. D}\ }\textbf {\bibinfo {volume} {51}},\
  \bibinfo {pages} {3357} (\bibinfo {year} {1995})},\ \Eprint
  {http://arxiv.org/abs/hep-ph/9403227} {arXiv:hep-ph/9403227} \BibitemShut
  {NoStop}%
\bibitem [{\citenamefont {Zhou}\ \emph {et~al.}(2010)\citenamefont {Zhou},
  \citenamefont {Yuan},\ and\ \citenamefont {Liang}}]{Zhou:2009jm}%
  \BibitemOpen
  \bibfield  {author} {\bibinfo {author} {\bibfnamefont {J.}~\bibnamefont
  {Zhou}}, \bibinfo {author} {\bibfnamefont {F.}~\bibnamefont {Yuan}}, \ and\
  \bibinfo {author} {\bibfnamefont {Z.-T.}\ \bibnamefont {Liang}},\ }\href
  {\doibase 10.1103/PhysRevD.81.054008} {\bibfield  {journal} {\bibinfo
  {journal} {Phys. Rev. D}\ }\textbf {\bibinfo {volume} {81}},\ \bibinfo
  {pages} {054008} (\bibinfo {year} {2010})},\ \Eprint
  {http://arxiv.org/abs/0909.2238} {arXiv:0909.2238 [hep-ph]} \BibitemShut
  {NoStop}%
\bibitem [{\citenamefont {Collins}(2002)}]{Collins:2002kn}%
  \BibitemOpen
  \bibfield  {author} {\bibinfo {author} {\bibfnamefont {J.~C.}\ \bibnamefont
  {Collins}},\ }\href {\doibase 10.1016/S0370-2693(02)01819-1} {\bibfield
  {journal} {\bibinfo  {journal} {Phys. Lett. B}\ }\textbf {\bibinfo {volume}
  {536}},\ \bibinfo {pages} {43} (\bibinfo {year} {2002})},\ \Eprint
  {http://arxiv.org/abs/hep-ph/0204004} {arXiv:hep-ph/0204004} \BibitemShut
  {NoStop}%
\bibitem [{\citenamefont {Collins}\ \emph {et~al.}(1985)\citenamefont
  {Collins}, \citenamefont {Soper},\ and\ \citenamefont
  {Sterman}}]{Collins1984TMD}%
  \BibitemOpen
  \bibfield  {author} {\bibinfo {author} {\bibfnamefont {J.~C.}\ \bibnamefont
  {Collins}}, \bibinfo {author} {\bibfnamefont {D.~E.}\ \bibnamefont {Soper}},
  \ and\ \bibinfo {author} {\bibfnamefont {G.~F.}\ \bibnamefont {Sterman}},\
  }\href {\doibase 10.1016/0550-3213(85)90479-1} {\bibfield  {journal}
  {\bibinfo  {journal} {Nucl. Phys. B}\ }\textbf {\bibinfo {volume} {250}},\
  \bibinfo {pages} {199} (\bibinfo {year} {1985})}\BibitemShut {NoStop}%
\bibitem [{\citenamefont {Collins}\ \emph {et~al.}(1983)\citenamefont
  {Collins}, \citenamefont {Soper},\ and\ \citenamefont
  {Sterman}}]{Collins1982factorization}%
  \BibitemOpen
  \bibfield  {author} {\bibinfo {author} {\bibfnamefont {J.~C.}\ \bibnamefont
  {Collins}}, \bibinfo {author} {\bibfnamefont {D.~E.}\ \bibnamefont {Soper}},
  \ and\ \bibinfo {author} {\bibfnamefont {G.~F.}\ \bibnamefont {Sterman}},\
  }\href {\doibase 10.1016/0550-3213(83)90062-7} {\bibfield  {journal}
  {\bibinfo  {journal} {Nucl. Phys. B}\ }\textbf {\bibinfo {volume} {223}},\
  \bibinfo {pages} {381} (\bibinfo {year} {1983})}\BibitemShut {NoStop}%
\bibitem [{\citenamefont {Catani}\ \emph {et~al.}(1990)\citenamefont {Catani},
  \citenamefont {Ciafaloni},\ and\ \citenamefont {Hautmann}}]{Catani1990gluon}%
  \BibitemOpen
  \bibfield  {author} {\bibinfo {author} {\bibfnamefont {S.}~\bibnamefont
  {Catani}}, \bibinfo {author} {\bibfnamefont {M.}~\bibnamefont {Ciafaloni}}, \
  and\ \bibinfo {author} {\bibfnamefont {F.}~\bibnamefont {Hautmann}},\ }\href
  {\doibase 10.1016/0370-2693(90)91601-7} {\bibfield  {journal} {\bibinfo
  {journal} {Phys. Lett. B}\ }\textbf {\bibinfo {volume} {242}},\ \bibinfo
  {pages} {97} (\bibinfo {year} {1990})}\BibitemShut {NoStop}%
\bibitem [{\citenamefont {Catani}\ \emph {et~al.}(1991)\citenamefont {Catani},
  \citenamefont {Ciafaloni},\ and\ \citenamefont
  {Hautmann}}]{Catani1990smallx}%
  \BibitemOpen
  \bibfield  {author} {\bibinfo {author} {\bibfnamefont {S.}~\bibnamefont
  {Catani}}, \bibinfo {author} {\bibfnamefont {M.}~\bibnamefont {Ciafaloni}}, \
  and\ \bibinfo {author} {\bibfnamefont {F.}~\bibnamefont {Hautmann}},\ }\href
  {\doibase 10.1016/0550-3213(91)90055-3} {\bibfield  {journal} {\bibinfo
  {journal} {Nucl. Phys. B}\ }\textbf {\bibinfo {volume} {366}},\ \bibinfo
  {pages} {135} (\bibinfo {year} {1991})}\BibitemShut {NoStop}%
\bibitem [{\citenamefont {Mulders}\ and\ \citenamefont
  {Tangerman}(1996)}]{Mulders1995treelevel}%
  \BibitemOpen
  \bibfield  {author} {\bibinfo {author} {\bibfnamefont {P.}~\bibnamefont
  {Mulders}}\ and\ \bibinfo {author} {\bibfnamefont {R.}~\bibnamefont
  {Tangerman}},\ }\href {\doibase 10.1016/0550-3213(95)00632-X} {\bibfield
  {journal} {\bibinfo  {journal} {Nucl. Phys. B}\ }\textbf {\bibinfo {volume}
  {461}},\ \bibinfo {pages} {197} (\bibinfo {year} {1996})},\ \bibinfo {note}
  {[Erratum: Nucl.Phys.B 484, 538--540 (1997)]},\ \Eprint
  {http://arxiv.org/abs/hep-ph/9510301} {arXiv:hep-ph/9510301} \BibitemShut
  {NoStop}%
\bibitem [{\citenamefont {Ji}\ \emph {et~al.}(2005)\citenamefont {Ji},
  \citenamefont {Ma},\ and\ \citenamefont {Yuan}}]{Ji:2004wu}%
  \BibitemOpen
  \bibfield  {author} {\bibinfo {author} {\bibfnamefont {X.-d.}\ \bibnamefont
  {Ji}}, \bibinfo {author} {\bibfnamefont {J.-p.}\ \bibnamefont {Ma}}, \ and\
  \bibinfo {author} {\bibfnamefont {F.}~\bibnamefont {Yuan}},\ }\href {\doibase
  10.1103/PhysRevD.71.034005} {\bibfield  {journal} {\bibinfo  {journal} {Phys.
  Rev. D}\ }\textbf {\bibinfo {volume} {71}},\ \bibinfo {pages} {034005}
  (\bibinfo {year} {2005})},\ \Eprint {http://arxiv.org/abs/hep-ph/0404183}
  {arXiv:hep-ph/0404183} \BibitemShut {NoStop}%
\bibitem [{\citenamefont {Anselmino}\ \emph {et~al.}(2007)\citenamefont
  {Anselmino}, \citenamefont {Boglione}, \citenamefont {D'Alesio},
  \citenamefont {Kotzinian}, \citenamefont {Murgia}, \citenamefont {Prokudin},\
  and\ \citenamefont {Turk}}]{Anselmino:2007fs}%
  \BibitemOpen
  \bibfield  {author} {\bibinfo {author} {\bibfnamefont {M.}~\bibnamefont
  {Anselmino}}, \bibinfo {author} {\bibfnamefont {M.}~\bibnamefont {Boglione}},
  \bibinfo {author} {\bibfnamefont {U.}~\bibnamefont {D'Alesio}}, \bibinfo
  {author} {\bibfnamefont {A.}~\bibnamefont {Kotzinian}}, \bibinfo {author}
  {\bibfnamefont {F.}~\bibnamefont {Murgia}}, \bibinfo {author} {\bibfnamefont
  {A.}~\bibnamefont {Prokudin}}, \ and\ \bibinfo {author} {\bibfnamefont
  {C.}~\bibnamefont {Turk}},\ }\href {\doibase 10.1103/PhysRevD.75.054032}
  {\bibfield  {journal} {\bibinfo  {journal} {Phys. Rev. D}\ }\textbf {\bibinfo
  {volume} {75}},\ \bibinfo {pages} {054032} (\bibinfo {year} {2007})},\
  \Eprint {http://arxiv.org/abs/hep-ph/0701006} {arXiv:hep-ph/0701006}
  \BibitemShut {NoStop}%
\bibitem [{\citenamefont {Anselmino}\ \emph {et~al.}(2011)\citenamefont
  {Anselmino}, \citenamefont {Boglione}, \citenamefont {D'Alesio},
  \citenamefont {Melis}, \citenamefont {Murgia}, \citenamefont {Nocera},\ and\
  \citenamefont {Prokudin}}]{Anselmino:2011ch}%
  \BibitemOpen
  \bibfield  {author} {\bibinfo {author} {\bibfnamefont {M.}~\bibnamefont
  {Anselmino}}, \bibinfo {author} {\bibfnamefont {M.}~\bibnamefont {Boglione}},
  \bibinfo {author} {\bibfnamefont {U.}~\bibnamefont {D'Alesio}}, \bibinfo
  {author} {\bibfnamefont {S.}~\bibnamefont {Melis}}, \bibinfo {author}
  {\bibfnamefont {F.}~\bibnamefont {Murgia}}, \bibinfo {author} {\bibfnamefont
  {E.}~\bibnamefont {Nocera}}, \ and\ \bibinfo {author} {\bibfnamefont
  {A.}~\bibnamefont {Prokudin}},\ }\href {\doibase 10.1103/PhysRevD.83.114019}
  {\bibfield  {journal} {\bibinfo  {journal} {Phys. Rev. D}\ }\textbf {\bibinfo
  {volume} {83}},\ \bibinfo {pages} {114019} (\bibinfo {year} {2011})},\
  \Eprint {http://arxiv.org/abs/1101.1011} {arXiv:1101.1011 [hep-ph]}
  \BibitemShut {NoStop}%
\bibitem [{\citenamefont {Anselmino}\ \emph {et~al.}(2013)\citenamefont
  {Anselmino}, \citenamefont {Boglione}, \citenamefont {D'Alesio},
  \citenamefont {Melis}, \citenamefont {Murgia},\ and\ \citenamefont
  {Prokudin}}]{Anselmino:2013vqa}%
  \BibitemOpen
  \bibfield  {author} {\bibinfo {author} {\bibfnamefont {M.}~\bibnamefont
  {Anselmino}}, \bibinfo {author} {\bibfnamefont {M.}~\bibnamefont {Boglione}},
  \bibinfo {author} {\bibfnamefont {U.}~\bibnamefont {D'Alesio}}, \bibinfo
  {author} {\bibfnamefont {S.}~\bibnamefont {Melis}}, \bibinfo {author}
  {\bibfnamefont {F.}~\bibnamefont {Murgia}}, \ and\ \bibinfo {author}
  {\bibfnamefont {A.}~\bibnamefont {Prokudin}},\ }\href {\doibase
  10.1103/PhysRevD.87.094019} {\bibfield  {journal} {\bibinfo  {journal} {Phys.
  Rev. D}\ }\textbf {\bibinfo {volume} {87}},\ \bibinfo {pages} {094019}
  (\bibinfo {year} {2013})},\ \Eprint {http://arxiv.org/abs/1303.3822}
  {arXiv:1303.3822 [hep-ph]} \BibitemShut {NoStop}%
\bibitem [{\citenamefont {Kotzinian}\ \emph {et~al.}(2006)\citenamefont
  {Kotzinian}, \citenamefont {Parsamyan},\ and\ \citenamefont
  {Prokudin}}]{Kotzinian:2006dw}%
  \BibitemOpen
  \bibfield  {author} {\bibinfo {author} {\bibfnamefont {A.}~\bibnamefont
  {Kotzinian}}, \bibinfo {author} {\bibfnamefont {B.}~\bibnamefont
  {Parsamyan}}, \ and\ \bibinfo {author} {\bibfnamefont {A.}~\bibnamefont
  {Prokudin}},\ }\href {\doibase 10.1103/PhysRevD.73.114017} {\bibfield
  {journal} {\bibinfo  {journal} {Phys. Rev. D}\ }\textbf {\bibinfo {volume}
  {73}},\ \bibinfo {pages} {114017} (\bibinfo {year} {2006})},\ \Eprint
  {http://arxiv.org/abs/hep-ph/0603194} {arXiv:hep-ph/0603194} \BibitemShut
  {NoStop}%
\bibitem [{\citenamefont {Lefky}\ and\ \citenamefont
  {Prokudin}(2015)}]{Lefky:2014eia}%
  \BibitemOpen
  \bibfield  {author} {\bibinfo {author} {\bibfnamefont {C.}~\bibnamefont
  {Lefky}}\ and\ \bibinfo {author} {\bibfnamefont {A.}~\bibnamefont
  {Prokudin}},\ }\href {\doibase 10.1103/PhysRevD.91.034010} {\bibfield
  {journal} {\bibinfo  {journal} {Phys. Rev. D}\ }\textbf {\bibinfo {volume}
  {91}},\ \bibinfo {pages} {034010} (\bibinfo {year} {2015})},\ \Eprint
  {http://arxiv.org/abs/1411.0580} {arXiv:1411.0580 [hep-ph]} \BibitemShut
  {NoStop}%
\bibitem [{\citenamefont {Jakob}\ \emph {et~al.}(1997)\citenamefont {Jakob},
  \citenamefont {Mulders},\ and\ \citenamefont {Rodrigues}}]{Jakob:1997wg}%
  \BibitemOpen
  \bibfield  {author} {\bibinfo {author} {\bibfnamefont {R.}~\bibnamefont
  {Jakob}}, \bibinfo {author} {\bibfnamefont {P.}~\bibnamefont {Mulders}}, \
  and\ \bibinfo {author} {\bibfnamefont {J.}~\bibnamefont {Rodrigues}},\ }\href
  {\doibase 10.1016/S0375-9474(97)00588-5} {\bibfield  {journal} {\bibinfo
  {journal} {Nucl. Phys. A}\ }\textbf {\bibinfo {volume} {626}},\ \bibinfo
  {pages} {937} (\bibinfo {year} {1997})},\ \Eprint
  {http://arxiv.org/abs/hep-ph/9704335} {arXiv:hep-ph/9704335} \BibitemShut
  {NoStop}%
\bibitem [{\citenamefont {Bacchetta}\ \emph {et~al.}(2008)\citenamefont
  {Bacchetta}, \citenamefont {Conti},\ and\ \citenamefont
  {Radici}}]{Bacchetta:2008af}%
  \BibitemOpen
  \bibfield  {author} {\bibinfo {author} {\bibfnamefont {A.}~\bibnamefont
  {Bacchetta}}, \bibinfo {author} {\bibfnamefont {F.}~\bibnamefont {Conti}}, \
  and\ \bibinfo {author} {\bibfnamefont {M.}~\bibnamefont {Radici}},\ }\href
  {\doibase 10.1103/PhysRevD.78.074010} {\bibfield  {journal} {\bibinfo
  {journal} {Phys. Rev. D}\ }\textbf {\bibinfo {volume} {78}},\ \bibinfo
  {pages} {074010} (\bibinfo {year} {2008})},\ \Eprint
  {http://arxiv.org/abs/0807.0323} {arXiv:0807.0323 [hep-ph]} \BibitemShut
  {NoStop}%
\bibitem [{\citenamefont {Avakian}\ \emph {et~al.}(2010)\citenamefont
  {Avakian}, \citenamefont {Efremov}, \citenamefont {Schweitzer},\ and\
  \citenamefont {Yuan}}]{Avakian:2010br}%
  \BibitemOpen
  \bibfield  {author} {\bibinfo {author} {\bibfnamefont {H.}~\bibnamefont
  {Avakian}}, \bibinfo {author} {\bibfnamefont {A.}~\bibnamefont {Efremov}},
  \bibinfo {author} {\bibfnamefont {P.}~\bibnamefont {Schweitzer}}, \ and\
  \bibinfo {author} {\bibfnamefont {F.}~\bibnamefont {Yuan}},\ }\href {\doibase
  10.1103/PhysRevD.81.074035} {\bibfield  {journal} {\bibinfo  {journal} {Phys.
  Rev. D}\ }\textbf {\bibinfo {volume} {81}},\ \bibinfo {pages} {074035}
  (\bibinfo {year} {2010})},\ \Eprint {http://arxiv.org/abs/1001.5467}
  {arXiv:1001.5467 [hep-ph]} \BibitemShut {NoStop}%
\bibitem [{\citenamefont {Cherednikov}\ \emph {et~al.}(2006)\citenamefont
  {Cherednikov}, \citenamefont {D'Alesio}, \citenamefont {Kochelev},\ and\
  \citenamefont {Murgia}}]{Cherednikov:2006zn}%
  \BibitemOpen
  \bibfield  {author} {\bibinfo {author} {\bibfnamefont {I.}~\bibnamefont
  {Cherednikov}}, \bibinfo {author} {\bibfnamefont {U.}~\bibnamefont
  {D'Alesio}}, \bibinfo {author} {\bibfnamefont {N.}~\bibnamefont {Kochelev}},
  \ and\ \bibinfo {author} {\bibfnamefont {F.}~\bibnamefont {Murgia}},\ }\href
  {\doibase 10.1016/j.physletb.2006.09.019} {\bibfield  {journal} {\bibinfo
  {journal} {Phys. Lett. B}\ }\textbf {\bibinfo {volume} {642}},\ \bibinfo
  {pages} {39} (\bibinfo {year} {2006})},\ \Eprint
  {http://arxiv.org/abs/hep-ph/0606238} {arXiv:hep-ph/0606238} \BibitemShut
  {NoStop}%
\bibitem [{\citenamefont {Yuan}(2003)}]{Yuan:2003wk}%
  \BibitemOpen
  \bibfield  {author} {\bibinfo {author} {\bibfnamefont {F.}~\bibnamefont
  {Yuan}},\ }\href {\doibase 10.1016/j.physletb.2003.09.052} {\bibfield
  {journal} {\bibinfo  {journal} {Phys. Lett. B}\ }\textbf {\bibinfo {volume}
  {575}},\ \bibinfo {pages} {45} (\bibinfo {year} {2003})},\ \Eprint
  {http://arxiv.org/abs/hep-ph/0308157} {arXiv:hep-ph/0308157} \BibitemShut
  {NoStop}%
\bibitem [{\citenamefont {Efremov}\ \emph {et~al.}(2009)\citenamefont
  {Efremov}, \citenamefont {Schweitzer}, \citenamefont {Teryaev},\ and\
  \citenamefont {Zavada}}]{Efremov:2009ze}%
  \BibitemOpen
  \bibfield  {author} {\bibinfo {author} {\bibfnamefont {A.}~\bibnamefont
  {Efremov}}, \bibinfo {author} {\bibfnamefont {P.}~\bibnamefont {Schweitzer}},
  \bibinfo {author} {\bibfnamefont {O.}~\bibnamefont {Teryaev}}, \ and\
  \bibinfo {author} {\bibfnamefont {P.}~\bibnamefont {Zavada}},\ }\href
  {\doibase 10.1103/PhysRevD.80.014021} {\bibfield  {journal} {\bibinfo
  {journal} {Phys. Rev. D}\ }\textbf {\bibinfo {volume} {80}},\ \bibinfo
  {pages} {014021} (\bibinfo {year} {2009})},\ \Eprint
  {http://arxiv.org/abs/0903.3490} {arXiv:0903.3490 [hep-ph]} \BibitemShut
  {NoStop}%
\bibitem [{\citenamefont {Pasquini}\ \emph {et~al.}(2008)\citenamefont
  {Pasquini}, \citenamefont {Cazzaniga},\ and\ \citenamefont
  {Boffi}}]{Pasquini:2008ax}%
  \BibitemOpen
  \bibfield  {author} {\bibinfo {author} {\bibfnamefont {B.}~\bibnamefont
  {Pasquini}}, \bibinfo {author} {\bibfnamefont {S.}~\bibnamefont {Cazzaniga}},
  \ and\ \bibinfo {author} {\bibfnamefont {S.}~\bibnamefont {Boffi}},\ }\href
  {\doibase 10.1103/PhysRevD.78.034025} {\bibfield  {journal} {\bibinfo
  {journal} {Phys. Rev. D}\ }\textbf {\bibinfo {volume} {78}},\ \bibinfo
  {pages} {034025} (\bibinfo {year} {2008})},\ \Eprint
  {http://arxiv.org/abs/0806.2298} {arXiv:0806.2298 [hep-ph]} \BibitemShut
  {NoStop}%
\bibitem [{\citenamefont {Maji}\ \emph {et~al.}(2016)\citenamefont {Maji},
  \citenamefont {Mondal}, \citenamefont {Chakrabarti},\ and\ \citenamefont
  {Teryaev}}]{Maji:2015vsa}%
  \BibitemOpen
  \bibfield  {author} {\bibinfo {author} {\bibfnamefont {T.}~\bibnamefont
  {Maji}}, \bibinfo {author} {\bibfnamefont {C.}~\bibnamefont {Mondal}},
  \bibinfo {author} {\bibfnamefont {D.}~\bibnamefont {Chakrabarti}}, \ and\
  \bibinfo {author} {\bibfnamefont {O.}~\bibnamefont {Teryaev}},\ }\href
  {\doibase 10.1007/JHEP01(2016)165} {\bibfield  {journal} {\bibinfo  {journal}
  {JHEP}\ }\textbf {\bibinfo {volume} {01}},\ \bibinfo {pages} {165} (\bibinfo
  {year} {2016})},\ \Eprint {http://arxiv.org/abs/1506.04560} {arXiv:1506.04560
  [hep-ph]} \BibitemShut {NoStop}%
\bibitem [{\citenamefont {Maji}\ and\ \citenamefont
  {Chakrabarti}(2017)}]{Maji:2017bcz}%
  \BibitemOpen
  \bibfield  {author} {\bibinfo {author} {\bibfnamefont {T.}~\bibnamefont
  {Maji}}\ and\ \bibinfo {author} {\bibfnamefont {D.}~\bibnamefont
  {Chakrabarti}},\ }\href {\doibase 10.1103/PhysRevD.95.074009} {\bibfield
  {journal} {\bibinfo  {journal} {Phys. Rev. D}\ }\textbf {\bibinfo {volume}
  {95}},\ \bibinfo {pages} {074009} (\bibinfo {year} {2017})},\ \Eprint
  {http://arxiv.org/abs/1702.04557} {arXiv:1702.04557 [hep-ph]} \BibitemShut
  {NoStop}%
\bibitem [{\citenamefont {Shi}\ and\ \citenamefont
  {Cloët}(2019)}]{Shi:2018zqd}%
  \BibitemOpen
  \bibfield  {author} {\bibinfo {author} {\bibfnamefont {C.}~\bibnamefont
  {Shi}}\ and\ \bibinfo {author} {\bibfnamefont {I.~C.}\ \bibnamefont
  {Cloët}},\ }\href {\doibase 10.1103/PhysRevLett.122.082301} {\bibfield
  {journal} {\bibinfo  {journal} {Phys. Rev. Lett.}\ }\textbf {\bibinfo
  {volume} {122}},\ \bibinfo {pages} {082301} (\bibinfo {year} {2019})},\
  \Eprint {http://arxiv.org/abs/1806.04799} {arXiv:1806.04799 [nucl-th]}
  \BibitemShut {NoStop}%
\bibitem [{\citenamefont {Shi}\ \emph {et~al.}(2020)\citenamefont {Shi},
  \citenamefont {Bednar}, \citenamefont {Cloët},\ and\ \citenamefont
  {Freese}}]{Shi:2020pqe}%
  \BibitemOpen
  \bibfield  {author} {\bibinfo {author} {\bibfnamefont {C.}~\bibnamefont
  {Shi}}, \bibinfo {author} {\bibfnamefont {K.}~\bibnamefont {Bednar}},
  \bibinfo {author} {\bibfnamefont {I.~C.}\ \bibnamefont {Cloët}}, \ and\
  \bibinfo {author} {\bibfnamefont {A.}~\bibnamefont {Freese}},\ }\href
  {\doibase 10.1103/PhysRevD.101.074014} {\bibfield  {journal} {\bibinfo
  {journal} {Phys. Rev. D}\ }\textbf {\bibinfo {volume} {101}},\ \bibinfo
  {pages} {074014} (\bibinfo {year} {2020})},\ \Eprint
  {http://arxiv.org/abs/2003.03037} {arXiv:2003.03037 [hep-ph]} \BibitemShut
  {NoStop}%
\bibitem [{\citenamefont {Musch}\ \emph {et~al.}(2011)\citenamefont {Musch},
  \citenamefont {Hagler}, \citenamefont {Negele},\ and\ \citenamefont
  {Schafer}}]{Musch:2010ka}%
  \BibitemOpen
  \bibfield  {author} {\bibinfo {author} {\bibfnamefont {B.~U.}\ \bibnamefont
  {Musch}}, \bibinfo {author} {\bibfnamefont {P.}~\bibnamefont {Hagler}},
  \bibinfo {author} {\bibfnamefont {J.~W.}\ \bibnamefont {Negele}}, \ and\
  \bibinfo {author} {\bibfnamefont {A.}~\bibnamefont {Schafer}},\ }\href
  {\doibase 10.1103/PhysRevD.83.094507} {\bibfield  {journal} {\bibinfo
  {journal} {Phys. Rev. D}\ }\textbf {\bibinfo {volume} {83}},\ \bibinfo
  {pages} {094507} (\bibinfo {year} {2011})},\ \Eprint
  {http://arxiv.org/abs/1011.1213} {arXiv:1011.1213 [hep-lat]} \BibitemShut
  {NoStop}%
\bibitem [{\citenamefont {Musch}\ \emph {et~al.}(2012)\citenamefont {Musch},
  \citenamefont {Hagler}, \citenamefont {Engelhardt}, \citenamefont {Negele},\
  and\ \citenamefont {Schafer}}]{Musch:2011er}%
  \BibitemOpen
  \bibfield  {author} {\bibinfo {author} {\bibfnamefont {B.}~\bibnamefont
  {Musch}}, \bibinfo {author} {\bibfnamefont {P.}~\bibnamefont {Hagler}},
  \bibinfo {author} {\bibfnamefont {M.}~\bibnamefont {Engelhardt}}, \bibinfo
  {author} {\bibfnamefont {J.}~\bibnamefont {Negele}}, \ and\ \bibinfo {author}
  {\bibfnamefont {A.}~\bibnamefont {Schafer}},\ }\href {\doibase
  10.1103/PhysRevD.85.094510} {\bibfield  {journal} {\bibinfo  {journal} {Phys.
  Rev. D}\ }\textbf {\bibinfo {volume} {85}},\ \bibinfo {pages} {094510}
  (\bibinfo {year} {2012})},\ \Eprint {http://arxiv.org/abs/1111.4249}
  {arXiv:1111.4249 [hep-lat]} \BibitemShut {NoStop}%
\bibitem [{\citenamefont {Ji}\ \emph {et~al.}(2015)\citenamefont {Ji},
  \citenamefont {Sun}, \citenamefont {Xiong},\ and\ \citenamefont
  {Yuan}}]{Ji:2014hxa}%
  \BibitemOpen
  \bibfield  {author} {\bibinfo {author} {\bibfnamefont {X.}~\bibnamefont
  {Ji}}, \bibinfo {author} {\bibfnamefont {P.}~\bibnamefont {Sun}}, \bibinfo
  {author} {\bibfnamefont {X.}~\bibnamefont {Xiong}}, \ and\ \bibinfo {author}
  {\bibfnamefont {F.}~\bibnamefont {Yuan}},\ }\href {\doibase
  10.1103/PhysRevD.91.074009} {\bibfield  {journal} {\bibinfo  {journal} {Phys.
  Rev. D}\ }\textbf {\bibinfo {volume} {91}},\ \bibinfo {pages} {074009}
  (\bibinfo {year} {2015})},\ \Eprint {http://arxiv.org/abs/1405.7640}
  {arXiv:1405.7640 [hep-ph]} \BibitemShut {NoStop}%
\bibitem [{\citenamefont {Yoon}\ \emph {et~al.}(2017)\citenamefont {Yoon},
  \citenamefont {Engelhardt}, \citenamefont {Gupta}, \citenamefont
  {Bhattacharya}, \citenamefont {Green}, \citenamefont {Musch}, \citenamefont
  {Negele}, \citenamefont {Pochinsky}, \citenamefont {Schäfer},\ and\
  \citenamefont {Syritsyn}}]{Yoon:2017qzo}%
  \BibitemOpen
  \bibfield  {author} {\bibinfo {author} {\bibfnamefont {B.}~\bibnamefont
  {Yoon}}, \bibinfo {author} {\bibfnamefont {M.}~\bibnamefont {Engelhardt}},
  \bibinfo {author} {\bibfnamefont {R.}~\bibnamefont {Gupta}}, \bibinfo
  {author} {\bibfnamefont {T.}~\bibnamefont {Bhattacharya}}, \bibinfo {author}
  {\bibfnamefont {J.~R.}\ \bibnamefont {Green}}, \bibinfo {author}
  {\bibfnamefont {B.~U.}\ \bibnamefont {Musch}}, \bibinfo {author}
  {\bibfnamefont {J.~W.}\ \bibnamefont {Negele}}, \bibinfo {author}
  {\bibfnamefont {A.~V.}\ \bibnamefont {Pochinsky}}, \bibinfo {author}
  {\bibfnamefont {A.}~\bibnamefont {Schäfer}}, \ and\ \bibinfo {author}
  {\bibfnamefont {S.~N.}\ \bibnamefont {Syritsyn}},\ }\href {\doibase
  10.1103/PhysRevD.96.094508} {\bibfield  {journal} {\bibinfo  {journal} {Phys.
  Rev. D}\ }\textbf {\bibinfo {volume} {96}},\ \bibinfo {pages} {094508}
  (\bibinfo {year} {2017})},\ \Eprint {http://arxiv.org/abs/1706.03406}
  {arXiv:1706.03406 [hep-lat]} \BibitemShut {NoStop}%
\bibitem [{\citenamefont {Constantinou}\ \emph {et~al.}(2020)\citenamefont
  {Constantinou} \emph {et~al.}}]{Lin:2020rut}%
  \BibitemOpen
  \bibfield  {author} {\bibinfo {author} {\bibfnamefont {M.}~\bibnamefont
  {Constantinou}} \emph {et~al.},\ }\href@noop {} {\  (\bibinfo {year}
  {2020})},\ \Eprint {http://arxiv.org/abs/2006.08636} {arXiv:2006.08636
  [hep-ph]} \BibitemShut {NoStop}%
\bibitem [{\citenamefont {Vary}\ \emph {et~al.}(2010)\citenamefont {Vary},
  \citenamefont {Honkanen}, \citenamefont {Li}, \citenamefont {Maris},
  \citenamefont {Brodsky}, \citenamefont {Harindranath}, \citenamefont
  {de~Teramond}, \citenamefont {Sternberg}, \citenamefont {Ng},\ and\
  \citenamefont {Yang}}]{Vary:2009gt}%
  \BibitemOpen
  \bibfield  {author} {\bibinfo {author} {\bibfnamefont {J.}~\bibnamefont
  {Vary}}, \bibinfo {author} {\bibfnamefont {H.}~\bibnamefont {Honkanen}},
  \bibinfo {author} {\bibfnamefont {J.}~\bibnamefont {Li}}, \bibinfo {author}
  {\bibfnamefont {P.}~\bibnamefont {Maris}}, \bibinfo {author} {\bibfnamefont
  {S.}~\bibnamefont {Brodsky}}, \bibinfo {author} {\bibfnamefont
  {A.}~\bibnamefont {Harindranath}}, \bibinfo {author} {\bibfnamefont
  {G.}~\bibnamefont {de~Teramond}}, \bibinfo {author} {\bibfnamefont
  {P.}~\bibnamefont {Sternberg}}, \bibinfo {author} {\bibfnamefont
  {E.}~\bibnamefont {Ng}}, \ and\ \bibinfo {author} {\bibfnamefont
  {C.}~\bibnamefont {Yang}},\ }\href {\doibase 10.1103/PhysRevC.81.035205}
  {\bibfield  {journal} {\bibinfo  {journal} {Phys. Rev. C}\ }\textbf {\bibinfo
  {volume} {81}},\ \bibinfo {pages} {035205} (\bibinfo {year} {2010})},\
  \Eprint {http://arxiv.org/abs/0905.1411} {arXiv:0905.1411 [nucl-th]}
  \BibitemShut {NoStop}%
\bibitem [{\citenamefont {Zhao}(2015)}]{Zhao:2014hpa}%
  \BibitemOpen
  \bibfield  {author} {\bibinfo {author} {\bibfnamefont {X.}~\bibnamefont
  {Zhao}},\ }\href {\doibase 10.1007/s00601-015-1003-y} {\bibfield  {journal}
  {\bibinfo  {journal} {Few Body Syst.}\ }\textbf {\bibinfo {volume} {56}},\
  \bibinfo {pages} {257} (\bibinfo {year} {2015})},\ \Eprint
  {http://arxiv.org/abs/1411.7748} {arXiv:1411.7748 [nucl-th]} \BibitemShut
  {NoStop}%
\bibitem [{\citenamefont {Zhao}\ \emph {et~al.}(2014)\citenamefont {Zhao},
  \citenamefont {Honkanen}, \citenamefont {Maris}, \citenamefont {Vary},\ and\
  \citenamefont {Brodsky}}]{zhao2014electrong2}%
  \BibitemOpen
  \bibfield  {author} {\bibinfo {author} {\bibfnamefont {X.}~\bibnamefont
  {Zhao}}, \bibinfo {author} {\bibfnamefont {H.}~\bibnamefont {Honkanen}},
  \bibinfo {author} {\bibfnamefont {P.}~\bibnamefont {Maris}}, \bibinfo
  {author} {\bibfnamefont {J.~P.}\ \bibnamefont {Vary}}, \ and\ \bibinfo
  {author} {\bibfnamefont {S.~J.}\ \bibnamefont {Brodsky}},\ }\href {\doibase
  10.1016/j.physletb.2014.08.020} {\bibfield  {journal} {\bibinfo  {journal}
  {Phys. Lett. B}\ }\textbf {\bibinfo {volume} {737}},\ \bibinfo {pages} {65}
  (\bibinfo {year} {2014})},\ \Eprint {http://arxiv.org/abs/1402.4195}
  {arXiv:1402.4195 [nucl-th]} \BibitemShut {NoStop}%
\bibitem [{\citenamefont {Wiecki}\ \emph {et~al.}(2015)\citenamefont {Wiecki},
  \citenamefont {Li}, \citenamefont {Zhao}, \citenamefont {Maris},\ and\
  \citenamefont {Vary}}]{Wiecki:2014ola}%
  \BibitemOpen
  \bibfield  {author} {\bibinfo {author} {\bibfnamefont {P.}~\bibnamefont
  {Wiecki}}, \bibinfo {author} {\bibfnamefont {Y.}~\bibnamefont {Li}}, \bibinfo
  {author} {\bibfnamefont {X.}~\bibnamefont {Zhao}}, \bibinfo {author}
  {\bibfnamefont {P.}~\bibnamefont {Maris}}, \ and\ \bibinfo {author}
  {\bibfnamefont {J.~P.}\ \bibnamefont {Vary}},\ }\href {\doibase
  10.1103/PhysRevD.91.105009} {\bibfield  {journal} {\bibinfo  {journal} {Phys.
  Rev. D}\ }\textbf {\bibinfo {volume} {91}},\ \bibinfo {pages} {105009}
  (\bibinfo {year} {2015})},\ \Eprint {http://arxiv.org/abs/1404.6234}
  {arXiv:1404.6234 [nucl-th]} \BibitemShut {NoStop}%
\bibitem [{\citenamefont {Li}\ \emph {et~al.}(2016)\citenamefont {Li},
  \citenamefont {Maris}, \citenamefont {Zhao},\ and\ \citenamefont
  {Vary}}]{LY2016}%
  \BibitemOpen
  \bibfield  {author} {\bibinfo {author} {\bibfnamefont {Y.}~\bibnamefont
  {Li}}, \bibinfo {author} {\bibfnamefont {P.}~\bibnamefont {Maris}}, \bibinfo
  {author} {\bibfnamefont {X.}~\bibnamefont {Zhao}}, \ and\ \bibinfo {author}
  {\bibfnamefont {J.~P.}\ \bibnamefont {Vary}},\ }\href {\doibase
  https://doi.org/10.1016/j.physletb.2016.04.065} {\bibfield  {journal}
  {\bibinfo  {journal} {Physics Letters B}\ }\textbf {\bibinfo {volume}
  {758}},\ \bibinfo {pages} {118 } (\bibinfo {year} {2016})}\BibitemShut
  {NoStop}%
\bibitem [{\citenamefont {Jia}\ and\ \citenamefont {Vary}(2019)}]{Jia:2018ary}%
  \BibitemOpen
  \bibfield  {author} {\bibinfo {author} {\bibfnamefont {S.}~\bibnamefont
  {Jia}}\ and\ \bibinfo {author} {\bibfnamefont {J.~P.}\ \bibnamefont {Vary}},\
  }\href {\doibase 10.1103/PhysRevC.99.035206} {\bibfield  {journal} {\bibinfo
  {journal} {Phys. Rev. C}\ }\textbf {\bibinfo {volume} {99}},\ \bibinfo
  {pages} {035206} (\bibinfo {year} {2019})},\ \Eprint
  {http://arxiv.org/abs/1811.08512} {arXiv:1811.08512 [nucl-th]} \BibitemShut
  {NoStop}%
\bibitem [{\citenamefont {Mondal}\ \emph {et~al.}(2020)\citenamefont {Mondal},
  \citenamefont {Xu}, \citenamefont {Lan}, \citenamefont {Zhao}, \citenamefont
  {Li}, \citenamefont {Chakrabarti},\ and\ \citenamefont {Vary}}]{Xu:2019xhk}%
  \BibitemOpen
  \bibfield  {author} {\bibinfo {author} {\bibfnamefont {C.}~\bibnamefont
  {Mondal}}, \bibinfo {author} {\bibfnamefont {S.}~\bibnamefont {Xu}}, \bibinfo
  {author} {\bibfnamefont {J.}~\bibnamefont {Lan}}, \bibinfo {author}
  {\bibfnamefont {X.}~\bibnamefont {Zhao}}, \bibinfo {author} {\bibfnamefont
  {Y.}~\bibnamefont {Li}}, \bibinfo {author} {\bibfnamefont {D.}~\bibnamefont
  {Chakrabarti}}, \ and\ \bibinfo {author} {\bibfnamefont {J.~P.}\ \bibnamefont
  {Vary}},\ }\href {\doibase 10.1103/PhysRevD.102.016008} {\bibfield  {journal}
  {\bibinfo  {journal} {Phys. Rev. D}\ }\textbf {\bibinfo {volume} {102}},\
  \bibinfo {pages} {016008} (\bibinfo {year} {2020})},\ \Eprint
  {http://arxiv.org/abs/1911.10913} {arXiv:1911.10913 [hep-ph]} \BibitemShut
  {NoStop}%
\bibitem [{\citenamefont {Lan}\ \emph {et~al.}(2019)\citenamefont {Lan},
  \citenamefont {Mondal}, \citenamefont {Jia}, \citenamefont {Zhao},\ and\
  \citenamefont {Vary}}]{Lan:2019vui}%
  \BibitemOpen
  \bibfield  {author} {\bibinfo {author} {\bibfnamefont {J.}~\bibnamefont
  {Lan}}, \bibinfo {author} {\bibfnamefont {C.}~\bibnamefont {Mondal}},
  \bibinfo {author} {\bibfnamefont {S.}~\bibnamefont {Jia}}, \bibinfo {author}
  {\bibfnamefont {X.}~\bibnamefont {Zhao}}, \ and\ \bibinfo {author}
  {\bibfnamefont {J.~P.}\ \bibnamefont {Vary}},\ }\href {\doibase
  10.1103/PhysRevLett.122.172001} {\bibfield  {journal} {\bibinfo  {journal}
  {Phys. Rev. Lett.}\ }\textbf {\bibinfo {volume} {122}},\ \bibinfo {pages}
  {172001} (\bibinfo {year} {2019})},\ \Eprint
  {http://arxiv.org/abs/1901.11430} {arXiv:1901.11430 [nucl-th]} \BibitemShut
  {NoStop}%
\bibitem [{\citenamefont {Chakrabarti}\ \emph {et~al.}(2014)\citenamefont
  {Chakrabarti}, \citenamefont {Zhao}, \citenamefont {Honkanen}, \citenamefont
  {Manohar}, \citenamefont {Maris},\ and\ \citenamefont
  {Vary}}]{Chakrabarti2014GPDinBLFQ}%
  \BibitemOpen
  \bibfield  {author} {\bibinfo {author} {\bibfnamefont {D.}~\bibnamefont
  {Chakrabarti}}, \bibinfo {author} {\bibfnamefont {X.}~\bibnamefont {Zhao}},
  \bibinfo {author} {\bibfnamefont {H.}~\bibnamefont {Honkanen}}, \bibinfo
  {author} {\bibfnamefont {R.}~\bibnamefont {Manohar}}, \bibinfo {author}
  {\bibfnamefont {P.}~\bibnamefont {Maris}}, \ and\ \bibinfo {author}
  {\bibfnamefont {J.}~\bibnamefont {Vary}},\ }\href {\doibase
  10.1103/PhysRevD.89.116004} {\bibfield  {journal} {\bibinfo  {journal} {Phys.
  Rev. D}\ }\textbf {\bibinfo {volume} {89}},\ \bibinfo {pages} {116004}
  (\bibinfo {year} {2014})},\ \Eprint {http://arxiv.org/abs/1403.0704}
  {arXiv:1403.0704 [hep-ph]} \BibitemShut {NoStop}%
\bibitem [{\citenamefont {Lan}\ \emph {et~al.}(2020)\citenamefont {Lan},
  \citenamefont {Mondal}, \citenamefont {Li}, \citenamefont {Li}, \citenamefont
  {Tang}, \citenamefont {Zhao},\ and\ \citenamefont {Vary}}]{Lan:2019img}%
  \BibitemOpen
  \bibfield  {author} {\bibinfo {author} {\bibfnamefont {J.}~\bibnamefont
  {Lan}}, \bibinfo {author} {\bibfnamefont {C.}~\bibnamefont {Mondal}},
  \bibinfo {author} {\bibfnamefont {M.}~\bibnamefont {Li}}, \bibinfo {author}
  {\bibfnamefont {Y.}~\bibnamefont {Li}}, \bibinfo {author} {\bibfnamefont
  {S.}~\bibnamefont {Tang}}, \bibinfo {author} {\bibfnamefont {X.}~\bibnamefont
  {Zhao}}, \ and\ \bibinfo {author} {\bibfnamefont {J.~P.}\ \bibnamefont
  {Vary}},\ }\href {\doibase 10.1103/PhysRevD.102.014020} {\bibfield  {journal}
  {\bibinfo  {journal} {Phys. Rev. D}\ }\textbf {\bibinfo {volume} {102}},\
  \bibinfo {pages} {014020} (\bibinfo {year} {2020})},\ \Eprint
  {http://arxiv.org/abs/1911.11676} {arXiv:1911.11676 [nucl-th]} \BibitemShut
  {NoStop}%
\bibitem [{\citenamefont {Zhao}\ \emph
  {et~al.}(2013{\natexlab{a}})\citenamefont {Zhao}, \citenamefont {Ilderton},
  \citenamefont {Maris},\ and\ \citenamefont {Vary}}]{xbzhao2013tBLFQ}%
  \BibitemOpen
  \bibfield  {author} {\bibinfo {author} {\bibfnamefont {X.}~\bibnamefont
  {Zhao}}, \bibinfo {author} {\bibfnamefont {A.}~\bibnamefont {Ilderton}},
  \bibinfo {author} {\bibfnamefont {P.}~\bibnamefont {Maris}}, \ and\ \bibinfo
  {author} {\bibfnamefont {J.~P.}\ \bibnamefont {Vary}},\ }\href {\doibase
  10.1103/physrevd.88.065014} {\bibfield  {journal} {\bibinfo  {journal}
  {Physical Review D}\ }\textbf {\bibinfo {volume} {88}} (\bibinfo {year}
  {2013}{\natexlab{a}}),\ 10.1103/physrevd.88.065014}\BibitemShut {NoStop}%
\bibitem [{\citenamefont {Zhao}\ \emph
  {et~al.}(2013{\natexlab{b}})\citenamefont {Zhao}, \citenamefont {Ilderton},
  \citenamefont {Maris},\ and\ \citenamefont {Vary}}]{Zhao:2013jia}%
  \BibitemOpen
  \bibfield  {author} {\bibinfo {author} {\bibfnamefont {X.}~\bibnamefont
  {Zhao}}, \bibinfo {author} {\bibfnamefont {A.}~\bibnamefont {Ilderton}},
  \bibinfo {author} {\bibfnamefont {P.}~\bibnamefont {Maris}}, \ and\ \bibinfo
  {author} {\bibfnamefont {J.~P.}\ \bibnamefont {Vary}},\ }\href {\doibase
  10.1016/j.physletb.2013.09.030} {\bibfield  {journal} {\bibinfo  {journal}
  {Phys. Lett. B}\ }\textbf {\bibinfo {volume} {726}},\ \bibinfo {pages} {856}
  (\bibinfo {year} {2013}{\natexlab{b}})},\ \Eprint
  {http://arxiv.org/abs/1309.5338} {arXiv:1309.5338 [nucl-th]} \BibitemShut
  {NoStop}%
\bibitem [{\citenamefont {Hu}\ \emph {et~al.}(2019)\citenamefont {Hu},
  \citenamefont {Ilderton},\ and\ \citenamefont {Zhao}}]{Hu:2019hjx}%
  \BibitemOpen
  \bibfield  {author} {\bibinfo {author} {\bibfnamefont {B.}~\bibnamefont
  {Hu}}, \bibinfo {author} {\bibfnamefont {A.}~\bibnamefont {Ilderton}}, \ and\
  \bibinfo {author} {\bibfnamefont {X.}~\bibnamefont {Zhao}},\ }\href@noop {}
  {\  (\bibinfo {year} {2019})},\ \Eprint {http://arxiv.org/abs/1911.12307}
  {arXiv:1911.12307 [nucl-th]} \BibitemShut {NoStop}%
\bibitem [{\citenamefont {Bacchetta}\ \emph {et~al.}(2016)\citenamefont
  {Bacchetta}, \citenamefont {Mantovani},\ and\ \citenamefont
  {Pasquini}}]{bacchetta2016electron}%
  \BibitemOpen
  \bibfield  {author} {\bibinfo {author} {\bibfnamefont {A.}~\bibnamefont
  {Bacchetta}}, \bibinfo {author} {\bibfnamefont {L.}~\bibnamefont
  {Mantovani}}, \ and\ \bibinfo {author} {\bibfnamefont {B.}~\bibnamefont
  {Pasquini}},\ }\href {\doibase 10.1103/physrevd.93.013005} {\bibfield
  {journal} {\bibinfo  {journal} {Physical Review D}\ }\textbf {\bibinfo
  {volume} {93}} (\bibinfo {year} {2016}),\
  10.1103/physrevd.93.013005}\BibitemShut {NoStop}%
\bibitem [{\citenamefont {Brodsky}\ \emph
  {et~al.}(2001{\natexlab{a}})\citenamefont {Brodsky}, \citenamefont {Hwang},
  \citenamefont {Ma},\ and\ \citenamefont {Schmidt}}]{Brodsky2000lightcone}%
  \BibitemOpen
  \bibfield  {author} {\bibinfo {author} {\bibfnamefont {S.~J.}\ \bibnamefont
  {Brodsky}}, \bibinfo {author} {\bibfnamefont {D.~S.}\ \bibnamefont {Hwang}},
  \bibinfo {author} {\bibfnamefont {B.-Q.}\ \bibnamefont {Ma}}, \ and\ \bibinfo
  {author} {\bibfnamefont {I.}~\bibnamefont {Schmidt}},\ }\href {\doibase
  10.1016/S0550-3213(00)00626-X} {\bibfield  {journal} {\bibinfo  {journal}
  {Nucl. Phys. B}\ }\textbf {\bibinfo {volume} {593}},\ \bibinfo {pages} {311}
  (\bibinfo {year} {2001}{\natexlab{a}})},\ \Eprint
  {http://arxiv.org/abs/hep-th/0003082} {arXiv:hep-th/0003082} \BibitemShut
  {NoStop}%
\bibitem [{\citenamefont {Brodsky}\ \emph
  {et~al.}(2001{\natexlab{b}})\citenamefont {Brodsky}, \citenamefont {Diehl},\
  and\ \citenamefont {Hwang}}]{Brodsky:2000xy}%
  \BibitemOpen
  \bibfield  {author} {\bibinfo {author} {\bibfnamefont {S.~J.}\ \bibnamefont
  {Brodsky}}, \bibinfo {author} {\bibfnamefont {M.}~\bibnamefont {Diehl}}, \
  and\ \bibinfo {author} {\bibfnamefont {D.~S.}\ \bibnamefont {Hwang}},\ }\href
  {\doibase 10.1016/S0550-3213(00)00695-7} {\bibfield  {journal} {\bibinfo
  {journal} {Nucl. Phys. B}\ }\textbf {\bibinfo {volume} {596}},\ \bibinfo
  {pages} {99} (\bibinfo {year} {2001}{\natexlab{b}})},\ \Eprint
  {http://arxiv.org/abs/hep-ph/0009254} {arXiv:hep-ph/0009254} \BibitemShut
  {NoStop}%
\bibitem [{\citenamefont {Brodsky}\ \emph {et~al.}(2007)\citenamefont
  {Brodsky}, \citenamefont {Chakrabarti}, \citenamefont {Harindranath},
  \citenamefont {Mukherjee},\ and\ \citenamefont {Vary}}]{Brodsky:2006ku}%
  \BibitemOpen
  \bibfield  {author} {\bibinfo {author} {\bibfnamefont {S.}~\bibnamefont
  {Brodsky}}, \bibinfo {author} {\bibfnamefont {D.}~\bibnamefont
  {Chakrabarti}}, \bibinfo {author} {\bibfnamefont {A.}~\bibnamefont
  {Harindranath}}, \bibinfo {author} {\bibfnamefont {A.}~\bibnamefont
  {Mukherjee}}, \ and\ \bibinfo {author} {\bibfnamefont {J.}~\bibnamefont
  {Vary}},\ }\href {\doibase 10.1103/PhysRevD.75.014003} {\bibfield  {journal}
  {\bibinfo  {journal} {Phys. Rev. D}\ }\textbf {\bibinfo {volume} {75}},\
  \bibinfo {pages} {014003} (\bibinfo {year} {2007})},\ \Eprint
  {http://arxiv.org/abs/hep-ph/0611159} {arXiv:hep-ph/0611159} \BibitemShut
  {NoStop}%
\bibitem [{\citenamefont {Dahiya}\ \emph {et~al.}(2007)\citenamefont {Dahiya},
  \citenamefont {Mukherjee},\ and\ \citenamefont {Ray}}]{Dahiya:2007is}%
  \BibitemOpen
  \bibfield  {author} {\bibinfo {author} {\bibfnamefont {H.}~\bibnamefont
  {Dahiya}}, \bibinfo {author} {\bibfnamefont {A.}~\bibnamefont {Mukherjee}}, \
  and\ \bibinfo {author} {\bibfnamefont {S.}~\bibnamefont {Ray}},\ }\href
  {\doibase 10.1103/PhysRevD.76.034010} {\bibfield  {journal} {\bibinfo
  {journal} {Phys. Rev. D}\ }\textbf {\bibinfo {volume} {76}},\ \bibinfo
  {pages} {034010} (\bibinfo {year} {2007})},\ \Eprint
  {http://arxiv.org/abs/0705.3580} {arXiv:0705.3580 [hep-ph]} \BibitemShut
  {NoStop}%
\bibitem [{\citenamefont {Kumar}\ and\ \citenamefont
  {Dahiya}(2015)}]{Kumar:2015fta}%
  \BibitemOpen
  \bibfield  {author} {\bibinfo {author} {\bibfnamefont {N.}~\bibnamefont
  {Kumar}}\ and\ \bibinfo {author} {\bibfnamefont {H.}~\bibnamefont {Dahiya}},\
  }\href {\doibase 10.1142/S0217751X15500104} {\bibfield  {journal} {\bibinfo
  {journal} {Int. J. Mod. Phys. A}\ }\textbf {\bibinfo {volume} {30}},\
  \bibinfo {pages} {1550010} (\bibinfo {year} {2015})},\ \Eprint
  {http://arxiv.org/abs/1501.04745} {arXiv:1501.04745 [hep-ph]} \BibitemShut
  {NoStop}%
\bibitem [{\citenamefont {Kumar}\ and\ \citenamefont
  {Mondal}(2018)}]{Kumar:2017xcm}%
  \BibitemOpen
  \bibfield  {author} {\bibinfo {author} {\bibfnamefont {N.}~\bibnamefont
  {Kumar}}\ and\ \bibinfo {author} {\bibfnamefont {C.}~\bibnamefont {Mondal}},\
  }\href {\doibase 10.1016/j.nuclphysb.2018.04.014} {\bibfield  {journal}
  {\bibinfo  {journal} {Nucl. Phys. B}\ }\textbf {\bibinfo {volume} {931}},\
  \bibinfo {pages} {226} (\bibinfo {year} {2018})},\ \Eprint
  {http://arxiv.org/abs/1705.03183} {arXiv:1705.03183 [hep-ph]} \BibitemShut
  {NoStop}%
\bibitem [{\citenamefont {Ji}\ and\ \citenamefont {Yuan}(2002)}]{jxd2002where}%
  \BibitemOpen
  \bibfield  {author} {\bibinfo {author} {\bibfnamefont {X.-d.}\ \bibnamefont
  {Ji}}\ and\ \bibinfo {author} {\bibfnamefont {F.}~\bibnamefont {Yuan}},\
  }\href {\doibase 10.1016/S0370-2693(02)02384-5} {\bibfield  {journal}
  {\bibinfo  {journal} {Phys. Lett. B}\ }\textbf {\bibinfo {volume} {543}},\
  \bibinfo {pages} {66} (\bibinfo {year} {2002})},\ \Eprint
  {http://arxiv.org/abs/hep-ph/0206057} {arXiv:hep-ph/0206057} \BibitemShut
  {NoStop}%
\bibitem [{\citenamefont {Belitsky}\ \emph {et~al.}(2003)\citenamefont
  {Belitsky}, \citenamefont {Ji},\ and\ \citenamefont
  {Yuan}}]{Belitsky:2002sm}%
  \BibitemOpen
  \bibfield  {author} {\bibinfo {author} {\bibfnamefont {A.~V.}\ \bibnamefont
  {Belitsky}}, \bibinfo {author} {\bibfnamefont {X.}~\bibnamefont {Ji}}, \ and\
  \bibinfo {author} {\bibfnamefont {F.}~\bibnamefont {Yuan}},\ }\href {\doibase
  10.1016/S0550-3213(03)00121-4} {\bibfield  {journal} {\bibinfo  {journal}
  {Nucl. Phys. B}\ }\textbf {\bibinfo {volume} {656}},\ \bibinfo {pages} {165}
  (\bibinfo {year} {2003})},\ \Eprint {http://arxiv.org/abs/hep-ph/0208038}
  {arXiv:hep-ph/0208038} \BibitemShut {NoStop}%
\bibitem [{\citenamefont {Brodsky}\ \emph {et~al.}(1998)\citenamefont
  {Brodsky}, \citenamefont {Pauli},\ and\ \citenamefont
  {Pinsky}}]{brodsky1998treatise}%
  \BibitemOpen
  \bibfield  {author} {\bibinfo {author} {\bibfnamefont {S.~J.}\ \bibnamefont
  {Brodsky}}, \bibinfo {author} {\bibfnamefont {H.-C.}\ \bibnamefont {Pauli}},
  \ and\ \bibinfo {author} {\bibfnamefont {S.~S.}\ \bibnamefont {Pinsky}},\
  }\href {\doibase https://doi.org/10.1016/S0370-1573(97)00089-6} {\bibfield
  {journal} {\bibinfo  {journal} {Physics Reports}\ }\textbf {\bibinfo {volume}
  {301}},\ \bibinfo {pages} {299 } (\bibinfo {year} {1998})}\BibitemShut
  {NoStop}%
\bibitem [{\citenamefont {Kogut}\ and\ \citenamefont
  {Soper}(1970)}]{kogut&soper1970QED_infi_mom}%
  \BibitemOpen
  \bibfield  {author} {\bibinfo {author} {\bibfnamefont {J.~B.}\ \bibnamefont
  {Kogut}}\ and\ \bibinfo {author} {\bibfnamefont {D.~E.}\ \bibnamefont
  {Soper}},\ }\href {\doibase 10.1103/PhysRevD.1.2901} {\bibfield  {journal}
  {\bibinfo  {journal} {Phys. Rev. D}\ }\textbf {\bibinfo {volume} {1}},\
  \bibinfo {pages} {2901} (\bibinfo {year} {1970})}\BibitemShut {NoStop}%
\bibitem [{\citenamefont {Maris}\ \emph {et~al.}(2013)\citenamefont {Maris},
  \citenamefont {Wiecki}, \citenamefont {Li}, \citenamefont {Zhao},\ and\
  \citenamefont {Vary}}]{maris2013bound}%
  \BibitemOpen
  \bibfield  {author} {\bibinfo {author} {\bibfnamefont {P.}~\bibnamefont
  {Maris}}, \bibinfo {author} {\bibfnamefont {P.}~\bibnamefont {Wiecki}},
  \bibinfo {author} {\bibfnamefont {Y.}~\bibnamefont {Li}}, \bibinfo {author}
  {\bibfnamefont {X.}~\bibnamefont {Zhao}}, \ and\ \bibinfo {author}
  {\bibfnamefont {J.~P.}\ \bibnamefont {Vary}},\ }\bibfield  {booktitle} {\emph
  {\bibinfo {booktitle} {{Proceedings, Conference on Modern approaches to
  nonperturbative gauge theories and their applications (Light Cone 2012):
  Cracow, Poland, July 8-13, 2012}}},\ }\href {\doibase
  10.5506/APhysPolBSupp.6.321} {\bibfield  {journal} {\bibinfo  {journal} {Acta
  Phys. Polon. Supp.}\ }\textbf {\bibinfo {volume} {6}},\ \bibinfo {pages}
  {321} (\bibinfo {year} {2013})}\BibitemShut {NoStop}%
\bibitem [{\citenamefont {Navr\'atil}\ \emph {et~al.}(2000)\citenamefont
  {Navr\'atil}, \citenamefont {Vary},\ and\ \citenamefont
  {Barrett}}]{vary2000NCFC}%
  \BibitemOpen
  \bibfield  {author} {\bibinfo {author} {\bibfnamefont {P.}~\bibnamefont
  {Navr\'atil}}, \bibinfo {author} {\bibfnamefont {J.~P.}\ \bibnamefont
  {Vary}}, \ and\ \bibinfo {author} {\bibfnamefont {B.~R.}\ \bibnamefont
  {Barrett}},\ }\href {\doibase 10.1103/PhysRevC.62.054311} {\bibfield
  {journal} {\bibinfo  {journal} {Phys. Rev. C}\ }\textbf {\bibinfo {volume}
  {62}},\ \bibinfo {pages} {054311} (\bibinfo {year} {2000})}\BibitemShut
  {NoStop}%
\bibitem [{\citenamefont {Maris}\ \emph {et~al.}(2009)\citenamefont {Maris},
  \citenamefont {Vary},\ and\ \citenamefont {Shirokov}}]{vary2009NCFC}%
  \BibitemOpen
  \bibfield  {author} {\bibinfo {author} {\bibfnamefont {P.}~\bibnamefont
  {Maris}}, \bibinfo {author} {\bibfnamefont {J.~P.}\ \bibnamefont {Vary}}, \
  and\ \bibinfo {author} {\bibfnamefont {A.~M.}\ \bibnamefont {Shirokov}},\
  }\href {\doibase 10.1103/PhysRevC.79.014308} {\bibfield  {journal} {\bibinfo
  {journal} {Phys. Rev. C}\ }\textbf {\bibinfo {volume} {79}},\ \bibinfo
  {pages} {014308} (\bibinfo {year} {2009})}\BibitemShut {NoStop}%
\bibitem [{\citenamefont {de~T\'eramond}\ and\ \citenamefont
  {Brodsky}(2009)}]{teramond2009holography}%
  \BibitemOpen
  \bibfield  {author} {\bibinfo {author} {\bibfnamefont {G.~F.}\ \bibnamefont
  {de~T\'eramond}}\ and\ \bibinfo {author} {\bibfnamefont {S.~J.}\ \bibnamefont
  {Brodsky}},\ }\href {\doibase 10.1103/PhysRevLett.102.081601} {\bibfield
  {journal} {\bibinfo  {journal} {Phys. Rev. Lett.}\ }\textbf {\bibinfo
  {volume} {102}},\ \bibinfo {pages} {081601} (\bibinfo {year}
  {2009})}\BibitemShut {NoStop}%
\bibitem [{\citenamefont {Soper}(1972)}]{soper1972infi_mom_helicity}%
  \BibitemOpen
  \bibfield  {author} {\bibinfo {author} {\bibfnamefont {D.~E.}\ \bibnamefont
  {Soper}},\ }\href {\doibase 10.1103/PhysRevD.5.1956} {\bibfield  {journal}
  {\bibinfo  {journal} {Phys. Rev. D}\ }\textbf {\bibinfo {volume} {5}},\
  \bibinfo {pages} {1956} (\bibinfo {year} {1972})}\BibitemShut {NoStop}%
\bibitem [{\citenamefont {Moshinsky}(1959)}]{Moshinsky:1959qbh}%
  \BibitemOpen
  \bibfield  {author} {\bibinfo {author} {\bibfnamefont {M.}~\bibnamefont
  {Moshinsky}},\ }\href {\doibase 10.1016/0029-5582(59)90143-9} {\bibfield
  {journal} {\bibinfo  {journal} {Nucl. Phys.}\ }\textbf {\bibinfo {volume}
  {13}},\ \bibinfo {pages} {104} (\bibinfo {year} {1959})}\BibitemShut
  {NoStop}%
\bibitem [{\citenamefont {Karmanov}\ \emph {et~al.}(2008)\citenamefont
  {Karmanov}, \citenamefont {Mathiot},\ and\ \citenamefont
  {Smirnov}}]{Karmanov2008systematic}%
  \BibitemOpen
  \bibfield  {author} {\bibinfo {author} {\bibfnamefont {V.}~\bibnamefont
  {Karmanov}}, \bibinfo {author} {\bibfnamefont {J.-F.}\ \bibnamefont
  {Mathiot}}, \ and\ \bibinfo {author} {\bibfnamefont {A.}~\bibnamefont
  {Smirnov}},\ }\href {\doibase 10.1103/PhysRevD.77.085028} {\bibfield
  {journal} {\bibinfo  {journal} {Phys. Rev. D}\ }\textbf {\bibinfo {volume}
  {77}},\ \bibinfo {pages} {085028} (\bibinfo {year} {2008})},\ \Eprint
  {http://arxiv.org/abs/0801.4507} {arXiv:0801.4507 [hep-th]} \BibitemShut
  {NoStop}%
\bibitem [{\citenamefont {Karmanov}\ \emph {et~al.}(2012)\citenamefont
  {Karmanov}, \citenamefont {Mathiot},\ and\ \citenamefont
  {Smirnov}}]{Karmanov2012abinitio}%
  \BibitemOpen
  \bibfield  {author} {\bibinfo {author} {\bibfnamefont {V.}~\bibnamefont
  {Karmanov}}, \bibinfo {author} {\bibfnamefont {J.}~\bibnamefont {Mathiot}}, \
  and\ \bibinfo {author} {\bibfnamefont {A.}~\bibnamefont {Smirnov}},\ }\href
  {\doibase 10.1103/PhysRevD.86.085006} {\bibfield  {journal} {\bibinfo
  {journal} {Phys. Rev. D}\ }\textbf {\bibinfo {volume} {86}},\ \bibinfo
  {pages} {085006} (\bibinfo {year} {2012})},\ \Eprint
  {http://arxiv.org/abs/1204.3257} {arXiv:1204.3257 [hep-th]} \BibitemShut
  {NoStop}%
\bibitem [{\citenamefont {Brodsky}\ \emph {et~al.}(2004)\citenamefont
  {Brodsky}, \citenamefont {Franke}, \citenamefont {Hiller}, \citenamefont
  {McCartor}, \citenamefont {Paston},\ and\ \citenamefont
  {Prokhvatilov}}]{Brodsky2004nonperturbative}%
  \BibitemOpen
  \bibfield  {author} {\bibinfo {author} {\bibfnamefont {S.}~\bibnamefont
  {Brodsky}}, \bibinfo {author} {\bibfnamefont {V.}~\bibnamefont {Franke}},
  \bibinfo {author} {\bibfnamefont {J.}~\bibnamefont {Hiller}}, \bibinfo
  {author} {\bibfnamefont {G.}~\bibnamefont {McCartor}}, \bibinfo {author}
  {\bibfnamefont {S.}~\bibnamefont {Paston}}, \ and\ \bibinfo {author}
  {\bibfnamefont {E.}~\bibnamefont {Prokhvatilov}},\ }\href {\doibase
  10.1016/j.nuclphysb.2004.10.027} {\bibfield  {journal} {\bibinfo  {journal}
  {Nucl. Phys. B}\ }\textbf {\bibinfo {volume} {703}},\ \bibinfo {pages} {333}
  (\bibinfo {year} {2004})},\ \Eprint {http://arxiv.org/abs/hep-ph/0406325}
  {arXiv:hep-ph/0406325} \BibitemShut {NoStop}%
\bibitem [{\citenamefont {Boer}\ and\ \citenamefont
  {Mulders}(1998)}]{boer1998todd}%
  \BibitemOpen
  \bibfield  {author} {\bibinfo {author} {\bibfnamefont {D.}~\bibnamefont
  {Boer}}\ and\ \bibinfo {author} {\bibfnamefont {P.~J.}\ \bibnamefont
  {Mulders}},\ }\href {\doibase 10.1103/PhysRevD.57.5780} {\bibfield  {journal}
  {\bibinfo  {journal} {Phys. Rev. D}\ }\textbf {\bibinfo {volume} {57}},\
  \bibinfo {pages} {5780} (\bibinfo {year} {1998})}\BibitemShut {NoStop}%
\bibitem [{\citenamefont {Sivers}(1990)}]{sivers1990ssa}%
  \BibitemOpen
  \bibfield  {author} {\bibinfo {author} {\bibfnamefont {D.}~\bibnamefont
  {Sivers}},\ }\href {\doibase 10.1103/PhysRevD.41.83} {\bibfield  {journal}
  {\bibinfo  {journal} {Phys. Rev. D}\ }\textbf {\bibinfo {volume} {41}},\
  \bibinfo {pages} {83} (\bibinfo {year} {1990})}\BibitemShut {NoStop}%
\end{thebibliography}
%

\end{document}